\documentclass[prd,twocolumn,superscriptaddress,preprintnumbers,nofootinbib,showpacs,floatfix]{revtex4}
\usepackage{graphicx}
\usepackage{amsmath}
\usepackage{amssymb}
\usepackage{color}
\usepackage{float}
\usepackage[tight]{subfigure}

\newcommand{\be}{\begin{equation}}
\newcommand{\ee}{\end{equation}}
\newcommand{\bea}{\begin{eqnarray}}
\newcommand{\eea}{\end{eqnarray}}
\def\met{\slash{\!\!\!\!E}_T}
\def\afb{A_{FB}^{t\bar t}}

\begin{document}

\preprint{ANL-HEP-PR-13-40, IIT-CAPP-13-05}

\title{LHC and Tevatron constraints on a $W^\prime$ model interpretation of the
top quark forward-backward asymmetry}

\author{Edmond L. Berger}
\email{berger@anl.gov}
\affiliation{High Energy Physics Division, Argonne National Laboratory, 
Argonne, IL 60439, USA}

\author{Zack Sullivan}
\email{Zack.Sullivan@IIT.edu}
\affiliation{Illinois Institute of Technology, Chicago, IL 60616-3793, USA}

\author{Hao Zhang}
\altaffiliation{Hao Zhang has moved to the Department of Physics,
  University of California, Santa Barbara}
\email{zhanghao@physics.ucsb.edu}
\affiliation{Illinois Institute of Technology, Chicago, IL 60616-3793, USA}
\affiliation{High Energy Physics Division, Argonne National Laboratory,
Argonne, IL 60439, USA}

\begin{abstract}
  Aspects of a flavor-changing $W^\prime$ model with right-handed
  couplings are addressed in this paper in light of Tevatron and LHC
  data.  Our fit to the Tevatron top-quark forward-backward asymmetry
  and the $t \bar{t}$ inclusive cross section includes higher-order
  loop effects in the effective interaction.  The higher order
  corrections change the best fit value of the $W^\prime$ effective
  coupling strength as a function of the $W^\prime$ mass.  The
  consistency of the model is checked against the shape of the $t
  \bar{t}$ invariant mass distribution.  We use these updated
  $W^\prime$ parameters to compute the expected contributions from
  $W^\prime t$ associated production and, for the first time,
  $W^\prime W^\prime$ pair production at the LHC.  We do a full Monte
  Carlo simulation of the $t \bar{t} X$ final state, including
  interference between the $tW^\prime$ induced $t\bar t j$ process and
  the standard model $t\bar tj$ process.  Interference
  effects are shown to be quantitatively important, particularly when
  the $W^\prime$ mass is large.  The jet multiplicity distribution in
  $t \bar{t}~\rm{jet}$ production at 8 TeV constrains the $W^\prime$
  model severely.
\end{abstract}


\maketitle

\section{Introduction}
\label{sec: intro}

Searches by the ATLAS and CMS Collaborations
\cite{CMS-PAS-B2G-12-010,ATLAS-CONF-2013-050} have placed significant
limits on the possible masses and coupling strengths of new charged
vector currents which couple to the third generation of quarks,
generically called $W^\prime$ bosons
\cite{Duffty:2012rf,Duffty:2013aba}.  While these measurements have
constrained a wide selection of models that go beyond the standard
model, there is a class of models that escapes the limits by
suppressing all flavor-changing couplings, except between the first
and third generation.  This particular class, in which a right-handed
$W^\prime$ boson couples a down quark to a top quark, has been
proposed
\cite{Cheung:2009ch,Barger:2010mw,Cao:2010zb,Cheung:2011qa,Shelton:2011hq,
  Gresham:2011dg,Barger:2011ih,Bhattacherjee:2011nr,Craig:2011an,Gresham:2011pa,
  Chen:2011mga,Krohn:2011tw,Gresham:2011fx,Cao:2011hr,Yan:2011tf,Berger:2011pu,Knapen:2011hu,Berger:2012nw,
  Duffty:2012zz,Adelman:2012py,Endo:2012mi,Berger:2012tj,Berger:2011xk} as a
possible explanation for anomalous measurements of the
forward-backward asymmetry in $t\bar t$ production ($A_{FB}^{t\bar
  t}$) by the CDF \cite{Aaltonen:2012it} and D0 Collaborations
\cite{Abazov:2011rq}.  In this paper we investigate whether the class
of models with a $W^\prime$--$t$--$d$ coupling strength that is
consistent with the Fermilab Tevatron anomaly can also be consistent
with data from the CERN Large Hadron Collider (LHC).

In a previous publication \cite{Duffty:2012zz} we considered the
leading-order (LO) correction to the forward-backward asymmetry due to
a new term in the Lagrangian of the form
\begin{equation}
  \mathcal{L}=\frac{g}{\sqrt{2}}V^\prime_{td}\bar{d}\gamma^{\mu}P_{R}tW^\prime_{\mu}
  +{\mathrm{h.c.}},\label{eqn:coupl}
\end{equation}
where $g$ is numerically equal to the standard model SU(2)$_{L}$ gauge
interaction coupling constant, and $V^\prime_{td}$ weights the
effective strength of the interaction.  In that paper we used the
first $0.7$~fb$^{-1}$ of data collected at 7~TeV by the ATLAS
Collaboration \cite{ATLAS-CONF-2011-100} to conclude $tW^\prime$ production
with a decay to $t\bar tj$ could be used to exclude much of the
interesting parameter space, and that with 5~fb$^{-1}$ of data the
entire parameter space might be excluded.  This conclusion was subject
to the caveat that the relevant parameter space was only determined at
leading order.

Both the ATLAS and CMS Collaborations reproduced our initial analysis
and published exclusion limits~\cite{Chatrchyan:2012su,Aad:2012em}.
However, there are large interference effects between $tW^\prime$
production and $t\bar tj$ that were not considered in the experimental
analyses.  The relevance of these effects is increased by the large
couplings necessary to explain the Tevatron anomaly,
$g_{\mathrm{eff}}=gV^\prime_{td} \sim 1$.  Large coupling leads to a
large width of the $W^\prime$ boson, and changes the observable signal
at the LHC.

In this paper we significantly improve our calculation of the relevant
parameter space for the class of models that satisfies the $t\bar t$
forward-backward asymmetry $A_{FB}^{t\bar t}$ measured by the CDF
Collaboration~ \cite{Aaltonen:2012it} and the $t\bar t$ inclusive
cross section.  In Sec.\ \ref{sec:nlo} we derive the contribution to
$A_{FB}^{t\bar t}$ at next-to-leading order (NLO) from $W^\prime$
bosons.  In Sec.\ \ref{sec:tev} we show that the range of effective
couplings $g_{\mathrm{eff}}$ changes from LO to NLO.  In Sec.\
\ref{sec:lhc} we discuss the contribution of $W^\prime$ bosons to
$t\bar t+\mathrm{n}j$ at the LHC, including full interference
effects, as well as the contribution of $W^\prime W^\prime$ production
and decay.  We show that a 20~fb$^{-1}$ measurement of $t\bar
t+\mathrm{n}j$ by the CMS Collaboration at 8~TeV excludes the region
of couplings $g_{\mathrm{eff}}$ consistent with the Tevatron anomaly.
We summarize our results in Sec.\ \ref{sec:concl}.  Within the mass
range $200 < m_{W^\prime} < 1100$~GeV, values of the coupling strength
$V^\prime_{td}$ large enough to accommodate $\afb$ observed at the Tevatron
are incompatible with a good fit to the multiplicity distribution at
the LHC.

Before proceeding, we comment briefly on indirect constraints on this
$W^\prime$ model from other than the collider observables we address
here.  A right-handed $W^\prime$ may be constrained by the ratio of
rare $B$ decays at the $2\sigma$ level \cite{Chen:2011mga}.  However,
the reach in these measurements is limited by theoretical uncertainty
in the matrix elements for $B$ decays \cite{Li:2005kt}.  While
additional constraints on low-mass $W^\prime$ bosons may be derived
from atomic parity violation \cite{Gresham:2012wc}, the direct
production limit we present from collider data is needed to exclude
this right-handed $W^\prime$ model.

\section{Tevatron physics}
In this section, we consider the influence of the $W^\prime$ model on
the $t\bar t$ inclusive total cross section and on the $t\bar t$
forward-backward asymmetry $A_{FB}^{t\bar t}$ at the Tevatron.  We fit
data on the cross section and $A_{FB}^{t\bar t}$ and determine the
best fit region of the parameters $(m_{W^\prime},g_R)$.  Consistency
with data on the $t\bar t$ invariant mass distribution is then
checked.
 
\subsection{Calculation of $\sigma_{t\bar t}$ and $A_{FB}^{t\bar t}$}
\label{sec:nlo}
\begin{figure*}[tp]
\includegraphics[scale=0.39,clip]{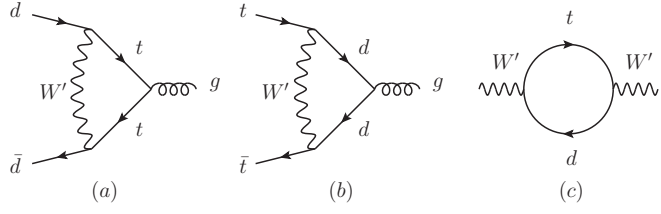}
\caption{(a) and (b) illustrate the $W^\prime$ loop correction to the
$d \bar{d} g$ vertex.  (c) shows the $t \bar{t}$ contribution to the
$W^\prime$ width. 
\label{fig:feyngraphs}}
\end{figure*}

Previous work~\cite{Duffty:2012zz,Berger:2012tj} shows that the best
fits to the Tevatron asymmetry $A_{FB}^{t\bar t}$ and the inclusive
cross section yield generally large values of the effective coupling
strength $g_R$, especially for heavy $W^\prime$ bosons which are not
excluded by direct observation.  Thus the
$\mathcal{O}\left(\alpha_R\right)$ ($\alpha_R\equiv g^2_R
V_{td}^{\prime 2}/\left(4\pi\right)$) effects might not be negligible.  We
discuss two places where $\mathcal{O}\left(\alpha_R\right)$ effects
play a role.  The first is the loop correction to the QCD vertex
$q\bar qg$, illustrated in Fig.\ \ref{fig:feyngraphs}.  We can express
the renormalized QCD vertex as
\begin{align}
-ig_sT^a_{ij}\biggl[&\gamma^\mu\left(1+\frac{\alpha_R}{4\pi}F_V^f\right)
+\gamma^\mu\gamma_5 \frac{\alpha_R}{4\pi}G_A^f\nonumber\\
&+\frac{\left(\bar p-p\right)^\mu}{2m_q}\frac{\alpha_R}{4\pi}F_M^f
+\left(\bar p+p\right)^\mu\gamma_5
\frac{\alpha_R}{4\pi}G_E^f\biggr],
\end{align}
where $p$ ($\bar p$) is the momentum of the quark (antiquark), and $f$
is the flavor index. The coefficients are non-zero for $f=t,d$.
Analytic results for $F_V^f, G_A^f$ and $F_M^f$ can be found in
Ref.~\cite{Beenakker:1993yr}.  Corrections to the total cross section
which are proportional to $G_E^f$ are all of order
$\mathcal{O}\left(\alpha_R^4\alpha_S^2\right)$ and highly
suppressed. They do not contribute to $\sigma_F-\sigma_B$.  Thus, we
will not consider them in this work.

The additional contribution to $A_{FB}^{t\bar t}$
is 
\begin{eqnarray}
\sigma_F-\sigma_B&=&\frac{\alpha_R^2\alpha_S^2\beta^2{\mathrm{Re}}\left(G_A^d\right){\mathrm{Re}}\left(G_A^t\right)}{18\pi s}
\nonumber\\
&&
+\mathcal{O}\left(\alpha_R^3\alpha_S^2\right),
\end{eqnarray}
where $\beta\equiv\sqrt{1-4m_t^2/s}$. The contribution to
$A_{FB}^{t\bar t}$ from the QCD vertex correction has been
investigated in Ref.~\cite{Gabrielli:2011jf}.

\begin{figure}[!htb]
\includegraphics[scale=0.4,clip]{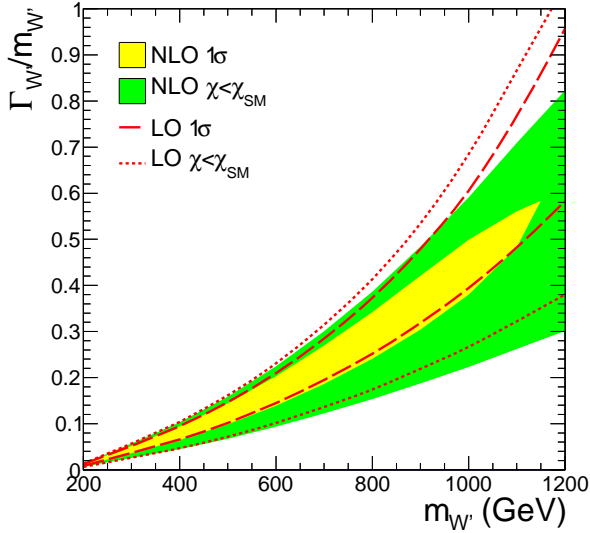}
\caption{The ratio of the width and the mass of the $W^\prime$ boson
determined from the parameters of our best fits at the Tevatron.   The values
of $\chi^2$ in the light-shaded (yellow) region are not greater than 1.  In
the dark shaded (green) region they are not greater than the standard model
(SM) value of $\chi^2$.   For comparison, we show LO results when the
$\mathcal{O}\left( \alpha_R\right)$ contributions are not included:  Between
the (red) dashed lines, $\chi^2$ is not greater than 1, and between the (red)
dotted line $\chi^2$ is not greater than its SM value. 
\label{fig:width} }
\end{figure}

The decay width of the $W^\prime$ is another place where
$\mathcal{O}\left(\alpha_R\right)$ effects are important for the LHC
phenomenology of the $W^\prime$ model.  The width is
\begin{eqnarray}
\Gamma_{W^\prime}=\frac{\alpha_Rm_{W^\prime}}{4}\left(1-r\right)\left(1+\frac{r}{2}\right)=\alpha_R\gamma\sim\mathcal{O}\left(\alpha_R\right),
\end{eqnarray}
where $r\equiv m_t^2/m_{W^\prime}^2$.  A numerical evaluation is shown
in Fig.\ \ref{fig:width}.

The standard model (SM) and new physics (NP) amplitudes are 
\begin{equation}
\mathcal{M}_{SM}\equiv \frac{\mathcal{M}_{SM}^{(0)}+\alpha_R \mathcal{M}_{SM}^{(1)}}{s},
\end{equation}
and
\begin{equation}
\mathcal{M}_{NP}\equiv \frac{\alpha_R\mathcal{M}_{NP}^{(1)}}{t-m_{W^\prime}^2+i\Gamma_{W^\prime} m_{W^\prime}} .
\end{equation}
To $\mathcal{O}\left(\alpha_R^2\right)$ in the numerator, the interference term is 
\begin{eqnarray}
2\mathrm{Re}\left(\mathcal{M}_{NP}\mathcal{M}_{SM}^*\right)&=&\frac{2\alpha_R\mathcal{M}_{NP}^{(1)}
\left(t-m_{W^\prime}^2\right)}
{s\left[\left(t-m_{W^\prime}^2\right)^2
+\alpha_R^2\gamma^2\right]}\nonumber\\
&&\times{\mathrm{Re}}\left(\mathcal{M}_{SM}^{(0)}
+\alpha_R \mathcal{M}_{SM}^{(1)}\right).
\end{eqnarray}

For the new physics term $\mathcal{M}_{NP}^*\mathcal{M}_{NP}$, it
suffices to replace $\left(t-m_{W^\prime}^2\right)$ with
$\left(t-m_{W^\prime}^2+i\Gamma_{W^\prime}m_{W^\prime}\right)$ to
include the finite width effect.

After including the $\mathcal{O}\left(\alpha_R\right)$ correction to
the QCD vertex, the non-zero helicity amplitudes can be written as
$\mathcal{M}\left(\lambda_q,\lambda_{\bar q}, \lambda_t,\lambda_{\bar
    t}\right)=4\pi\alpha_St^a_{c_3c_4}t^a_{c_2c_1}\mathcal{M}_{SM}^{\left(\lambda_q,\lambda_{\bar
      q}, \lambda_t,\lambda_{\bar
      t}\right)}+4\pi\alpha_R\delta_{c_3c_1}\delta_{c_4c_2}\mathcal{M}_{NP}^{\left(\lambda_q,\lambda_{\bar
      q}, \lambda_t,\lambda_{\bar t}\right)}$, where
\begin{eqnarray}
\mathcal{M}_{SM}^{\left(+-++\right)}&=&-
\biggl[1+\frac{\alpha_R}{4\pi}\left(F_V^t+F_V^q+G_A^q
+\frac{\beta^2F_M^t}{1-\beta^2}\right)\nonumber\\&&
+\left(\frac{\alpha_R}{4\pi}\right)^2\left(F_V^q+G_A^q\right)
\left(F_V^t
+\frac{\beta^2F_M^t}{1-\beta^2}\right)\biggr]\nonumber\\&&\times\sqrt{1-\beta^2}\sin\theta,
\end{eqnarray}
\begin{eqnarray}
\mathcal{M}_{SM}^{\left(-+++\right)}&=&-
\biggl[1+\frac{\alpha_R}{4\pi}
\left(F_V^t+F_V^q-G_A^q
+\frac{\beta^2F_M^t}{1-\beta^2}\right)\nonumber\\&&
+\left(\frac{\alpha_R}{4\pi}\right)^2\left(F_V^q-G_A^q\right)
\left(F_V^t
+\frac{\beta^2F_M^t}{1-\beta^2}\right)\biggr]\nonumber\\&&\times\sqrt{1-\beta^2}\sin\theta,
\end{eqnarray}
\begin{eqnarray}
\mathcal{M}_{SM}^{\left(+-+-\right)}&=&-
\biggl[1+\frac{\alpha_R}{4\pi}
\left(F_V^t+F_V^q+G_A^q
+\beta G_A^t\right)\nonumber\\&&
+\left(\frac{\alpha_R}{4\pi}\right)^2\left(F_V^q+G_A^q\right)
\left(F_V^t
+\beta G_A^t\right)\biggr]\nonumber\\&&\times\left(1+\cos\theta\right),
\end{eqnarray}
\begin{eqnarray}
\mathcal{M}_{SM}^{\left(-++-\right)}&=&
\biggl[1+\frac{\alpha_R}{4\pi}
\left(F_V^t+F_V^q-G_A^q
+\beta G_A^t\right)\nonumber\\&&
+\left(\frac{\alpha_R}{4\pi}\right)^2\left(F_V^q-G_A^q\right)
\left(F_V^t
+\beta G_A^t\right)\biggr]\nonumber\\&&\times\left(1-\cos\theta\right),
\end{eqnarray}
\begin{eqnarray}
\mathcal{M}_{SM}^{\left(+--+\right)}&=&
\biggl[1+\frac{\alpha_R}{4\pi}
\left(F_V^t+F_V^q+G_A^q
-\beta G_A^t\right)\nonumber\\&&
+\left(\frac{\alpha_R}{4\pi}\right)^2\left(F_V^q+G_A^q\right)
\left(F_V^t
-\beta G_A^t\right)\biggr]\nonumber\\&&\times\left(1-\cos\theta\right),
\end{eqnarray}
\begin{eqnarray}
\mathcal{M}_{SM}^{\left(-+-+\right)}&=&-
\biggl[1+\frac{\alpha_R}{4\pi}
\left(F_V^t+F_V^q-G_A^q
-\beta G_A^t\right)\nonumber\\&&
+\left(\frac{\alpha_R}{4\pi}\right)^2\left(F_V^q-G_A^q\right)
\left(F_V^t
-\beta G_A^t\right)\biggr]\nonumber\\&&\times\left(1+\cos\theta\right),
\end{eqnarray}
\begin{eqnarray}
\mathcal{M}_{SM}^{\left(+---\right)}&=&
\biggl[1+\frac{\alpha_R}{4\pi}
\left(F_V^t+F_V^q+G_A^q
+\frac{\beta^2F_M^t}{1-\beta^2}\right)\nonumber\\&&
+\left(\frac{\alpha_R}{4\pi}\right)^2\left(F_V^q+G_A^q\right)
\left(F_V^t
+\frac{\beta^2F_M^t}{1-\beta^2}\right)\biggr]\nonumber\\&&\times\sqrt{1-\beta^2}\sin\theta,
\end{eqnarray}
\begin{eqnarray}
\mathcal{M}_{SM}^{\left(+---\right)}&=&
\biggl[1+\frac{\alpha_R}{4\pi}
\left(F_V^t+F_V^q-G_A^q
+\frac{\beta^2F_M^t}{1-\beta^2}\right)\nonumber\\&&
+\left(\frac{\alpha_R}{4\pi}\right)^2\left(F_V^q-G_A^q\right)
\left(F_V^t
+\frac{\beta^2F_M^t}{1-\beta^2}\right)\biggr]\nonumber\\&&\times\sqrt{1-\beta^2}\sin\theta .
\end{eqnarray}

The symbol $\theta$ denotes the angle between the 3-momentum of the
initial state quark and the final state top quark in the
center-of-mass frame.  Explicit expressions for the new physics
amplitudes are
 
\begin{eqnarray}
\mathcal{M}_{NP}^{\left(+-++\right)}&=&
\frac{\left(1-\beta^2+8r_{W^\prime}\right)\sqrt{1-\beta^2}\sin\theta}
{8r_{W^\prime}\left(1+\beta^2-2\beta\cos\theta+4r_{W^\prime}\right)},\\
\mathcal{M}_{NP}^{\left(+-+-\right)}&=&
\frac{\left[\left(1-\beta\right)^2+8r_{W^\prime}\right]
\left(1+\beta\right)}{8r_{W^\prime}\left(1+\beta^2-2\beta\cos\theta+4r_{W^\prime}\right)}\nonumber\\&&\times
\left(1+\cos\theta\right),\\
\mathcal{M}_{NP}^{\left(+--+\right)}&=&-
\frac{\left[\left(1+\beta\right)^2+8r_{W^\prime}\right]
\left(1-\beta\right)}{8r_{W^\prime}\left(1+\beta^2-2\beta\cos\theta+4r_{W^\prime}\right)}
\nonumber\\&&\times\left(1-\cos\theta\right),\\
\mathcal{M}_{NP}^{\left(+---\right)}&=&-
\frac{\left(1-\beta^2+8r_{W^\prime}\right)\sqrt{1-\beta^2}\sin\theta}
{8r_{W^\prime}\left(1+\beta^2-2\beta\cos\theta+4r_{W^\prime}\right)},
\end{eqnarray} 
where $r_{W^\prime} \equiv m_{W^\prime}^2/s$. After integration over the azimuthal angle, the cross section can be written as 
\begin{equation}
\frac{d\sigma}{d\cos\theta}=\frac{\beta}{32\pi s}\left(\frac{1}{2}\times\frac{1}{2}\times\frac{1}{3}
\times\frac{1}{3}\right)\left|\mathcal{M}\right|^2.
\end{equation}

We evaluate $\sigma_{t\bar t}$ and $A_{FB}^{t\bar t}$ using our
analytic results for the squared amplitudes and the MSTW2008 parton
distribution functions \cite{Martin:2009iq}.  To include the NLO QCD
and NNLO QCD contribution to $\sigma_{t\bar t}$ in the SM, and the NLO
QCD SM contribution to $A_{FB}^{t\bar t}$, we remove the
$\mathcal{O}\left(\alpha_S^2\alpha_R^0\right)$ portion of our result
and substitute the NNLO QCD SM contribution for $\sigma_{t\bar t}$ and
the NLO QCD term for $A_{FB}^{t\bar t}$. A complete NLO QCD
calculation of this process is presented in \cite{Yan:2011tf}.

\subsection{Fit to the Tevatron asymmetry data}
\label{sec:tev}

\begin{figure*}[tbh]
\subfigure[]{\includegraphics[scale=0.29,clip]{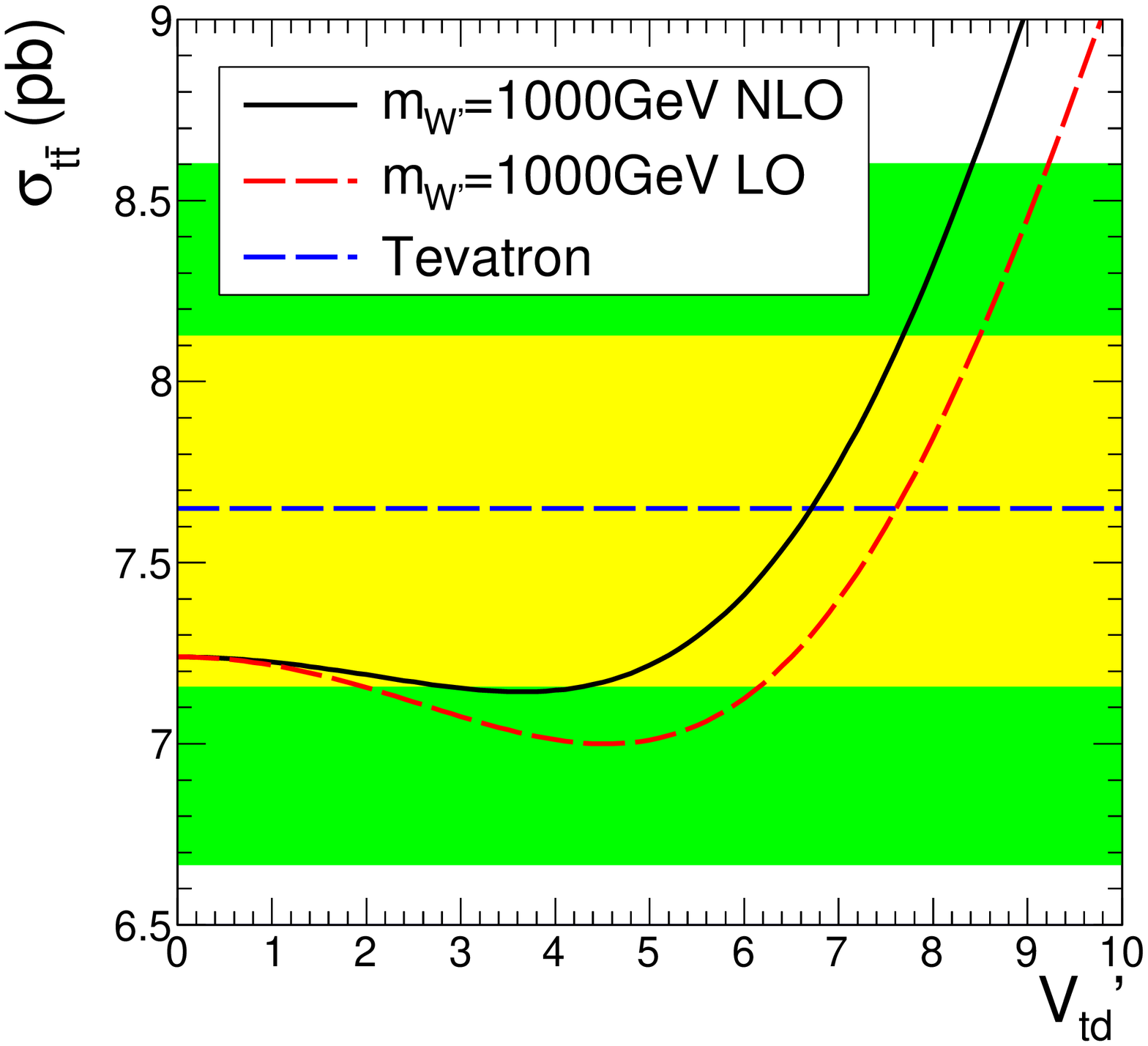}}
\subfigure[]{\includegraphics[scale=0.29,clip]{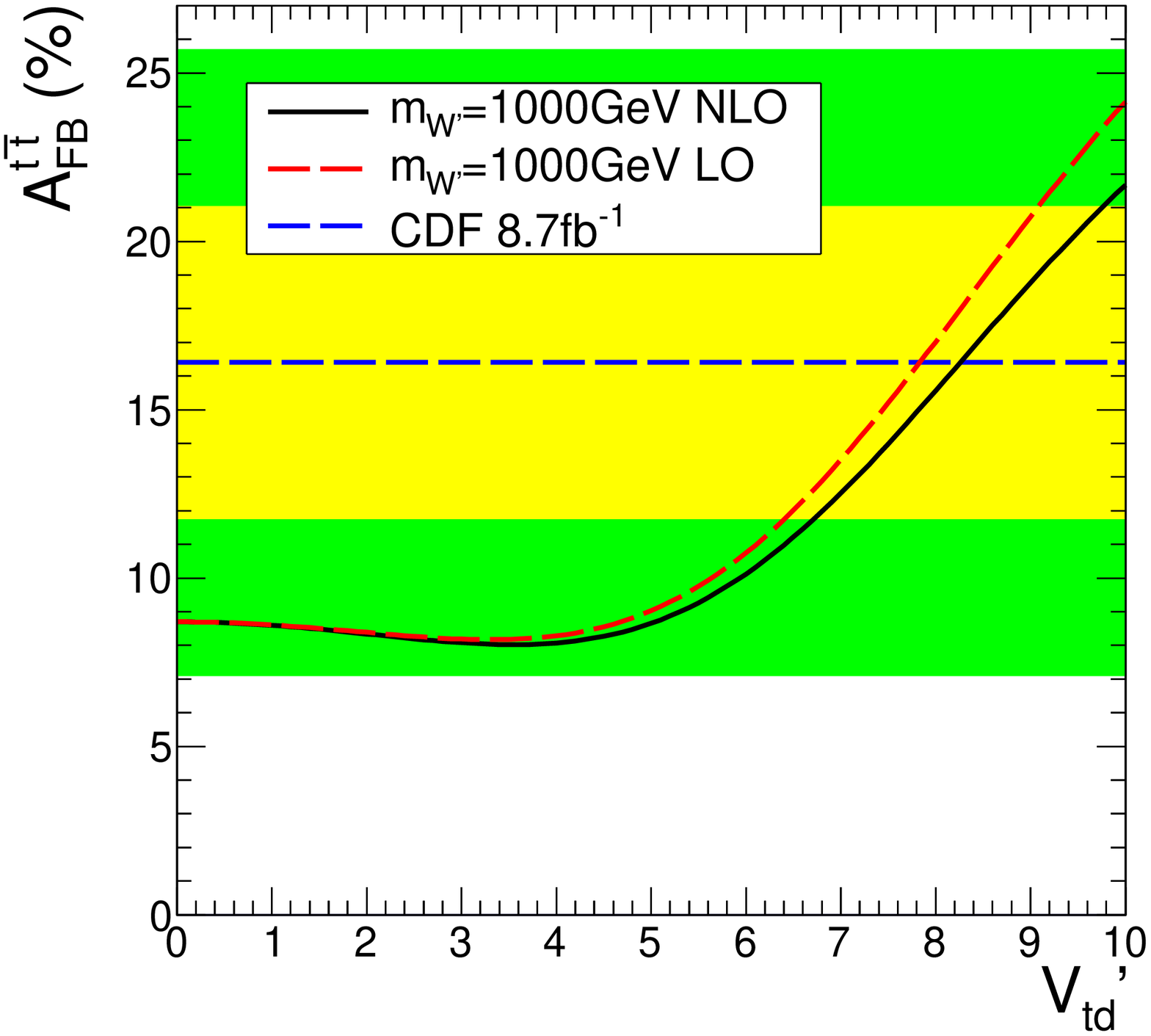}}
\subfigure[]{\includegraphics[scale=0.29,clip]{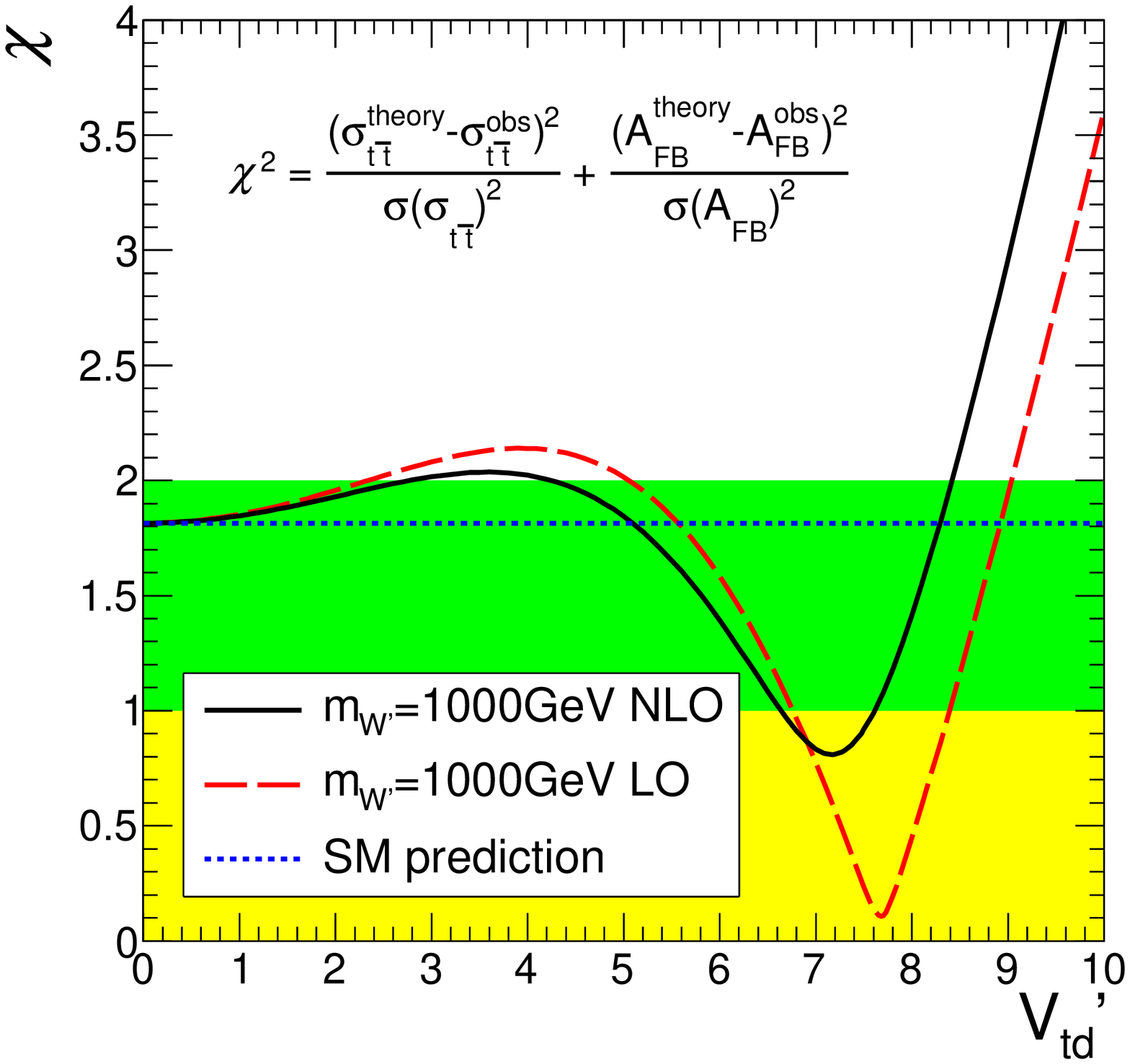}}
\caption{Results of our fit to (a) the inclusive total cross section
$\sigma_{t\bar t}$, and (b) the forward-backward asymmetry $\afb$.
In (c) we show the values of $\chi$ from our fit for a 1000 GeV
$W^\prime$ boson.  The light-shading (yellow) shows the 1$\sigma$
region from our combined fits.  The dark-shading (green) shows the 2$\sigma$
region.  The (black) solid line is the result with NLO vertex
corrections included, while the (red) dashed line is the result
without the NLO vertex corrections.  The horizontal (blue) dashed lines in
(a) and (b) denote the central values of the Tevatron data.  The
(blue) dotted line in (c) shows the value of $\chi$ for the SM.
\label{fig:1000gevtev} }
\end{figure*}

Among the top quark observables at the Tevatron affected by the
$W^\prime$ model contributions, we choose to determine our parameters
from data on the inclusive cross section $\sigma_{t\bar t}$ and the
asymmetry $A_{FB}^{t\bar t}$.  We use the latest measurement of
$\sigma_{t\bar t}$ at the Tevatron \cite{D0-NOTE-6363}:
\be
\sigma_{t\bar t}=7.65\pm0.2~({\mathrm{stat.}})\pm0.36~({\mathrm{syst.}})~{\mathrm{pb}} .
\ee
The corresponding (partial) NNLO SM QCD result is
$7.24^{+0.24}_{-0.27}~{\mathrm{pb}}$, whereas our
$\mathcal{O}\left(\alpha_S^2\alpha_R^0\right)$ result is $6.64$ pb.
The latest measurement of the asymmetry from the CDF collaboration is
$A_{FB}^{t\bar t} = (16.4\pm 4.7)\%$ \cite{Aaltonen:2012it} while the
SM prediction (QCD+EW) is $(8.7\pm 1.0)\%$ \cite{Kuhn:2011ri}.  For
the calculation of $\chi^2$, we combine all these uncertainties
treating them as uncorrelated.

We use the result for a 1000 GeV $W^\prime$ as an example to show the
effect of the vertex correction most clearly.  As shown in Fig.\
\ref{fig:1000gevtev}(a), the vertex correction increases the
predicted total cross section, making the best fit value of $V_{td}^\prime$
smaller than in the LO fits.  The definition of $\afb$
\be
\afb=\frac{\sigma_F-\sigma_B}{\sigma_F+\sigma_B}=\frac{\Delta\sigma}{\sigma_{\mathrm{tot}}} ,
\ee
shows that the corrections of $\Delta\sigma$ and $\sigma_{\mathrm{tot}}$
both contribute to the correction of $\afb$. We have
\bea
\afb\left(NLO\right)&=&\frac{\Delta\sigma+\delta\Delta\sigma}{\sigma_{\mathrm{tot}}
+\delta\sigma_{\mathrm{tot}}}\nonumber\\&
\simeq&\afb\left(1+\frac{\delta\Delta\sigma}{\Delta\sigma}\right)
\left(1-\frac{\delta\sigma_{\mathrm{tot}}}{\sigma_{\mathrm{tot}}}\right).
\eea
In the $W^\prime$ model, $\delta\Delta\sigma$ is of $\mathcal{O}
\left(\alpha_R^2\alpha_S^2\right)$ which is tiny, and we see that
$\delta\sigma_{\mathrm{tot}}$ is significant from
Fig.~\ref{fig:1000gevtev}(a). Thus the NLO $\afb$ is smaller than the
LO prediction (Fig.~\ref{fig:1000gevtev}(b)).

\begin{figure*}[tbhp]
\subfigure[]{\includegraphics[scale=0.39,clip]{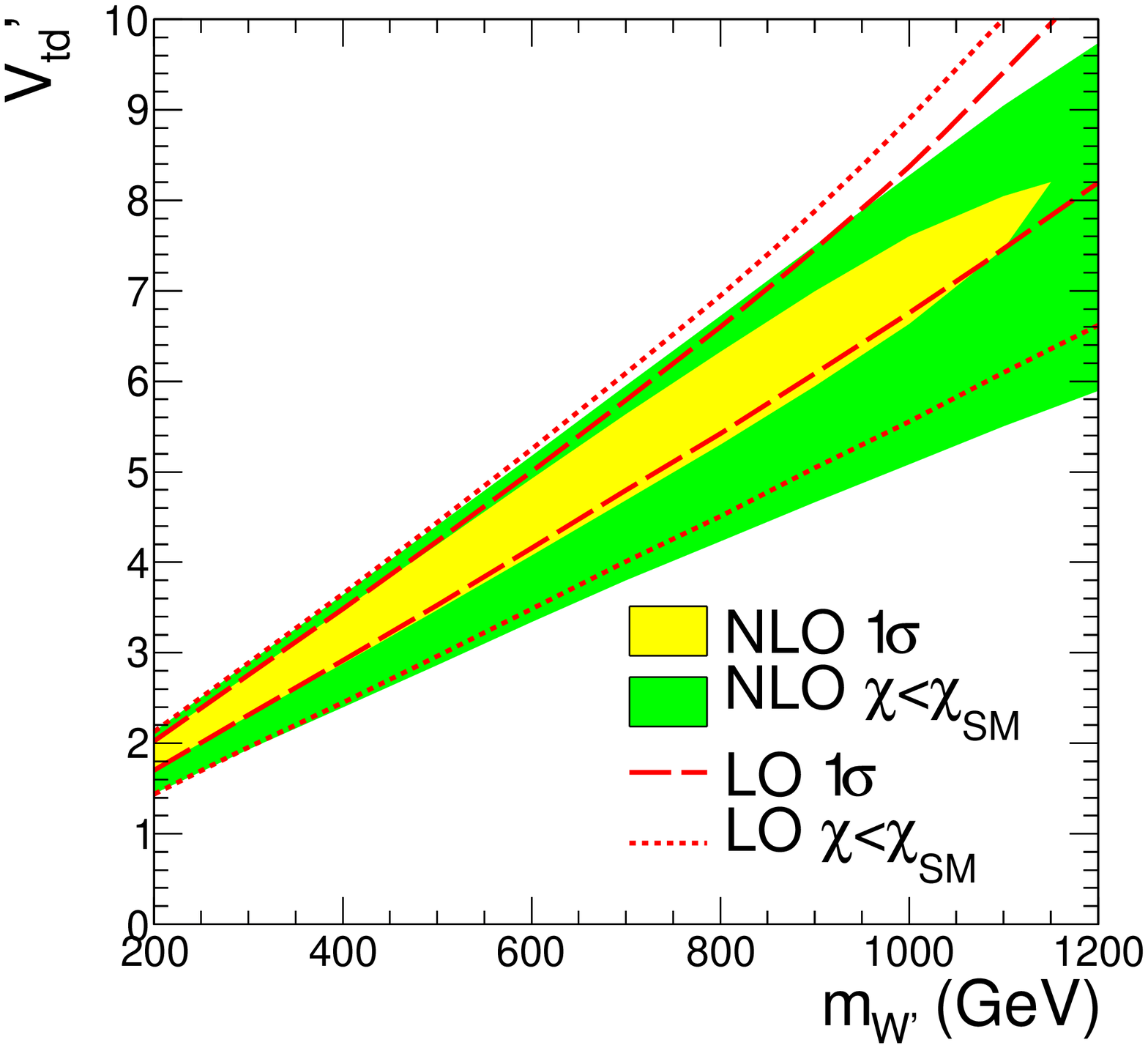}}
\subfigure[]{\includegraphics[scale=0.39,clip]{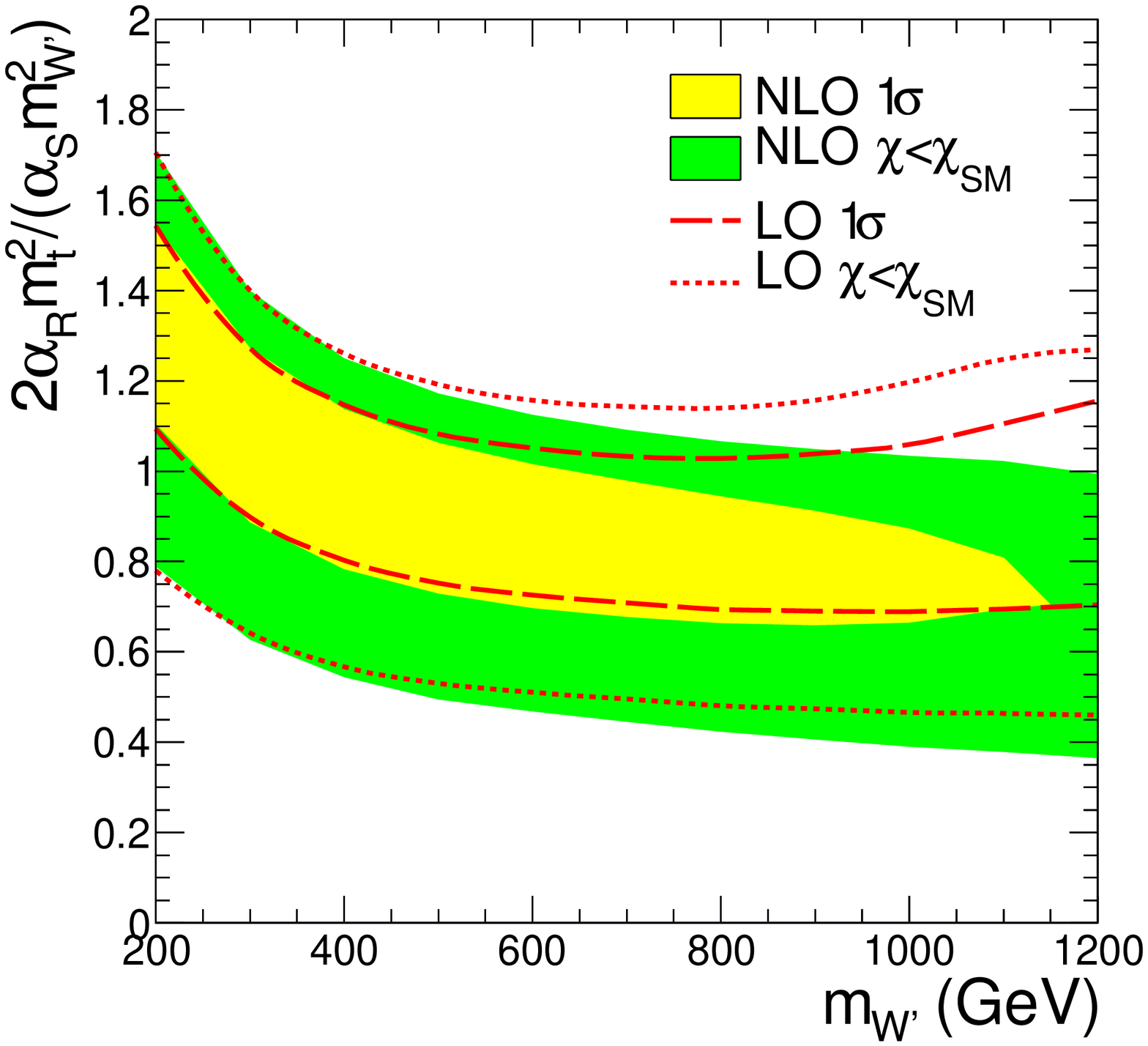}}
\caption{(a) The parameter $V^\prime_{td}$ obtained from our fits to
$\sigma_{t\bar t}$ and $A_{FB}^t$ is shown as a function of the $W^\prime$
mass.  (b) The coefficient $2\alpha_Rm_t^2/
\left(\alpha_Sm_{W^\prime}^2\right)$ obtained from our fits is shown as a
function of the  $W^\prime$ mass.   Values of $\chi^2$ in the light-shaded 
(yellow) region remain less than 1.   Values of $\chi^2$ in the dark-shaded
(green) region are not greater than the SM $\chi^2$.   For comparison, we
show the results without NLO $\mathcal{O}\left(\alpha_R\right)$ contributions:
In the region between the (red) dashed lines $\chi^2$ is not greater than 1,
and in the region between the (red) dotted lines $\chi^2$ is not 
greater than the SM $\chi^2$. 
\label{fig:chi2_tev} }
\end{figure*}

The values of $\chi$ from the combined fit to $\sigma_{t\bar t}$ and
$\afb$ are shown in Fig.\ \ref{fig:1000gevtev}(c) for $m_{W^\prime} =
1$~TeV.  Results for other values of the $W^\prime$ mass are
qualitatively similar.  In Fig.\ \ref{fig:chi2_tev}(a) we plot the
allowed region $V^\prime_{td}$ as a function of the $W^\prime$ mass.

\subsection{The $t\bar t$ mass distribution at the Tevatron}

The distribution in the ${t\bar t}$ invariant mass at the Tevatron
provides a potentially strong constraint on the $W^\prime$ model
because the prediction of the $W^\prime$ model at high $m_{t\bar t}$
(last few bins of data) is much higher than the data
\cite{Aaltonen:2009iz}.  However, in Ref.\ \cite{Gresham:2011pa}, the
authors argue that it is not accurate to compare with the unfolded
experimental result because there is a non-negligible difference
between the cut acceptance in the $W^\prime$ model and the SM.  This
difference can reduce the tension between the $W^\prime$ model and
data on the $m_{t\bar t}$ distribution.

In this work, we examine the consistency of our expectations with data
on the distribution in $m_{t\bar t}$.  We consider both the absolute
cross section $d\sigma/d m_{t\bar t}$ and the mass distribution
normalized by the integrated cross section.  This latter {\em shape}
distribution is arguably more pertinent because our parameters,
determined from fits to data on the integrated cross section, already
include information on the integrated cross section.  We select two
values of the $W^\prime$ mass and use the parameters from our best fit
to compute the $m_{t\bar t}$ distribution.  One value is a light
$W^\prime$ ($m_{W^\prime}=500$ GeV, $g_R=3.8$), and the other is a
heavy $W^\prime$ ($m_{W^\prime}=1000$ GeV, $g_R=7.0$).

\begin{figure*}[!tbhp]
\includegraphics[scale=0.35,clip]{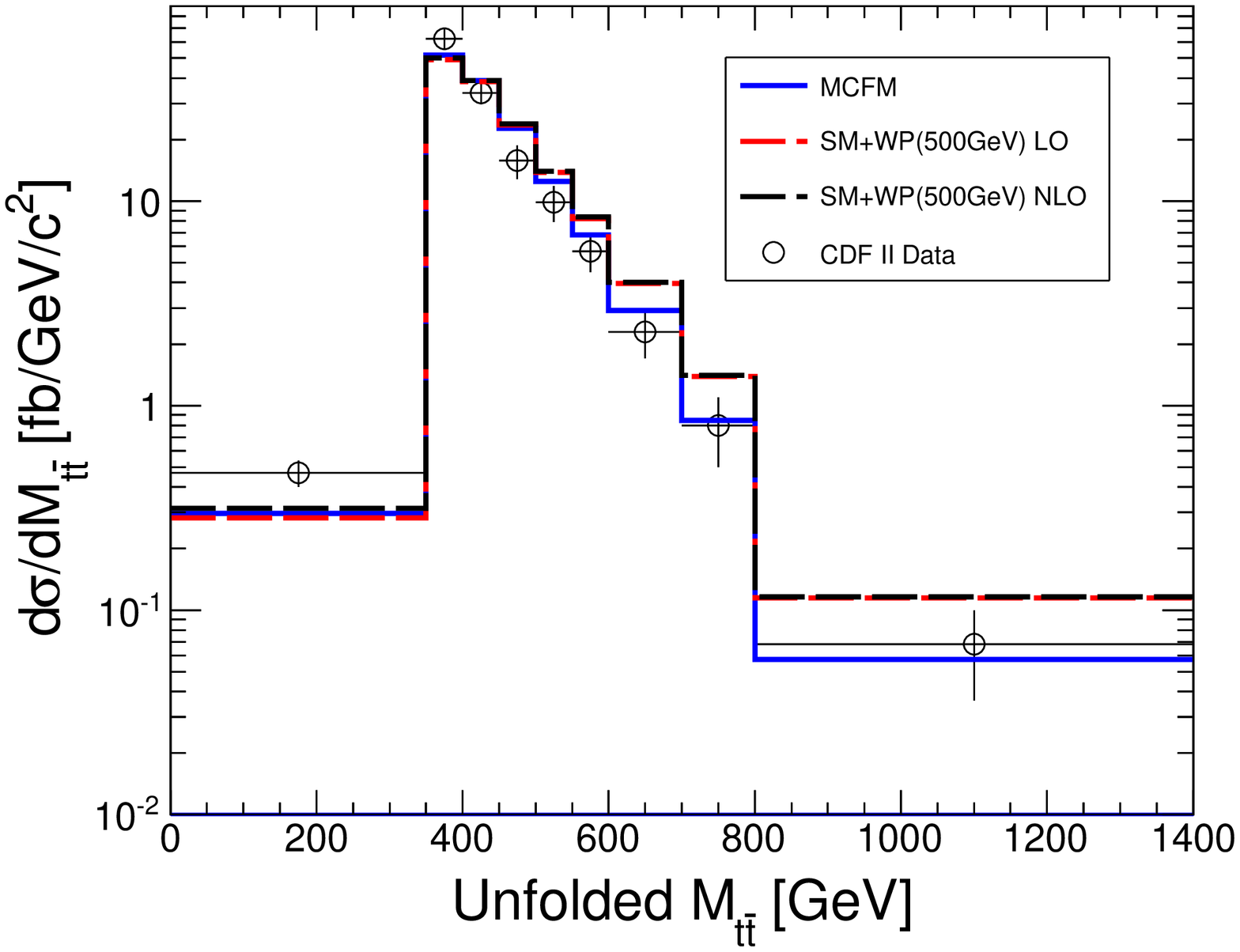}~~~~~~~~~~~
\includegraphics[scale=0.35,clip]{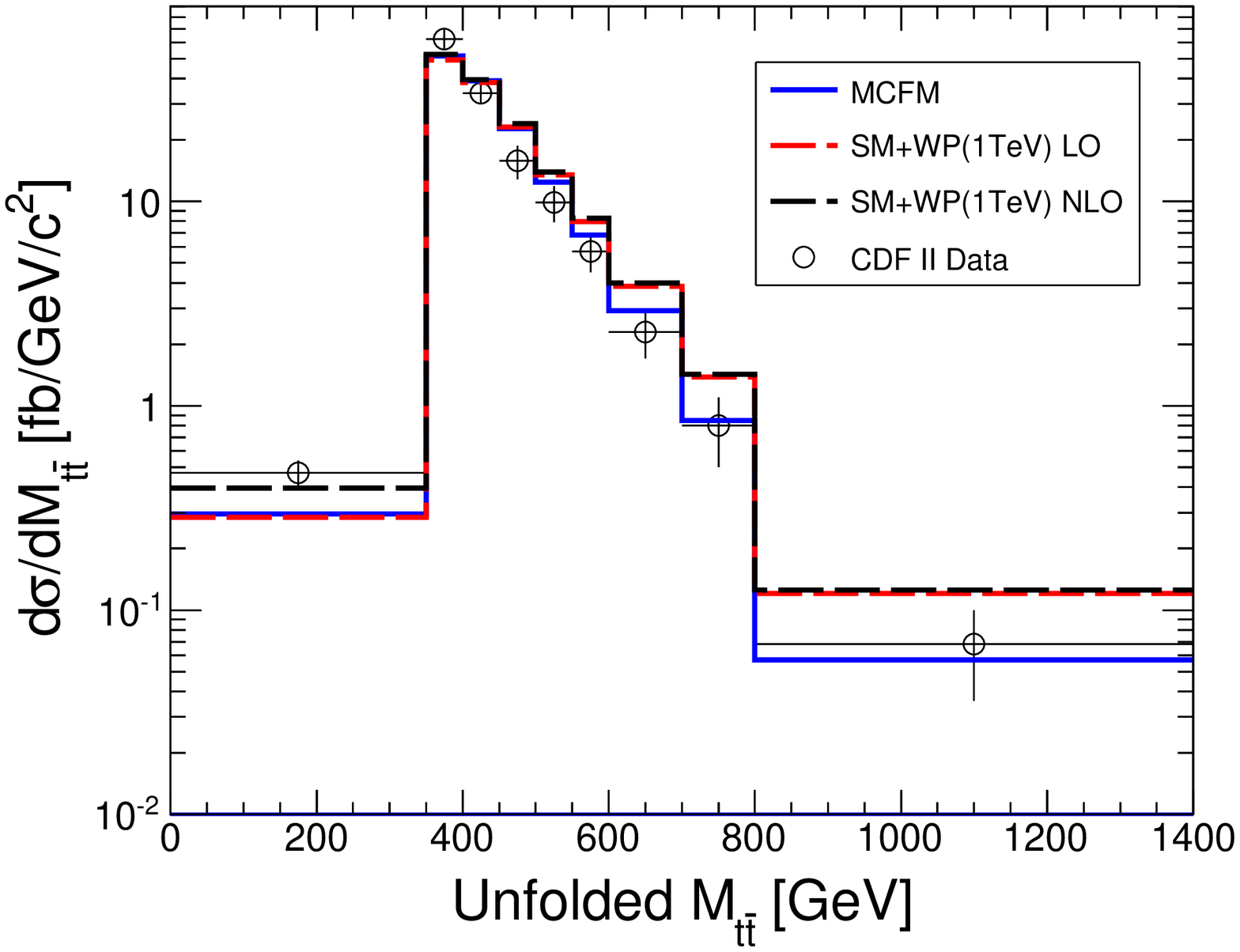}\\
\includegraphics[scale=0.35,clip]{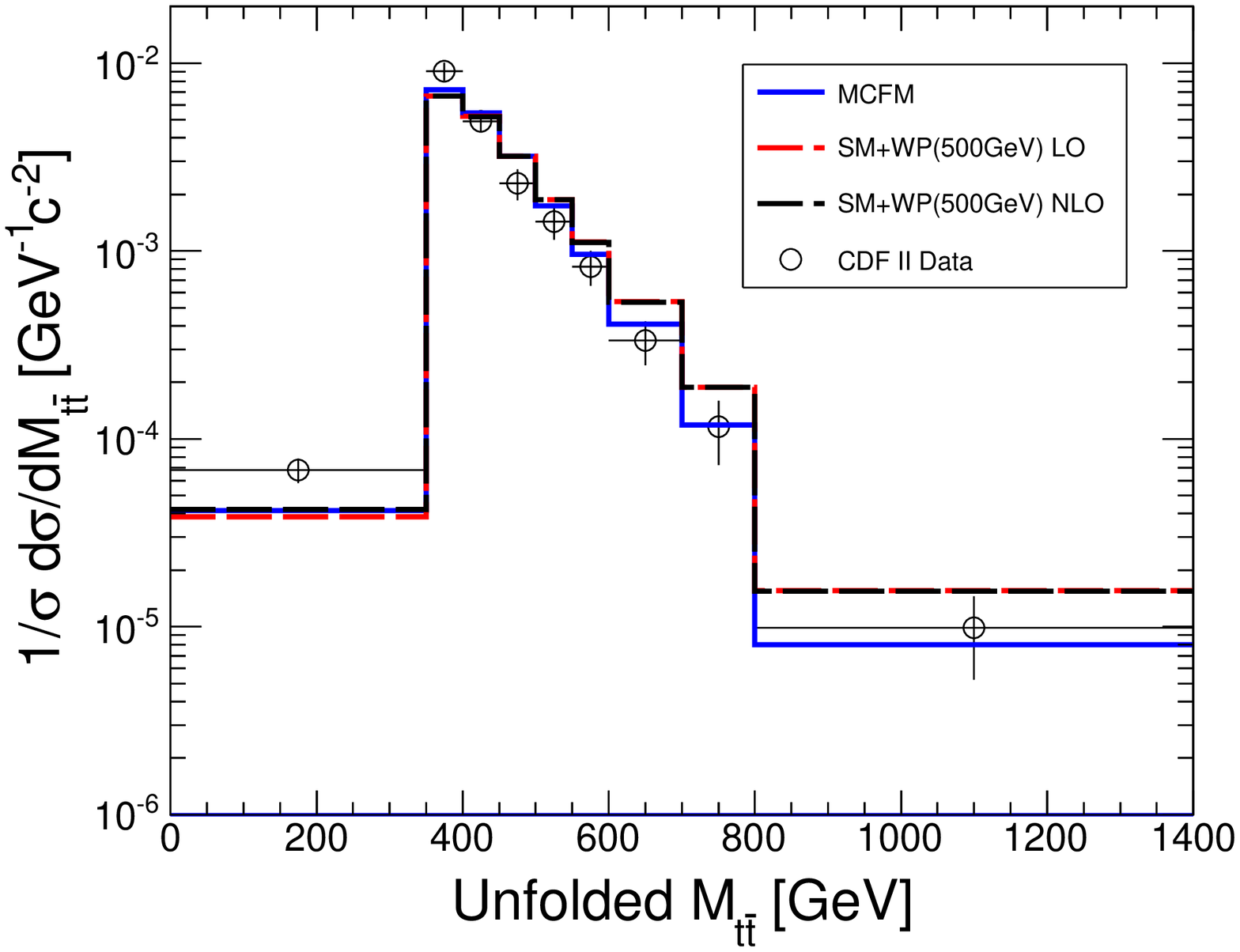}~~~~~~~~~~~
\includegraphics[scale=0.35,clip]{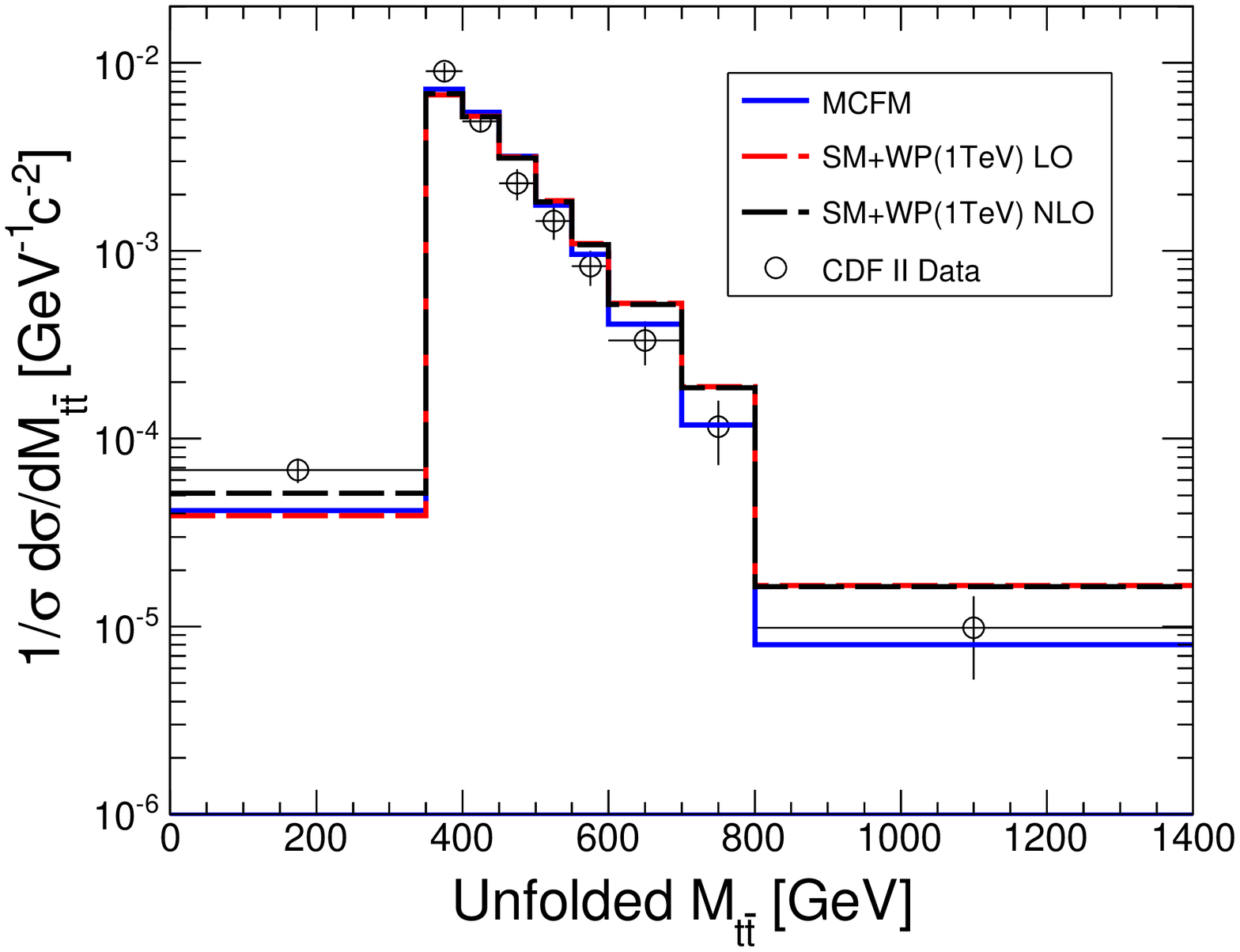}
\caption{The $m_{t\bar t}$ distribution (upper two panels) and the 
normalized $m_{t\bar t}$ distribution (lower two panels) from 
the SM and the $W^\prime$ model. {\it{Left}}:
($m_{W^\prime}=500$ GeV, $g_R=3.8$); {\it{Right}}:
($m_{W^\prime}=1000$ GeV, $g_R=7.0$).  The (blue) solid line is the SM NLO
QCD prediction obtained from the \textsc{MCFM} \protect{\cite{mcfm}} code. 
The lighter dashed (red) line shows the $W^\prime$ model prediction without
the NLO vertex correction, and the darker dashed (black) line shows 
the $W^\prime$ model prediction with the NLO vertex correction included.  The
pure SM part of the $W^\prime$ model prediction is corrected to the NLO
QCD level.  The circles denote the Tevatron data along with their
uncertainties.
\label{fig:mttunfold} }
\end{figure*}

First, we compare the theoretical prediction with the unfolded
Tevatron data (Fig.\ \ref{fig:mttunfold}).  Values of chi-squared per
degree of freedom for the absolute cross section
($\chi^2/\mathrm{d.o.f.}$) and for the normalized distribution
($\chi_N^2/\mathrm{d.o.f.}$) are shown in Table~\ref{table:chimtt}.
Compared with the unfolded data, the $W^\prime$ model prediction in
the high $m_{t\bar t}$ region is not as good as the SM prediction, but
the difference for a heavy $W^\prime$ is not sufficient to exclude a
heavy $W^\prime$ from Tevatron data alone, $\chi^2/\mathrm{d.o.f.} =
2.1$ in the $W^\prime$ case compared with $1.6$ in the SM.  We note
that a heavy $W^\prime$ boson fits the shape of the distribution
(normalized distribution) better than it fits the absolute
distribution, $\chi^2/\mathrm{d.o.f.} = 2.1$ vs
$\chi^2/\mathrm{d.o.f.} = 3.6$.  Moreover, the $W^\prime$ vertex
correction relaxes the constraint from the shape of the $m_{t\bar t}$
distribution, $\chi^2/\mathrm{d.o.f.} = 2.1$ vs
$\chi^2/\mathrm{d.o.f.} = 2.7$.

\begin{table}[!htbp]
\caption{Chi squared per degree of freedom for the $m_{t\bar t}$ distribution
 at the Tevatron.}
\begin{center}
\begin{tabular}{l|c|c}
 \hline \hline
 & $\chi^2/\mathrm{d.o.f.}$ & $\chi_N^2/\mathrm{d.o.f.}$ \\ \hline \hline
SM & 1.6 & 1.6 \\ \hline
500 GeV $W^\prime$ (LO) & 3.7 & 2.8 \\ \hline
500 GeV $W^\prime$ (NLO) & 3.8 & 2.6 \\ \hline
1 TeV $W^\prime$ (LO) & 3.4 & 2.7 \\ \hline
1 TeV $W^\prime$ (NLO) & 3.6 & 2.1 \\ \hline\hline
\end{tabular}
\end{center}
\label{table:chimtt}
\end{table}

Before turning to constraints from LHC data, we consider the role of
the difference in cut acceptance between the SM and the $W^\prime$
model~\cite{Gresham:2011pa}.  This difference arises partially because
the angular distribution of the top quark in the $W^\prime$ model
behaves like $\left(1+\cos\theta\right)^2$, wheres in the SM it
behaves like $\left(1+\cos^2\theta\right)$.  More top quarks are
expected in the large (positive) rapidity region in the $W^\prime$
case compared with the SM.  The charged lepton from the top-quark
decay will have nearly the same rapidity for an energetic top-quark
owing to the right-handed coupling of the $W^\prime$
model~\cite{Krohn:2011tw,Berger:2011pu,Berger:2012nw,Berger:2012tj}.
On the other hand, these events will be suppressed by the small
charged-lepton rapidity cut $|\eta_\ell|<1.0$ at Tevatron.

A simple analytic analysis is helpful for understanding the behavior
of the cut acceptance.  In the large $m_{t\bar t}$ region, $\beta\to
0$, and the squared-amplitude from the $d\bar d$ initial state behaves
as
\bea
&\propto&\left(1+\cos^2\theta\right)-\frac{\alpha_Rs}{2\alpha_Sm_{W^\prime}^2}
\frac{\left(1+\cos\theta\right)^2}{1+\frac{s}{2m_{W^\prime}^2}\left(1-\cos\theta\right)}
\nonumber\\&&+\frac{9}{8}\left(\frac{\alpha_Rs}{2\alpha_Sm_{W^\prime}^2}\right)^2
\frac{\left(1+\cos\theta\right)^2}{\left[1+\frac{s}{2m_{W^\prime}^2}\left(1-\cos\theta\right)\right]^2}.
\label{eq:cutacc}
\eea
We show the $W^\prime$ mass dependence of
$\alpha_Rs/\left(2\alpha_Sm_{W^\prime}^2\right)$ for $s=4m_t^2$ in
Fig.~\ref{fig:chi2_tev}(b).  Using parameters from our
best fits, we see that the coefficient is nearly independent of the
mass of the $W^\prime$.  It depends primarily on the center mass
energy.  Since the quadratic term gives a positive contribution which
grows faster than the linear term, the contribution from the
$W^\prime$ model is more significant in the large $m_{t\bar t}$ region
than in the small $m_{t\bar t}$ region (c.f., Fig.\ \ref{fig:mttunfold}).

\begin{figure}[!htb]
\includegraphics[scale=0.4,clip]{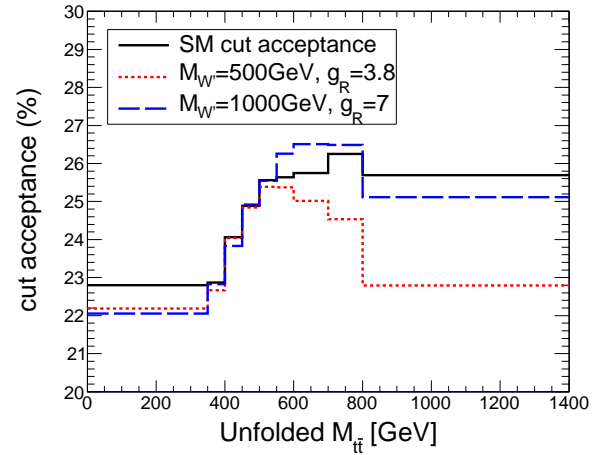}
\caption{The cut acceptance for $t\bar t$ events as a function of
$m_{t\bar t}$.  The solid (black) line is the SM value.   Acceptances in
the $W^\prime$ model are shown for a 500 GeV $W^\prime$ dotted (red) line,
and a 1 TeV $W^\prime$ dashed (blue) line.
\label{fig:acc} }
\end{figure}

To illustrate the effects of cut acceptance, we perform a simple
parton level simulation whose results are shown Fig.\ \ref{fig:acc}.
We use \textsc{MadGraph5/MadEvent} \cite{Alwall:2011uj} to generate parton
level $t\bar t$ events and decay the (anti-)top-quarks respecting
their helicity information.  We include the following energy smearing
effects for jets
\be
\frac{\delta E}{E}=0.1\oplus\frac{1}{\sqrt{E_T/{\mathrm{GeV}}}},
\ee
and charged leptons
\be
\frac{\delta E}{E}=0.02\oplus\frac{0.135}{\sqrt{E_T/{\mathrm{GeV}}}}.
\ee
The $b$-tagging efficiency is taken from \textsc{PGS4}~\cite{pgs} as a
function of the transverse energy and the rapidity of the $b$-quark.
The difference between the cut acceptances of the SM and the
$W^\prime$ model partially protects the $W^\prime$ model from the
constraints of the $m_{t\bar t}$ distribution for a 500~GeV
$W^\prime$.  However, the difference is not great for a heavy
$W^\prime$.  We also checked the contribution from the $t\bar t j$
final state and found it to be negligibly small at the Tevatron as
expected.

The message we draw is that while the shape of the invariant mass
distribution favors the pure SM relative to a model that includes a
$W^\prime$, this constraint is not decisive with Tevatron data.

\begin{figure*}[!htp]
\subfigure[]{\includegraphics[scale=0.29,clip]{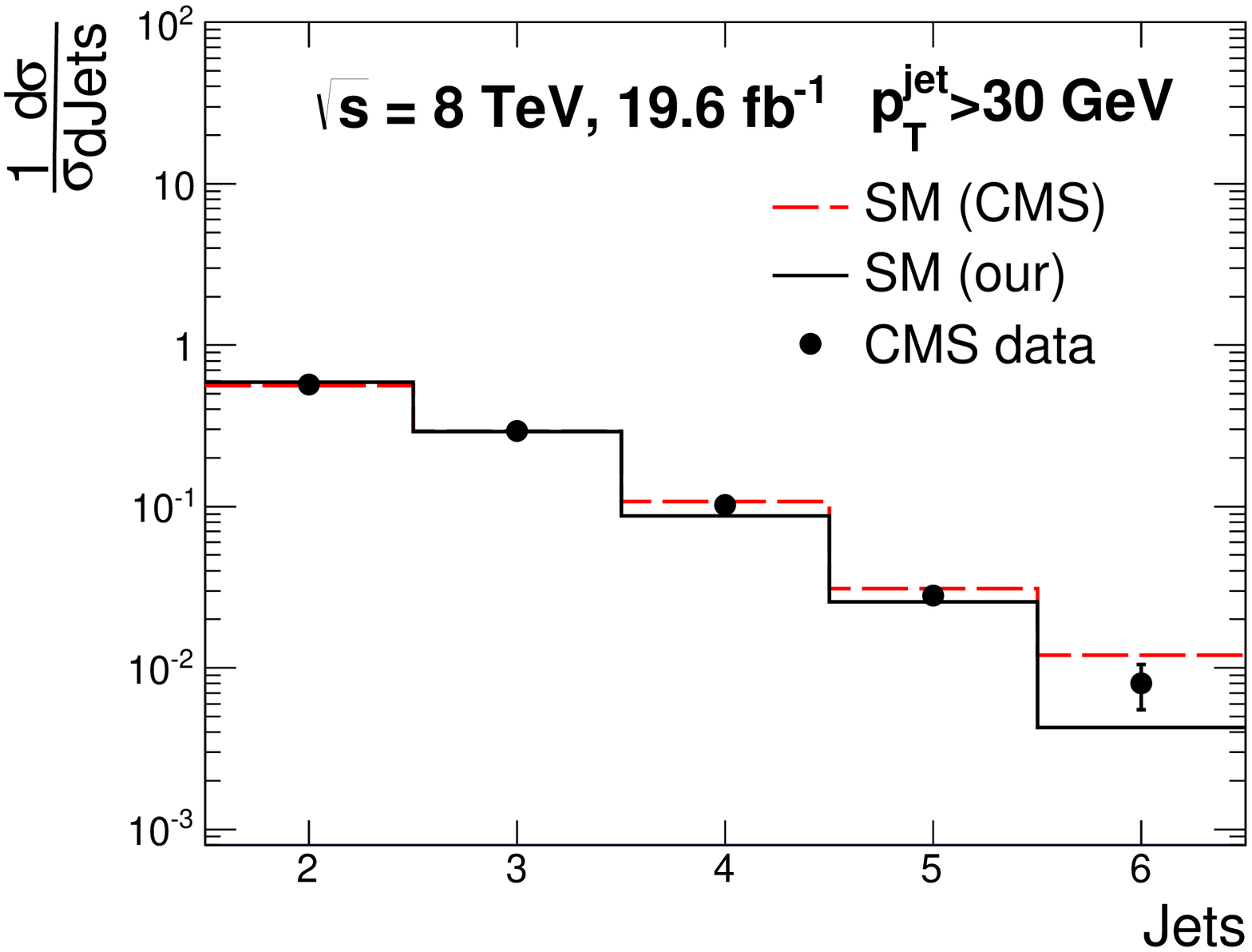}}
\subfigure[]{\includegraphics[scale=0.29,clip]{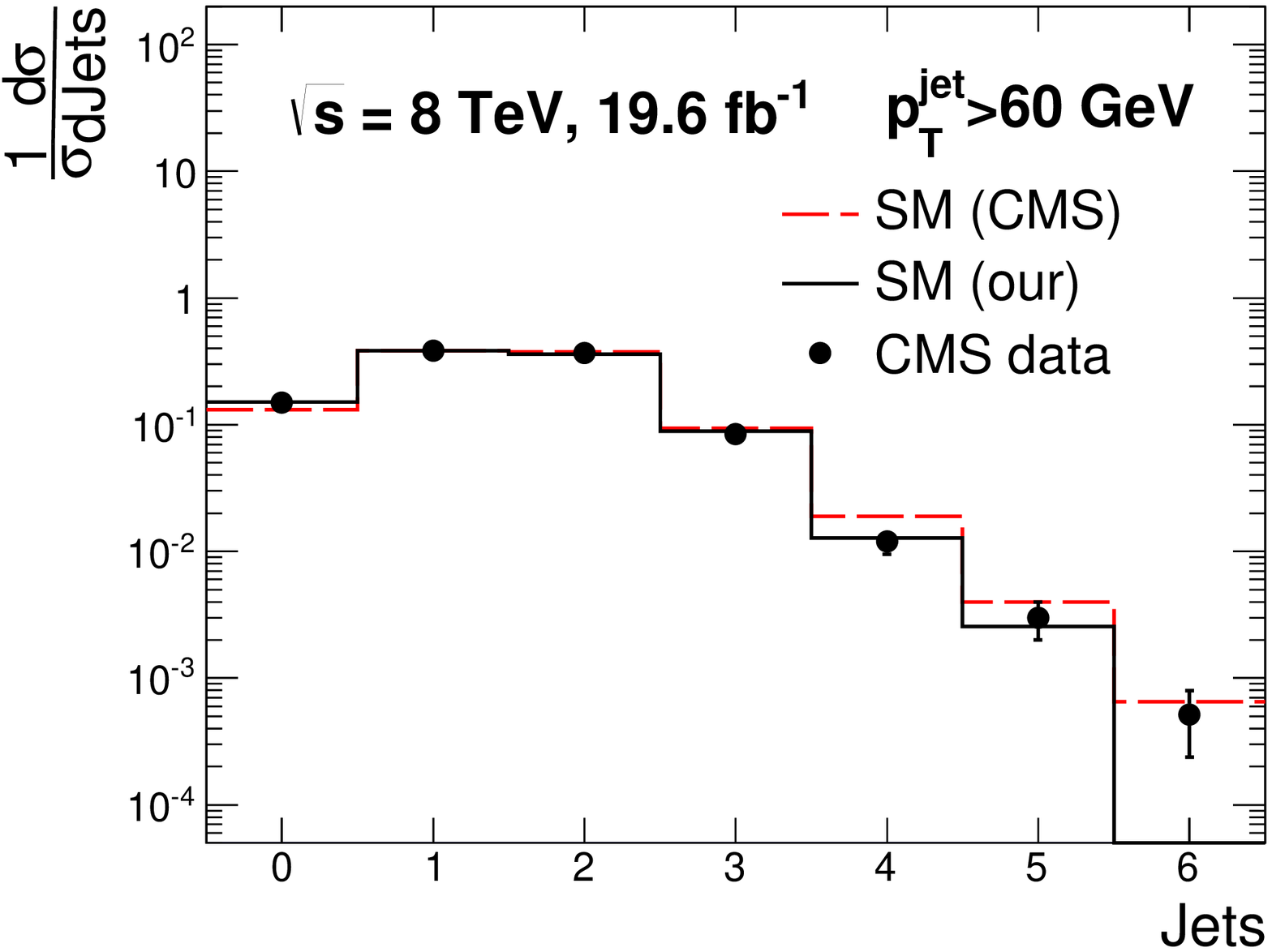}}
\subfigure[]{\includegraphics[scale=0.29,clip]{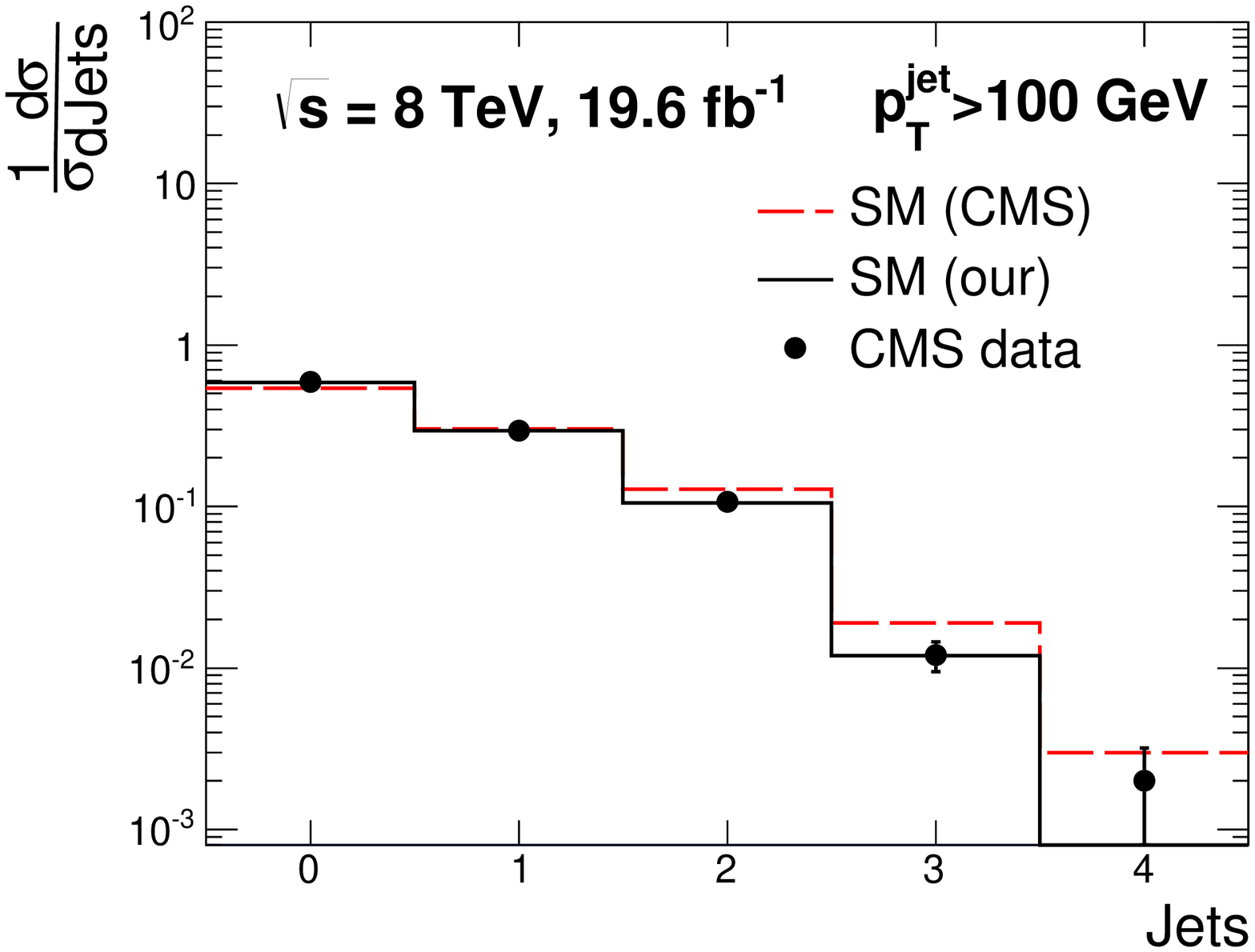}}
\caption{Normalized cross section as a function of jet-multiplicity
for jets with (a) $p_T>30$ GeV, (b) $p_T>60$ GeV, and (c)
$p_T>100$ GeV.   The data are from the
CMS collaboration~\protect{\cite{CMS-PAS-TOP-12-041}}.  
The dashed (red) lines are the results of the simulation shown in the
CMS paper, whereas the solid (black) line is our simulation.  
\label{fig:cms_check} }
\end{figure*}

\section{$W^\prime$ and $t\bar tj$ at LHC}
\label{sec:lhc}

Having determined parameters of the $W^\prime$ model that are
consistent with Tevatron data, we turn to an examination of the
viability of the model at the LHC.  We use data on the multiplicity of
jets in $t \bar{t}$ events as our principal observable.  In the
$W^\prime$ model, the associated production of a top-quark and a
$W^\prime$, with $W^\prime \rightarrow d \bar{t}$ contributes to the
jet multiplicity along with SM QCD production of $t\bar
t+\mathrm{n}j$.  This contribution was proposed in
\cite{Knapen:2011hu,Duffty:2012zz,Endo:2012mi} and studied in data at
7 TeV \cite{Chatrchyan:2012su,Aad:2012em}.  We pay particular
attention to the region of large $W^\prime$ mass where the coupling
strength $g_R$ and $W^\prime$ width are large (c.f., Fig.\
\ref{fig:width}).  Owing to the broad width, interference between the
amplitudes for $tW^\prime$ associated production and SM production of
$t\bar t+j$ is not negligible \cite{Duffty:2012zz,Endo:2012mi}.
Interference has not yet been included in experimental analyses
\cite{Chatrchyan:2012su,Aad:2012em}. In this study, we also include
for the first time the contribution to the jet multiplicity
distribution in $t\bar t+\mathrm{n}j$ from $W^\prime$ pair production
with, again, $W^\prime \rightarrow d \bar{t}$.  It is important to
include all of the contributions from the $W^\prime$ model to achieve
a good estimation of the jet-multiplicity.  In particular, the
$t$-channel $W^\prime$ exchange process has a non-negligible influence
on the $t\bar t+0j$ cross section.

\subsection{Normalized jet-multiplicity}

The normalized jet-multiplicity distribution in $t\bar t +X$ events is
presented by the CMS collaboration in Figure 2 of their
paper~\cite{CMS-PAS-TOP-12-041}.  Our first task is to verify the
accuracy of our simulation of SM $t\bar t+X$ production by comparing
our simulation with that of CMS.  We generate parton level $t\bar
t+{\mathrm{n}}~j$ events to n$=2$ using
\textsc{MadGraph5/MadEvent}~\cite{Alwall:2011uj}.  The generated
events are subsequently processed with
\textsc{PYTHIA6.4}~\cite{Sjostrand:2006za} for fragmentation and
hadronization using the MLM prescription~\cite{Mangano:2006rw} for
matching of jets with parton showers.  We perform a detector
simulation using the \textsc{PGS4} code~\cite{pgs}.

Following the CMS cuts, muon candidates are required to have a
transverse momentum $p_T>20$ GeV within a pseudorapidity region
$|\eta|<2.4$ and to be isolated with $I_\mathrm{rel}<0.15$.  The
quantity $I_\mathrm{rel}$ is the sum of the transverse momenta of all
neutral and charged reconstructed objects, except the muon itself,
inside a cone of size $\Delta
R\equiv\sqrt{\Delta\eta^2+\Delta\phi^2}<0.3$, divided by the muon
transverse momentum.  Electron candidates are required to have a
transverse energy $E_T>20$ GeV within a pseudorapidity region
$|\eta|<2.4$ and to be isolated with $I_\mathrm{rel}<0.15$.  Jets are
reconstructed using the anti-$k_T$ clustering algorithm with $R=0.5$
and required to have a transverse momentum $p_T>30$ GeV within a
pseudorapidity region $|\eta|<2.4$.  The \textsc{PGS4}
$b$-tagging~\cite{pgs} efficiency is re-weighted to a maximum of 80\%
to mimic the $b$-tagging efficiency of CMS.

Signal events are required to have at least two isolated leptons with
opposite electric charge (electrons or muons), and two jets, at least
one of which is identified as a $b$-jet.  Events with a lepton pair
invariant mass smaller than 20 GeV are removed to suppress events from
heavy flavor decays.  In the $\mu^+\mu^-$ and $e^+e^-$ channels, the
dilepton invariant mass is required to be outside a $Z$-boson mass
window of $91\pm 15$ GeV, and the missing transverse energy $\met$ is
required to be larger than 40 GeV.

\begin{table}[tbp]
\caption{Standard model values of chi-squared for the normalized jet-multiplicity distribution in $t\bar t+ X$ events 
at 8 TeV for different values of the jet transverse momentum cut. }
\begin{center}
\begin{tabular}{l|c|c|c}
 \hline \hline
 & $p^{\mathrm{jet}}_T>30$ GeV & $p^{\mathrm{jet}}_T>60$ GeV & $p^{\mathrm{jet}}_T>100$ GeV \\ \hline \hline
$\chi^2/{\mathrm{d.o.f}}$ (ours)& 0.6 & 0.06 & 0.2 \\ \hline
$\chi^2/{\mathrm{d.o.f}}$ (CMS)& 0.2 & 1.6 & 4.5 \\ \hline\hline
\end{tabular}
\end{center}
\label{table:chism}
\end{table}

\begin{figure*}[tbp]
\includegraphics[scale=0.49,clip]{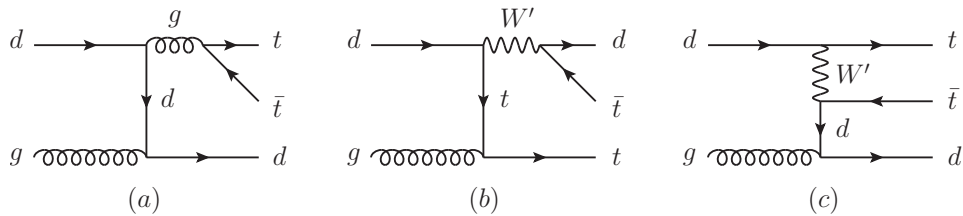}
\caption{(a) Example of standard model production of $t \bar{t} d$ production in a $d g$ interaction; (b) $W^\prime$ model 
production of $t \bar{t} d$; (c) $W^\prime$ exchange contribution to  production of $t \bar{t} d$.   Interference occurs between 
the processes (a) and (b), and between (a) and (c).   
\label{fig:intdiag}}
\end{figure*}

The results of our SM simulation are shown in Fig.\
\ref{fig:cms_check} and compared with the CMS simulation and data.
Our simulation agrees with the simulation by the CMS collaboration,
and it agrees well with the data, except in the 6 jet bins at
$p_T^{\rm jet} = 30$ and $60$~GeV.  We attribute this difference to
the fact that we generate only up to $t\bar t + 2j$ events at 
parton level.  Thus, there are at most 4 jets in our parton level
events.  To calculate the value of $\chi^2$ of the SM simulation, we
estimate the theoretical uncertainty from the differences between
predictions obtained with different event generators and choices of
hard scales~\cite{CMS-PAS-TOP-12-041}.  Treating the experimental and
theoretical uncertainties as uncorrelated, we obtain the values of the
SM $\chi^2$ from our simulation shown in Table~\ref{table:chism}.  The
comparison of $\chi^2$ values shows that our simulation is as good as
the CMS simulation.  (For the samples with $p_T>60$ GeV and $p_T>100$
GeV, our values of ${\chi}^2$ are in fact better.)

Having established the reliability of our simulation code, we generate
events from the $W^\prime$ model following the same method used for
the SM events.  At the parton level, we generate all $t\bar
t+{\mathrm{n}}j$ processes including the interference between the SM
$t\bar t+{\mathrm{n}}j$ process and inclusive $tW^\prime$ associated
production.  We generate parton level events to n=2.  Examples of some
of the processes that we compute are shown in Fig.~\ref{fig:intdiag}.
We remark that contributions from the $W^{\prime +}W^{\prime -}$ channel
are also included.  We examine the entire mass range $200 <
m_{W^\prime} < 1100$~GeV, bearing in mind that a very light $W^\prime$
has been excluded in prior studies of Tevatron \cite{Aaltonen:2012qn}
and 7 TeV LHC data \cite{Chatrchyan:2012su,Aad:2012em}.  We are also
aware that an extremely heavy $W^\prime$ (heavier than 1 TeV) is not
consistent with the Tevatron $t\bar t$ observables (c.f., Fig.\
\ref{fig:chi2_tev}).

\begin{figure*}[!htb]
\includegraphics[scale=0.29,clip]{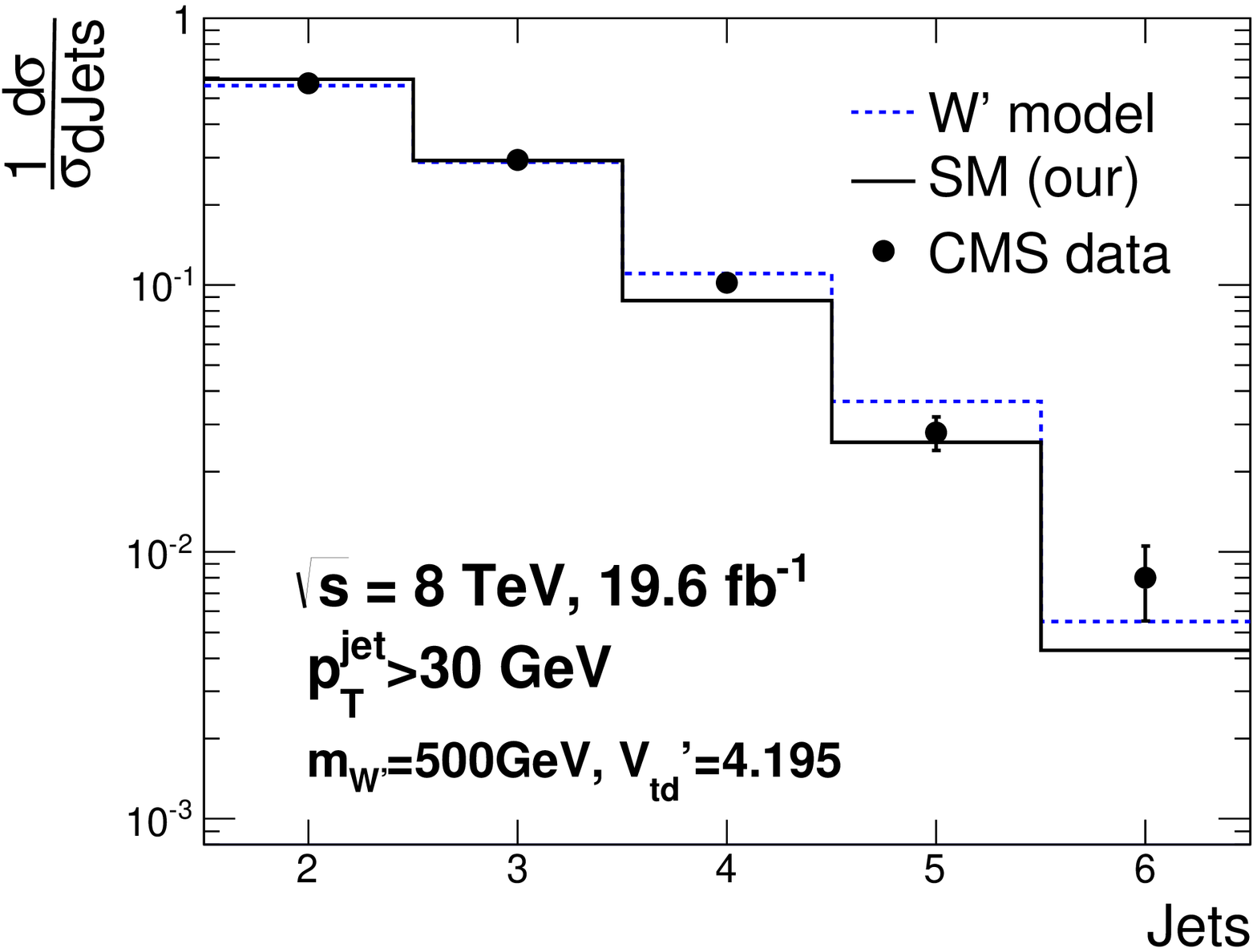}
\includegraphics[scale=0.29,clip]{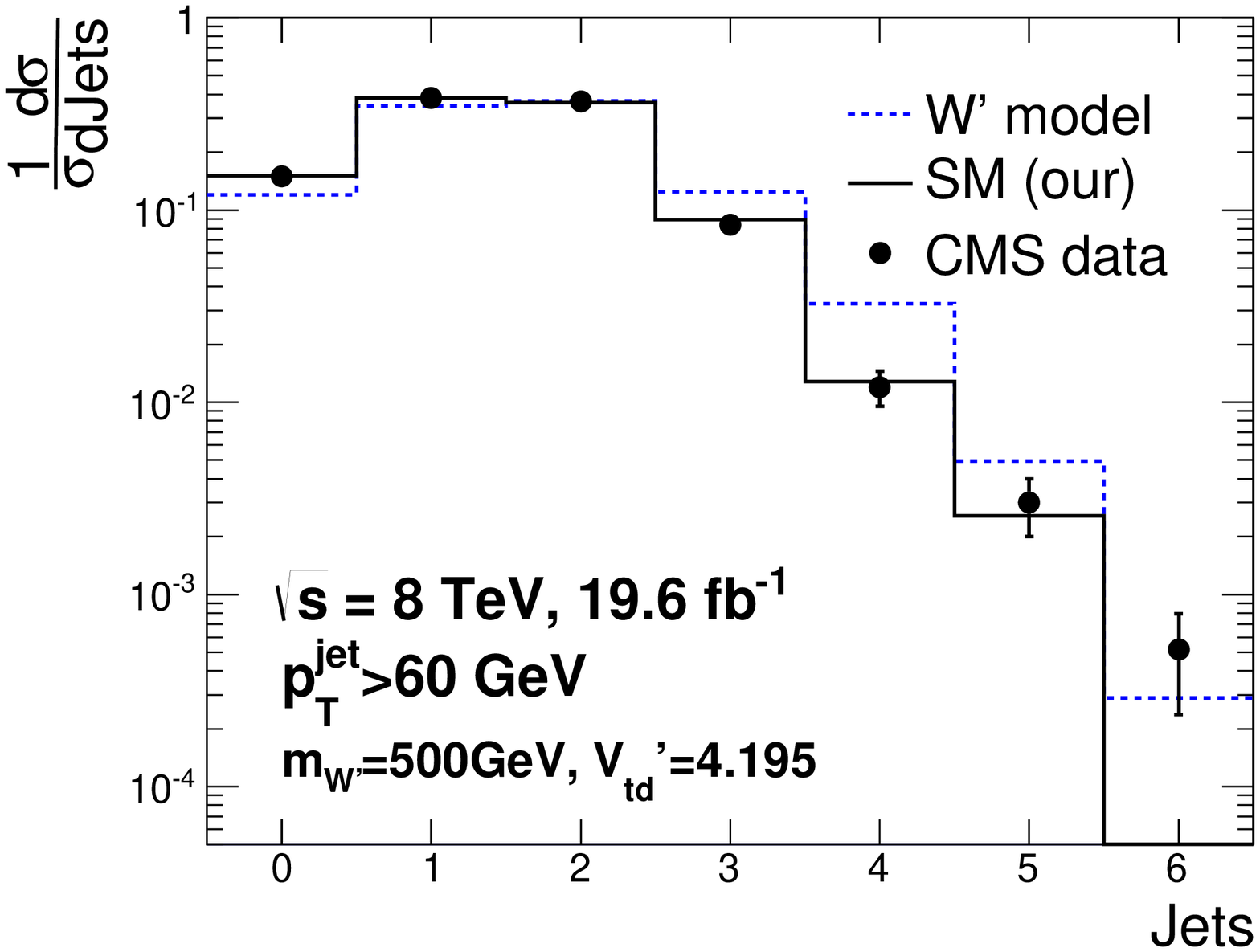}
\includegraphics[scale=0.29,clip]{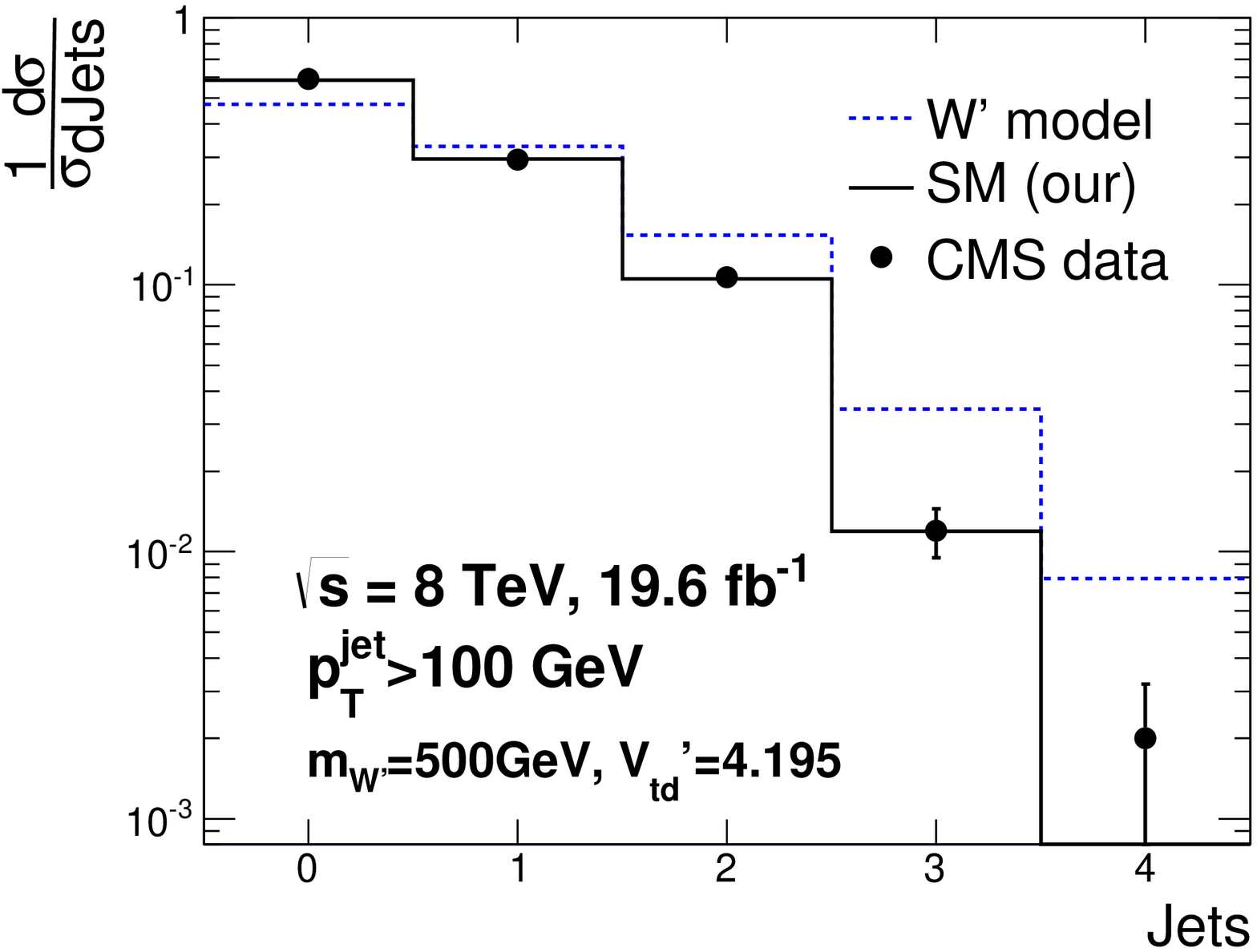}\\
\includegraphics[scale=0.29,clip]{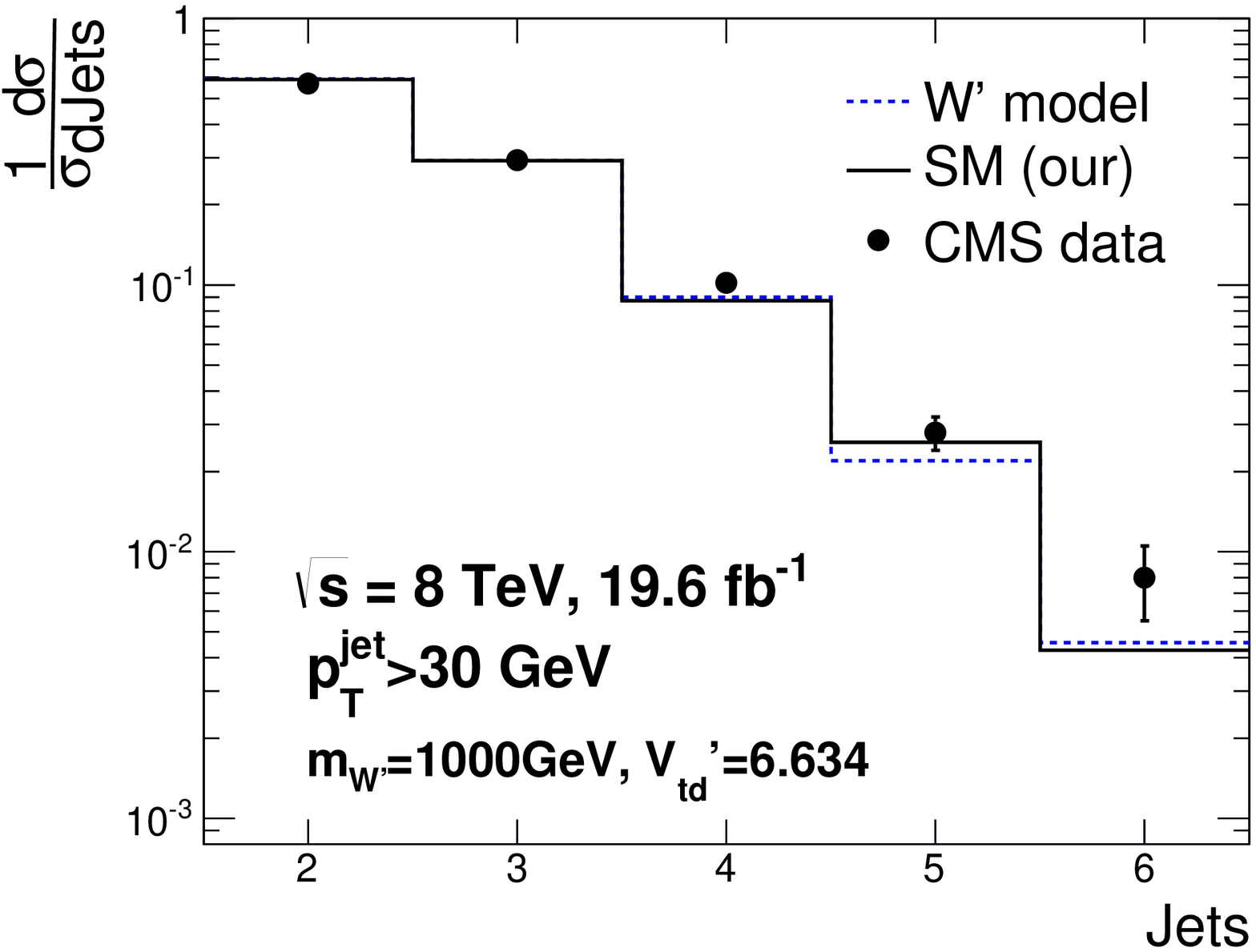}
\includegraphics[scale=0.29,clip]{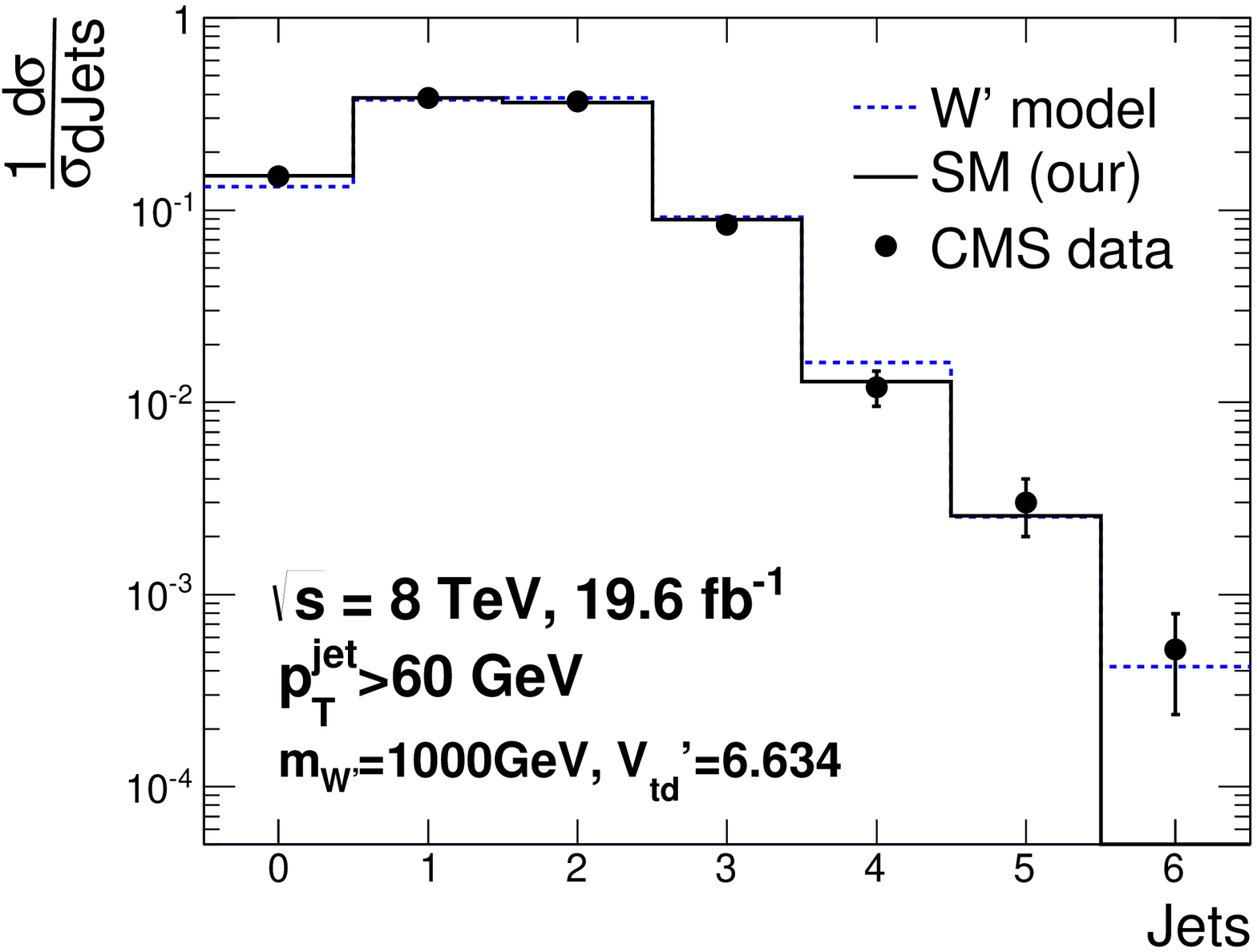}
\includegraphics[scale=0.29,clip]{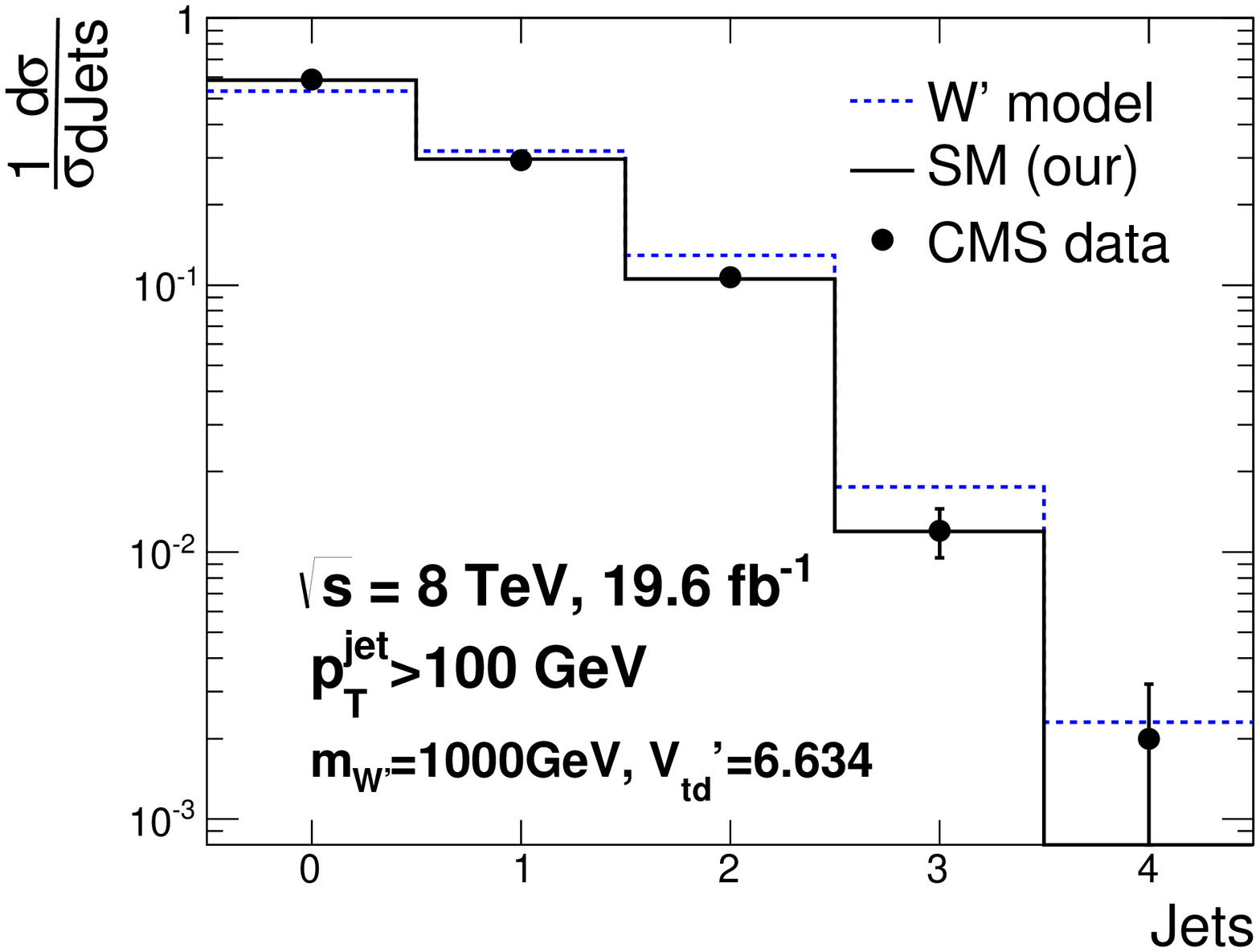}
\caption{Normalized cross sections as a function of jet-multiplicity for jets with $p_T>30$ GeV 
(left panels), $p_T>60$ GeV (middle panels), and $p_T>100$ GeV (right panels).  
The figures in the upper row are the results for $m_{W^\prime}=500$ GeV and $V_{td}^\prime=4.195$.
The figures in the lower row are the results for $m_{W^\prime}=1$ TeV and $V_{td}^\prime=6.634$.  
The (black) solid line is our SM simulation.  The (blue) dotted lines represent our $W^\prime$ model 
results. 
\label{fig:xyzz} }
\end{figure*}

\begin{figure*}[tp]
\subfigure[]{\includegraphics[scale=0.4,clip]{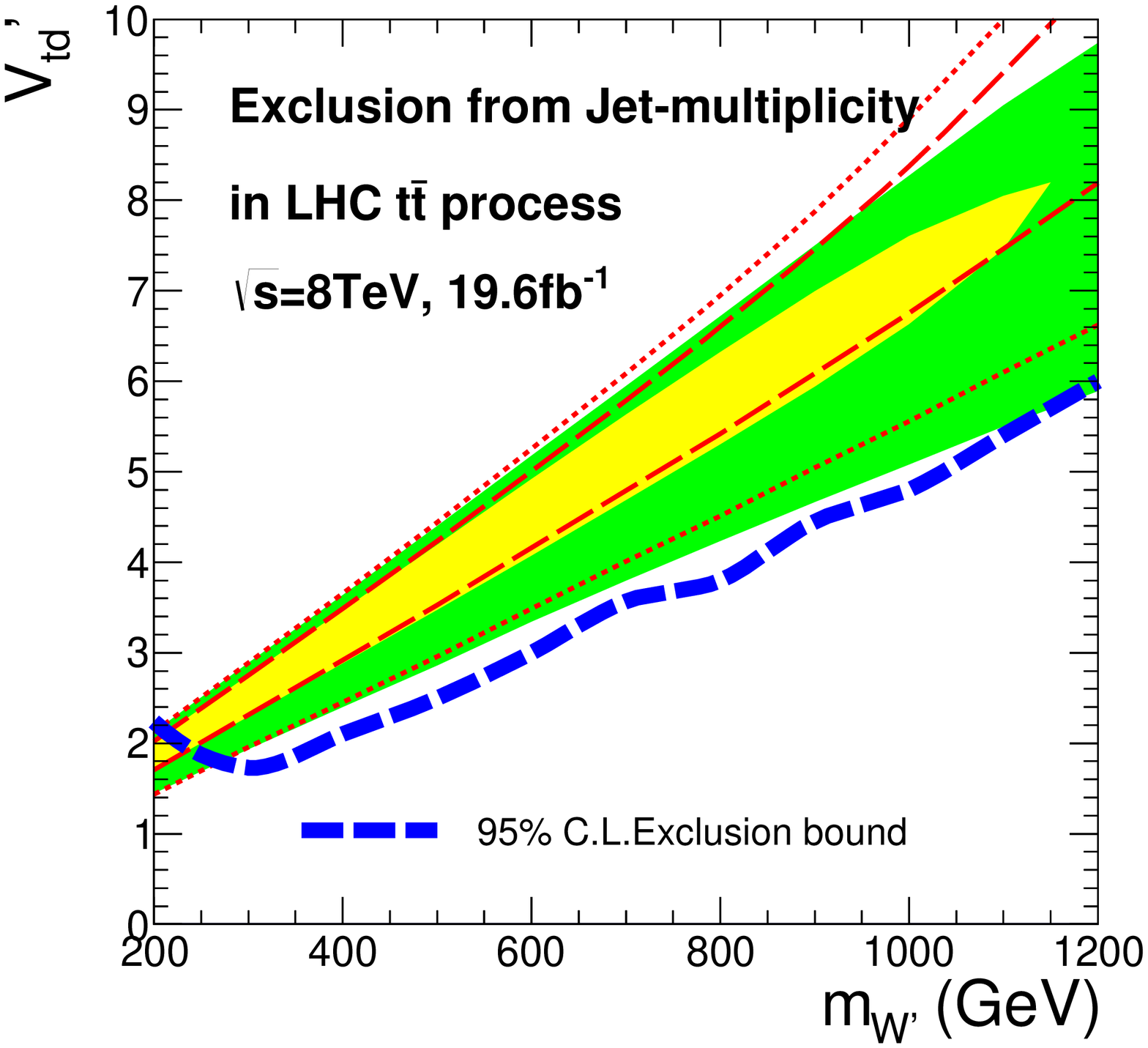}}
\subfigure[]{\includegraphics[scale=0.4,clip]{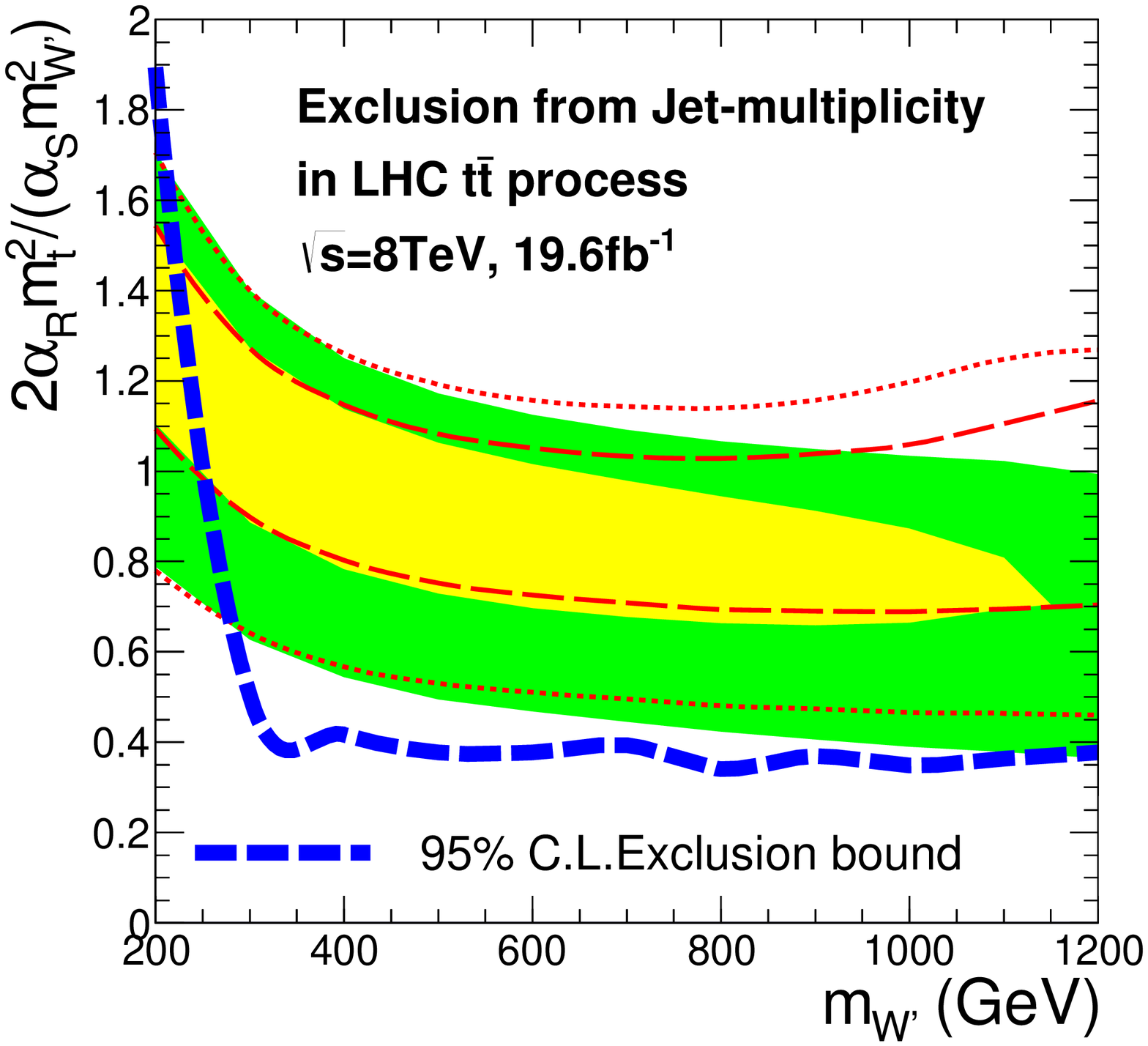}}
\caption{(a) The 95\% exclusion bound in the parameter space of $V^\prime_{td}$
vs $m_{W^\prime}$ from $\chi^2/{\mathrm{d.o.f}}$ fits to the LHC
jet-multiplicity distribution
in the inclusive $t\bar tX$ process at 8 TeV. The (blue) dashed line 
is the 95\% exclusion bound. The dotted and dashed (red) lines and the
light (yellow) and dark (green) shaded regions have the same meanings as
those in Fig.\ \protect{\ref{fig:chi2_tev}}.
(b) The 95\% exclusion bound shown for $2\alpha_Rm_t^2
/\left(\alpha_Sm_{W^\prime}^2\right)$ vs $m_{W^\prime}$.
\label{fig:chi_ttj} }
\end{figure*}

An examination of Fig.\ \ref{fig:xyzz} shows qualitatively that the
$W^\prime$ model agrees less well with the CMS normalized data than
with the SM.  In order to make this conclusion more quantitative for
$W^\prime$ boson masses in the range $200 < m_{W^\prime} < 1100$~GeV,
we perform fits in the space of $V^\prime_{td}$ vs $m_{W^\prime}$, and
compute the resulting values of $\chi^2/{\mathrm{d.o.f}}$ in each bin of
the normalized multiplicity distribution for each of the three
values of $p^{\mathrm{jet}}_T>30$ GeV, $p^{\mathrm{jet}}_T>60$ GeV and
$p^{\mathrm{jet}}_T>100$ GeV.  
In our calculation of $\chi^2/{\mathrm{d.o.f}}$ for
$p^{\mathrm{jet}}_T>60$ GeV, we use only the bins with jet number up to
5, but including the 6 jet bin does not affect the final results.  We
use the set of $\chi^2$ values at each $W^\prime$ mass to determine
the $95\%$ confidence level exclusion lines shown in
Fig.~\ref{fig:chi_ttj}.  These results show that the $W^\prime$ model
is disfavored by more than $2\sigma$ at the LHC if we use the
parameter space determined in our fits to the Tevatron data and the
$W^\prime$ boson is heavier than 300 GeV.  In addition (not shown)
most points in the best-fit region (light-shaded yellow region of
Fig.~\ref{fig:chi_ttj}) for explaining $\afb$ at the Tevatron are
excluded by 15--25$\sigma$.

For light $W^\prime$ whose mass is $\sim200$ GeV, the normalized
jet-multiplicity is not a good observable for testing the $W^\prime$
model.  For such a light $W^\prime$ boson, the narrow width
approximation is good enough, and $tj$ resonance searches can be used
at the LHC and the
Tevatron~\cite{Aaltonen:2012qn,Chatrchyan:2012su,Aad:2012em}.

\subsection{Jet-multiplicity distribution: \\ Effects of interference and
$W^\prime W^\prime$ pair-production}

\begin{figure*}[!hbtp]
\includegraphics[scale=0.29,clip]{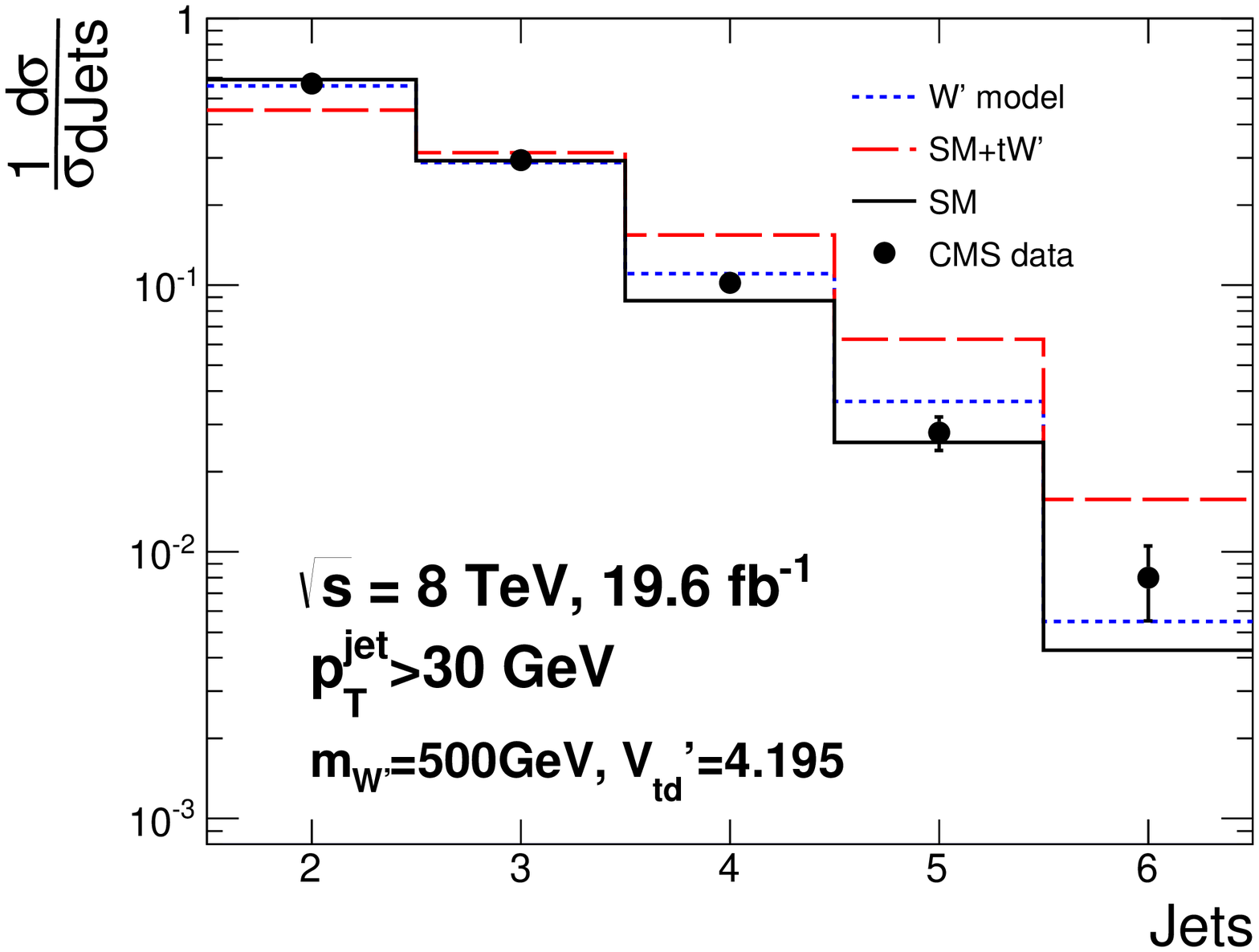}
\includegraphics[scale=0.29,clip]{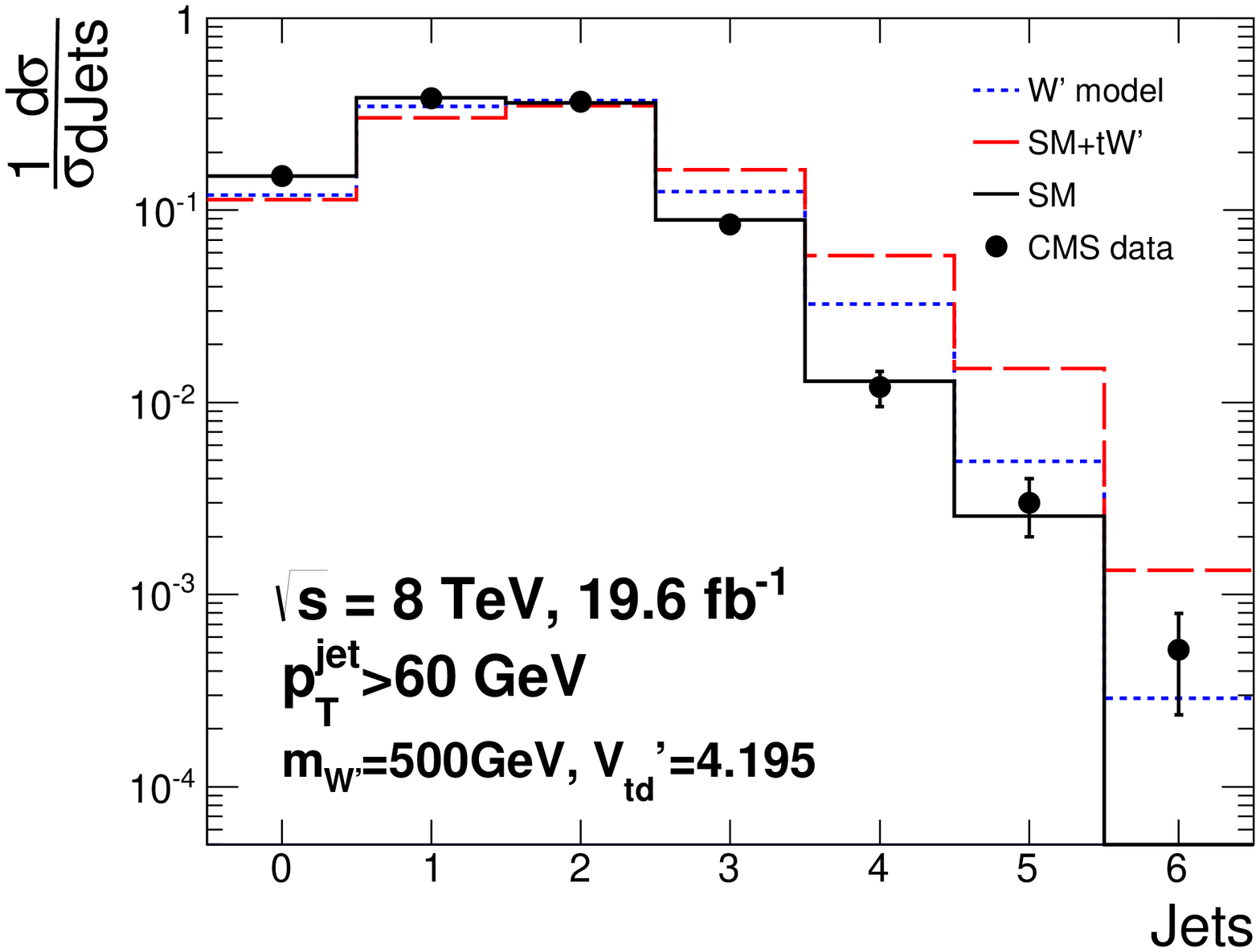}
\includegraphics[scale=0.29,clip]{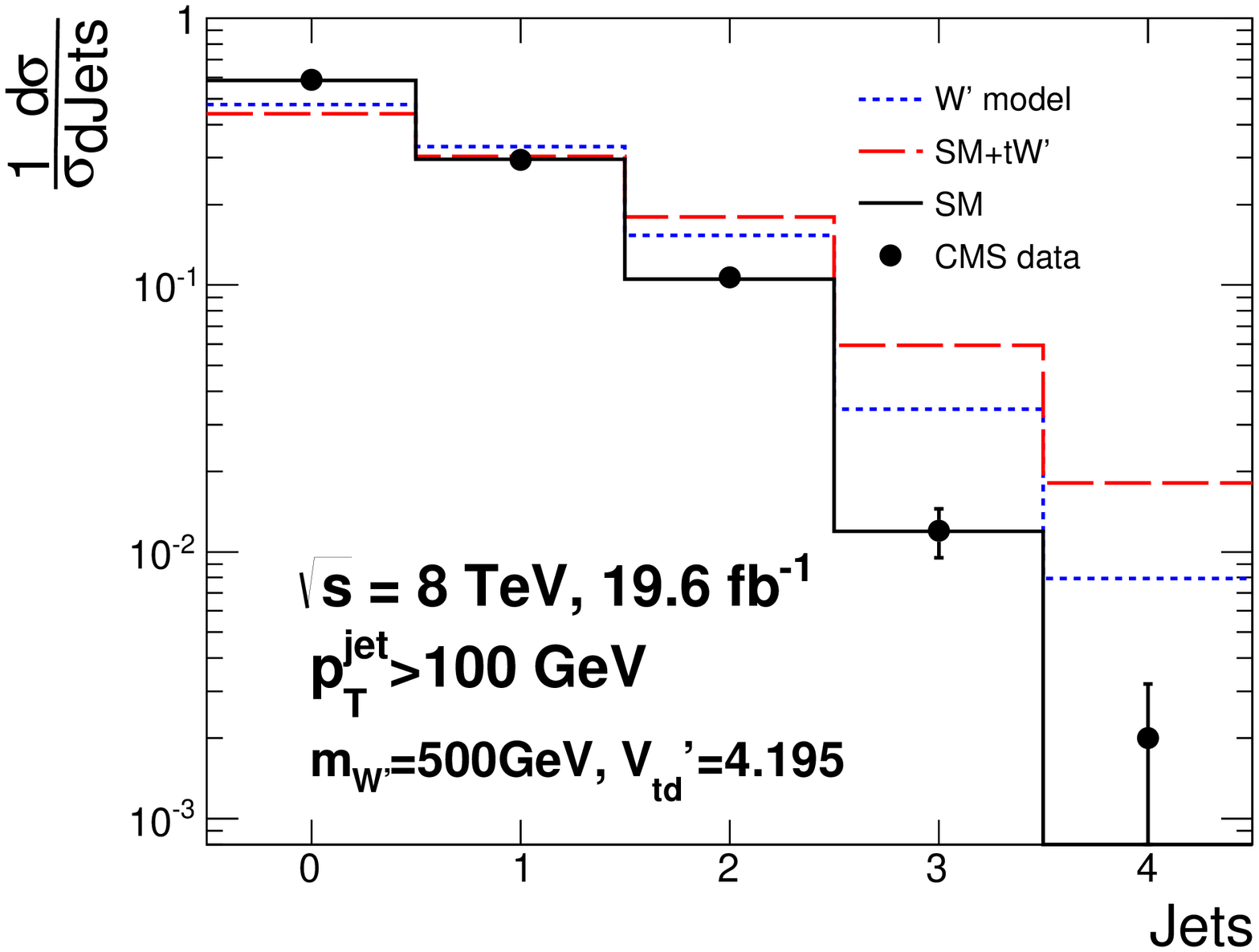}\\
\includegraphics[scale=0.29,clip]{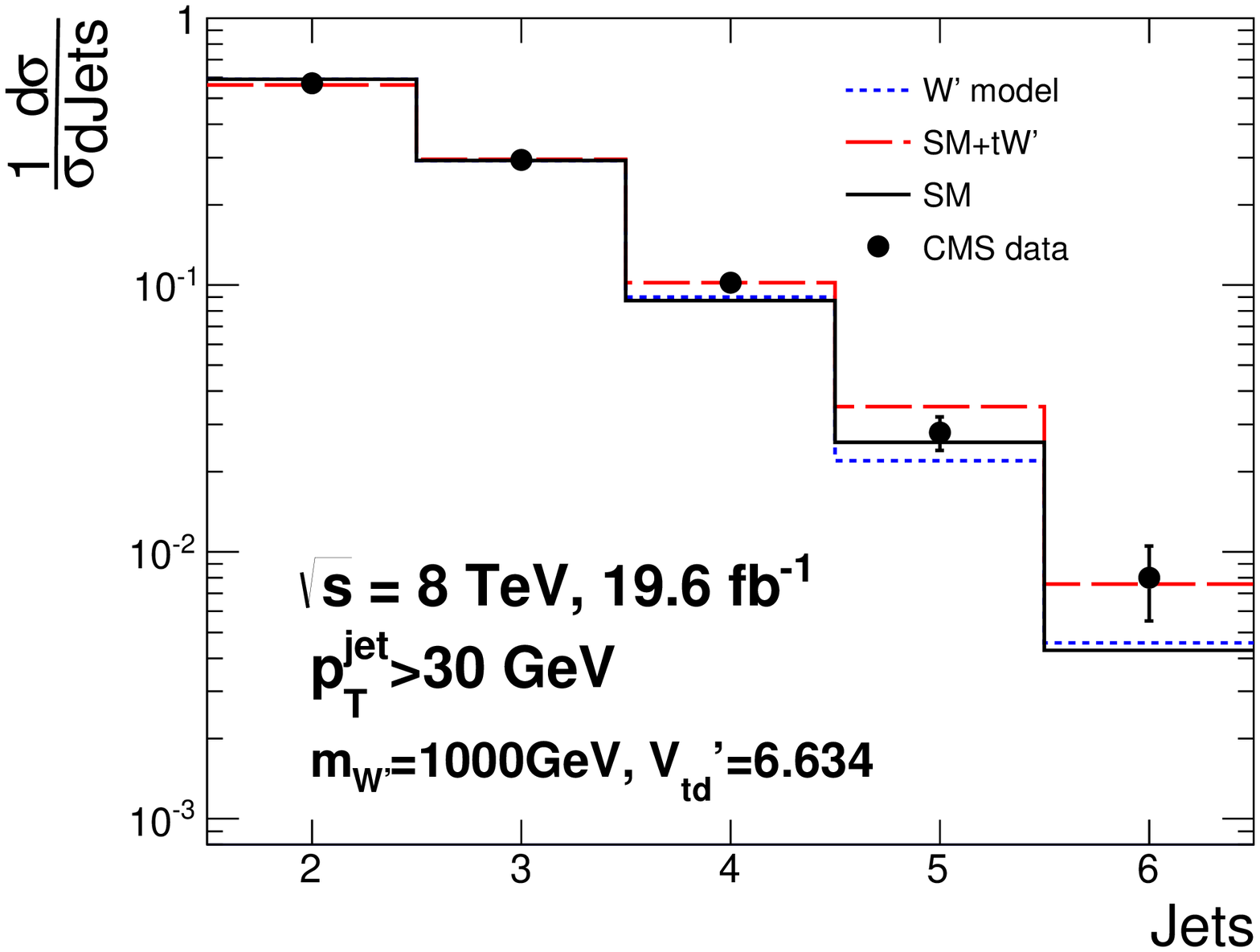}
\includegraphics[scale=0.29,clip]{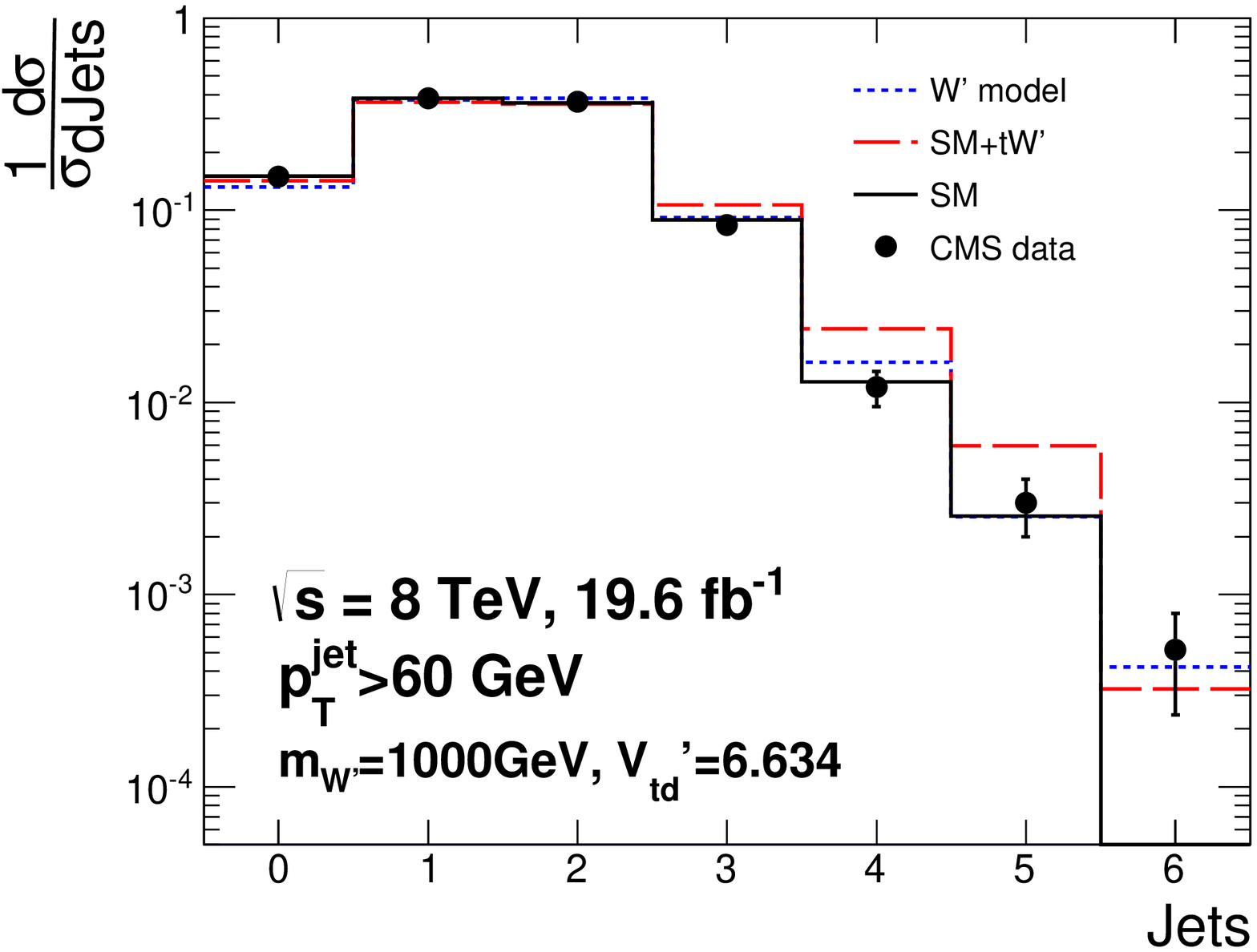}
\includegraphics[scale=0.29,clip]{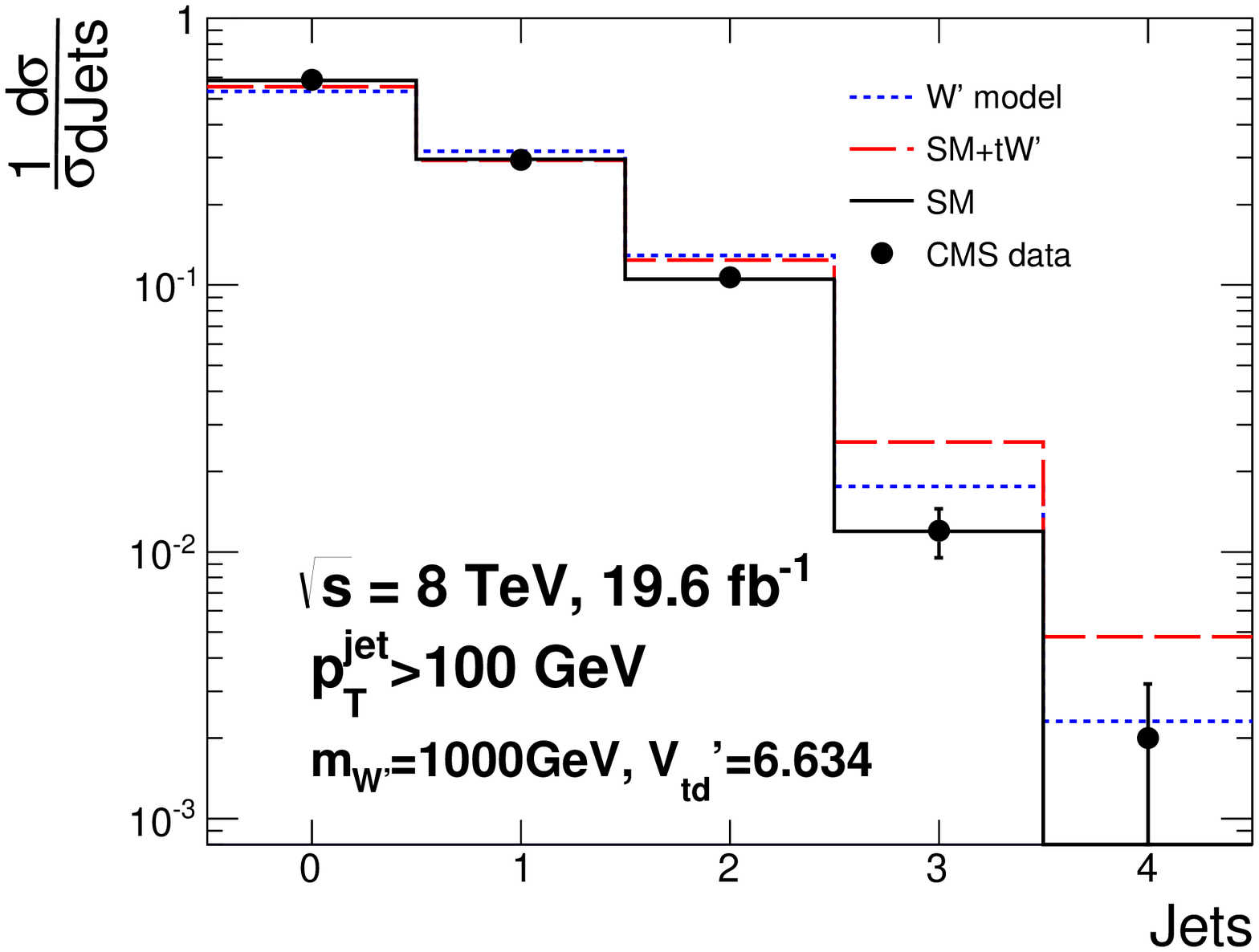}
\caption{The normalized distribution as a function of jet-multiplicity
for jets with $p_T>30$ GeV (left panels), $p_T>60$ GeV (middle panels),
$p_T>100$ GeV (right panels).   The figures in the upper row are the results
for $m_{W^\prime}=500$ GeV  and $V_{td}^\prime=4.195$.  The figures in the
lower row are the results for $m_{W^\prime}=1$ TeV and $V_{td}^\prime=6.634$.
The (red) dashed lines represent the results without the interference 
between $tW^\prime+X$ and SM $t\bar t +X$ production.  The (blue) dotted
lines are the results with interference included. 
\label{fig:wpint_dis} }
\end{figure*}

Both interference and $W^\prime W^\prime$ pair production contribute
in each multiplicity bin.  In order to isolate the effects from the
interference of $tW^\prime$ associated production with SM $t\bar t+X$
production, and the effects from $W^{\prime +}W^{\prime -}$ pair
production, we show figures in which we focus separately on each of
these two contributions.
We choose two values of $m_{W^\prime}$ with their associated values of
$V_{td}^\prime$.  Our benchmark points are $m_{W^\prime}=500$ GeV,
$V_{td}^\prime=4.195$ and $m_{W^\prime}=1$ TeV and $V_{td}^\prime=6.634$, typical
of a light $W^\prime$ and a heavy $W^\prime$, respectively.  The
results for $tW^\prime$ associated production without $W^\prime
W^\prime$ pair-production are shown in Fig.\ \ref{fig:wpint_dis}.

To obtain results that represent the \textit{incoherent} sum of the SM and
$tW^\prime$ processes, we generate parton-level
$tW^\prime+{\mathrm{n}}j$ events to n$=1$, and decay the $W^\prime$ and
$t$ ($\bar t$).  Contributions from $t$-channel $W^\prime$ exchange
processes are included when there is $W^\prime$ in the final state.
After showering, hadronization, and event selection, we then add the
SM $t \bar{t}+{\mathrm{n}}j$ contribution to the $tW^\prime+{\mathrm{n}}j$
result.

Our Fig.\ \ref{fig:wpint_dis} shows that the difference between the
incoherent SM+$tW^\prime$ result and the full $W^\prime$ model result
is significant.  There are two reasons for this difference.  First, as
$m_{W^\prime}$ increases, the width of the $W^\prime$ increases, and
interference between the $tW^\prime$ and the SM $t\bar t+j$ processes
grows in importance.  Second, the full result contains the
contribution from the $t$-channel $W^\prime$ exchange contribution to
the $t\bar t$ production process, a contribution which is not small at
the LHC.  In Fig.\ \ref{fig:wpint_dis} we see the complete $W^\prime$
model result is smaller than the incoherent SM+$tW^\prime$ result and
that it agrees better with the data.  Therefore, the strength of the
signal may be overestimated if the incoherent sum of SM+$tW^\prime$ is
used as an approximation.

\begin{figure*}[!hbtp]
\includegraphics[scale=0.29,clip]{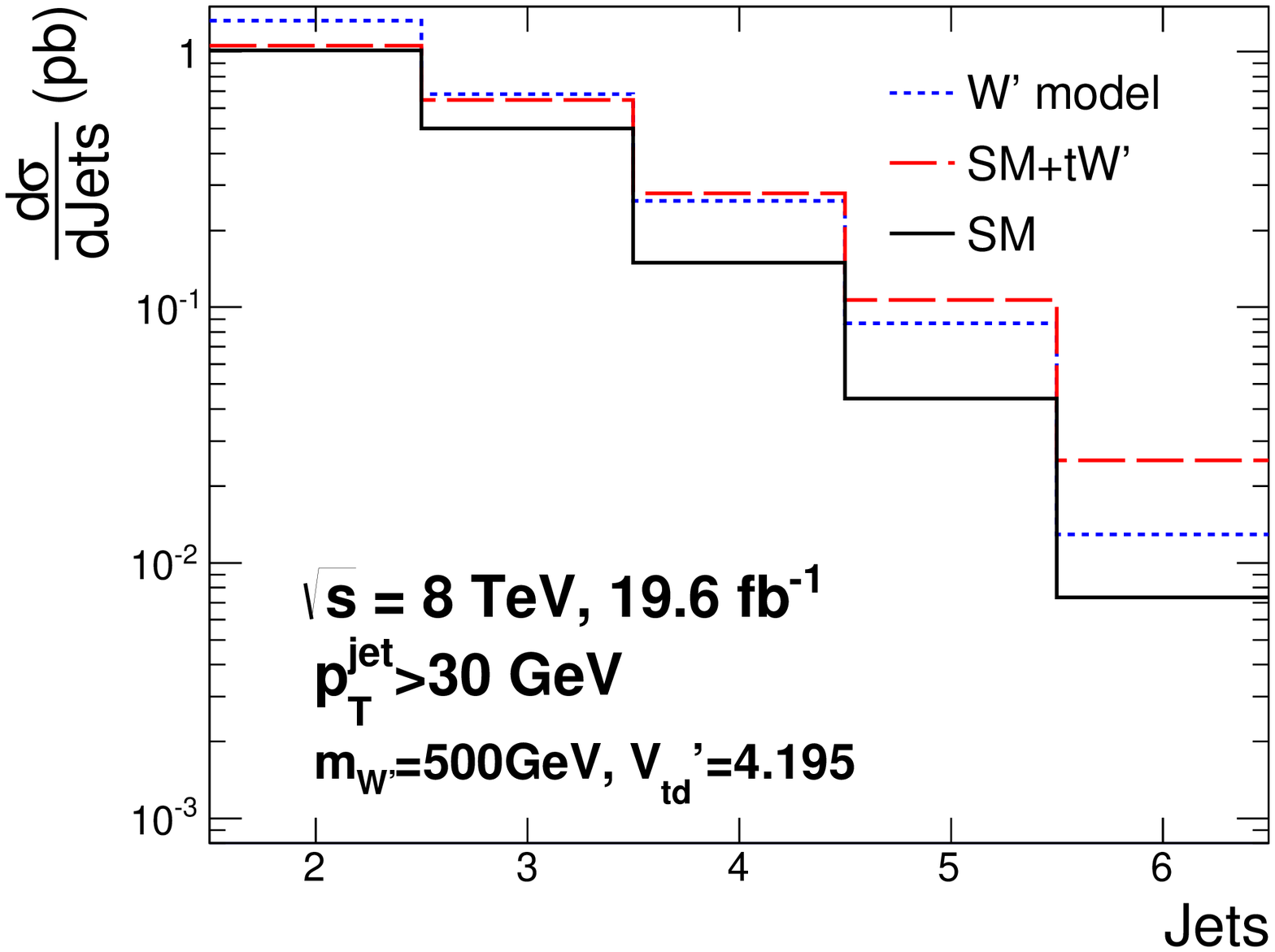}
\includegraphics[scale=0.29,clip]{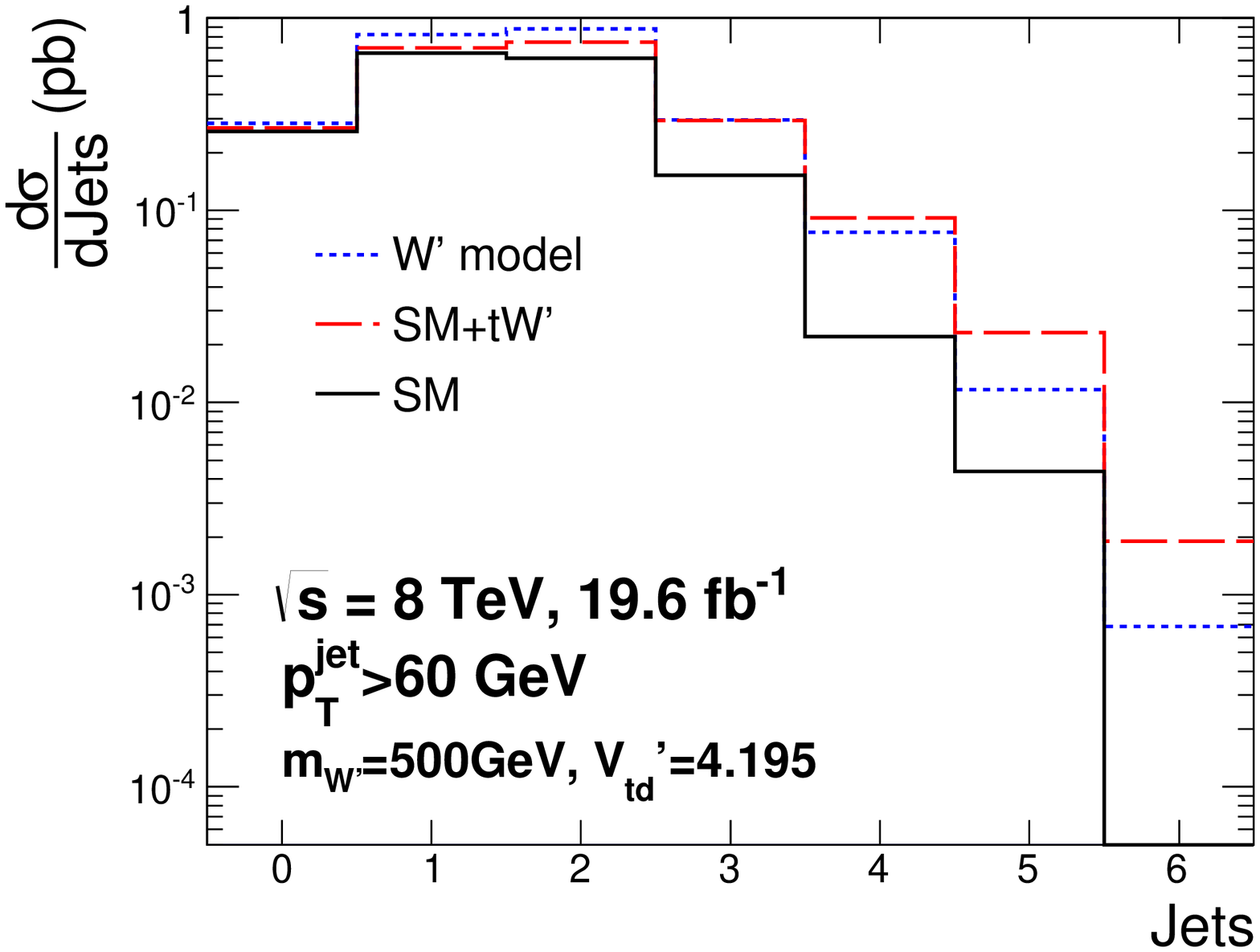}
\includegraphics[scale=0.29,clip]{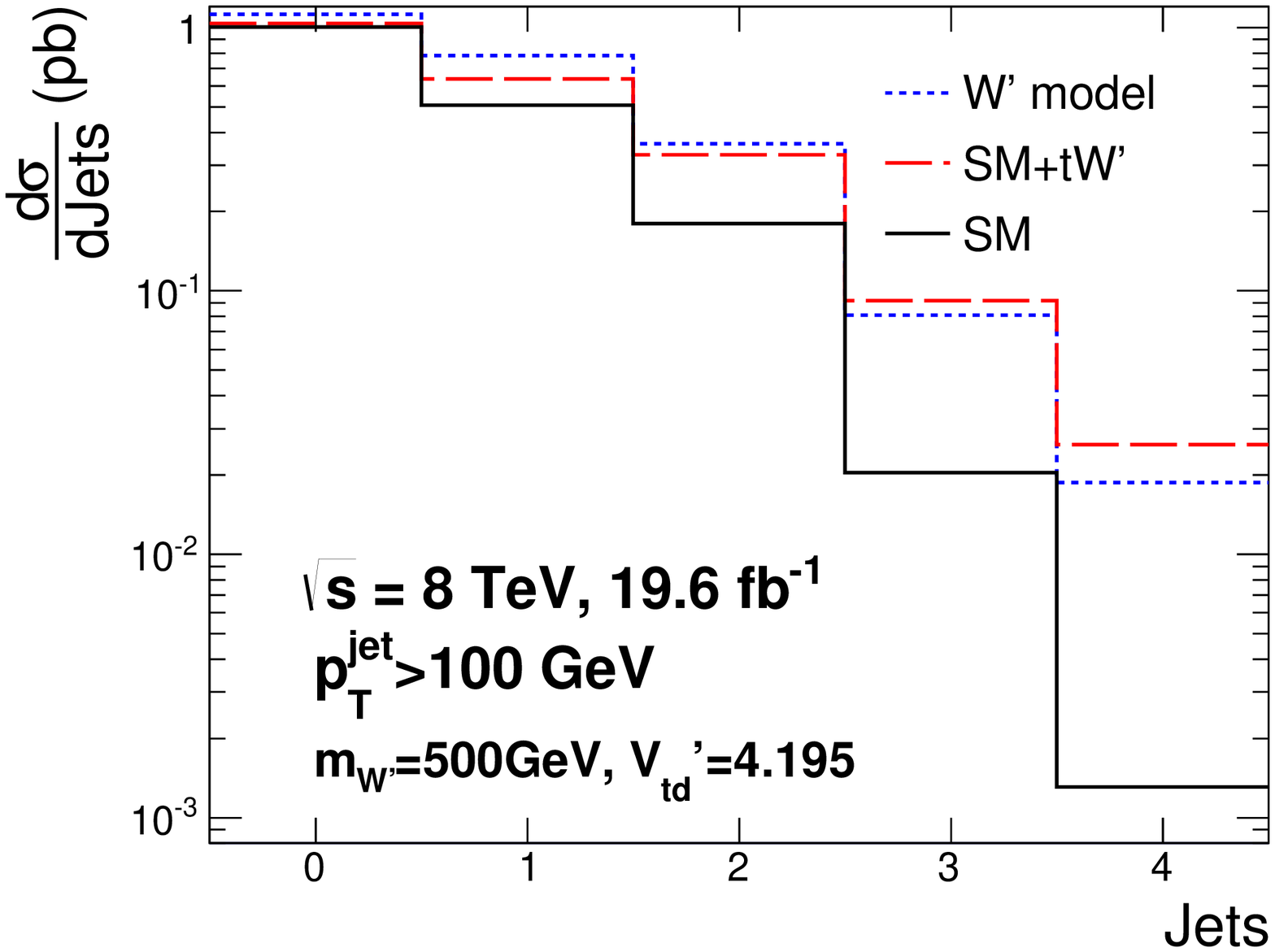}\\
\includegraphics[scale=0.29,clip]{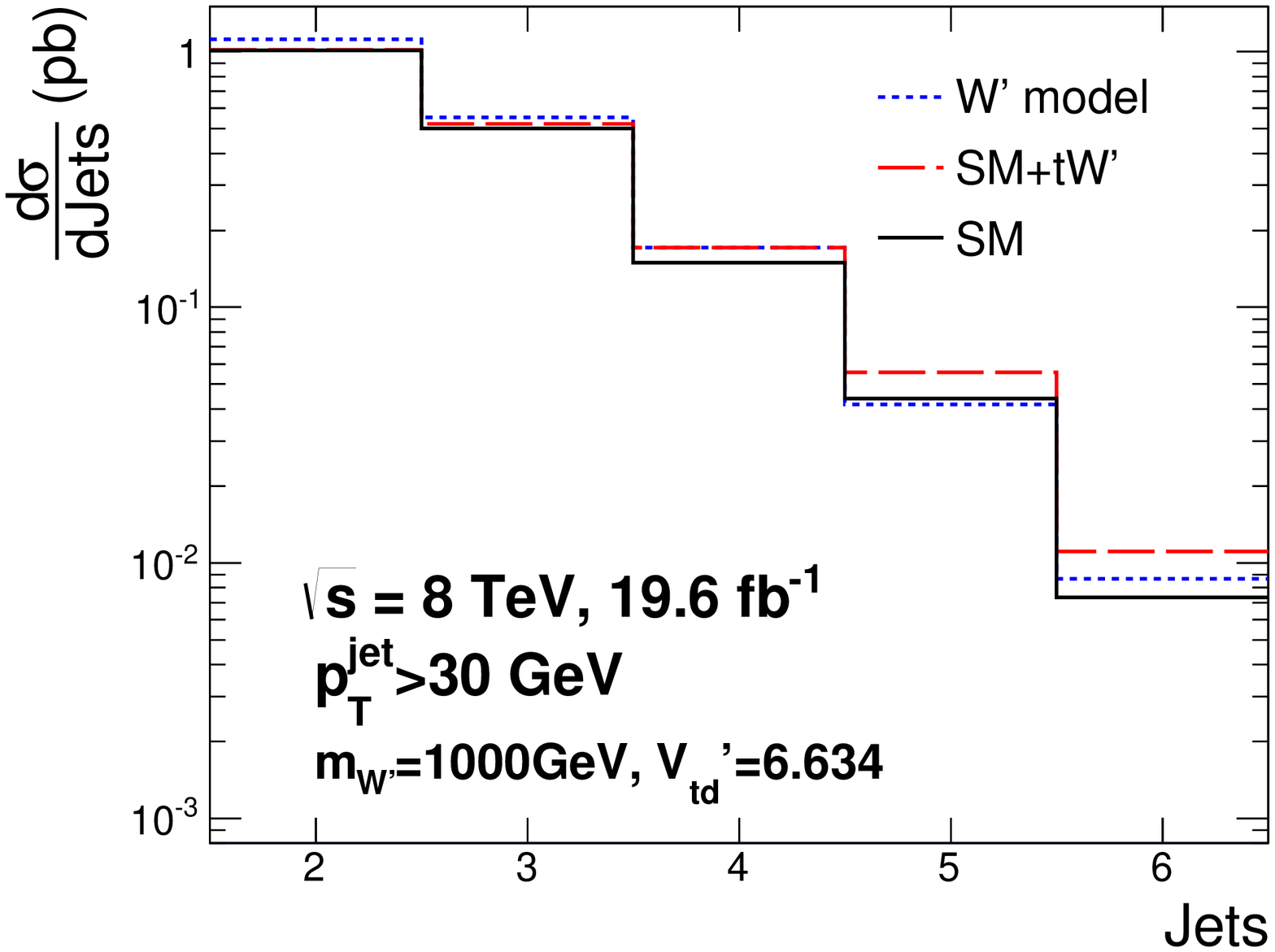}
\includegraphics[scale=0.29,clip]{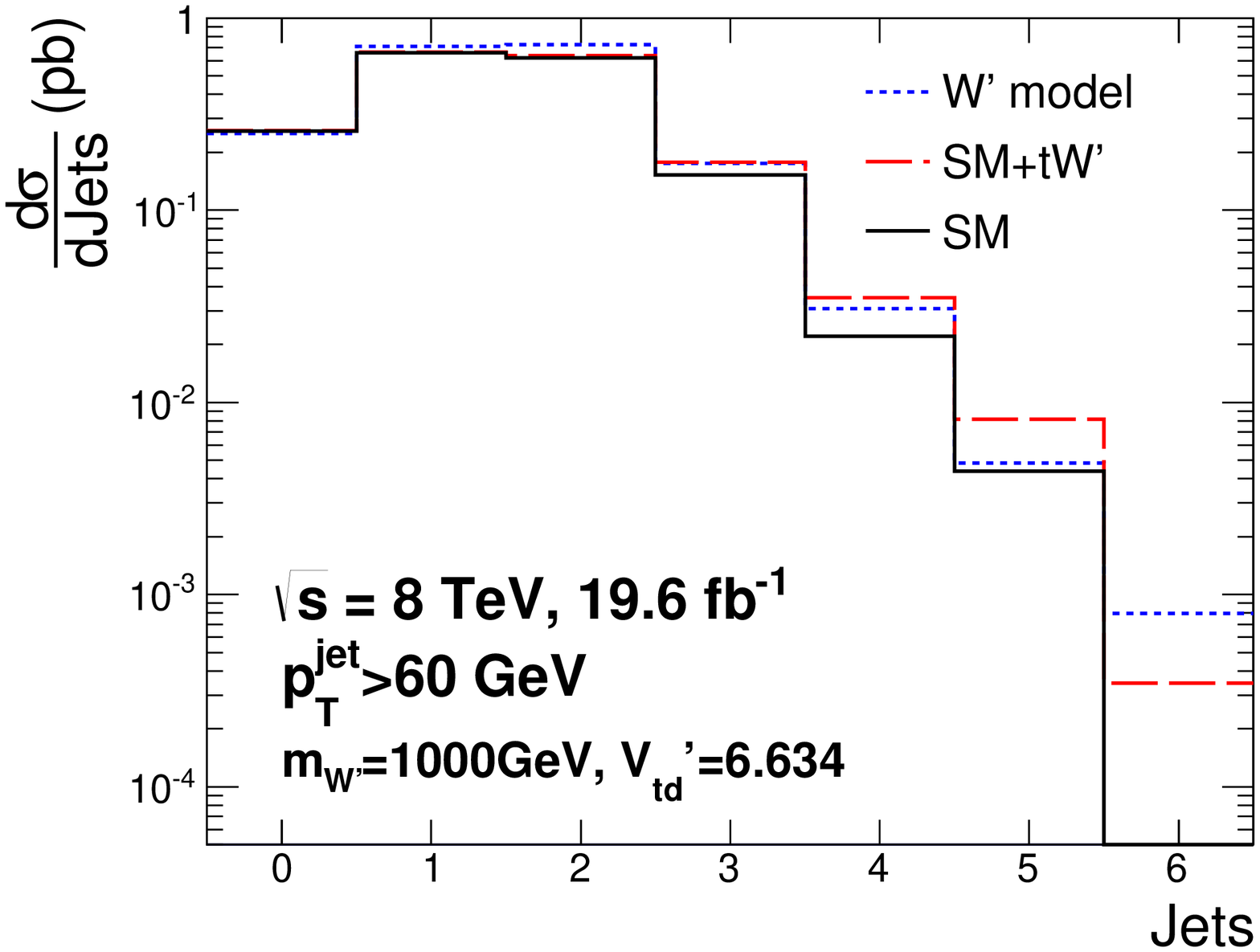}
\includegraphics[scale=0.29,clip]{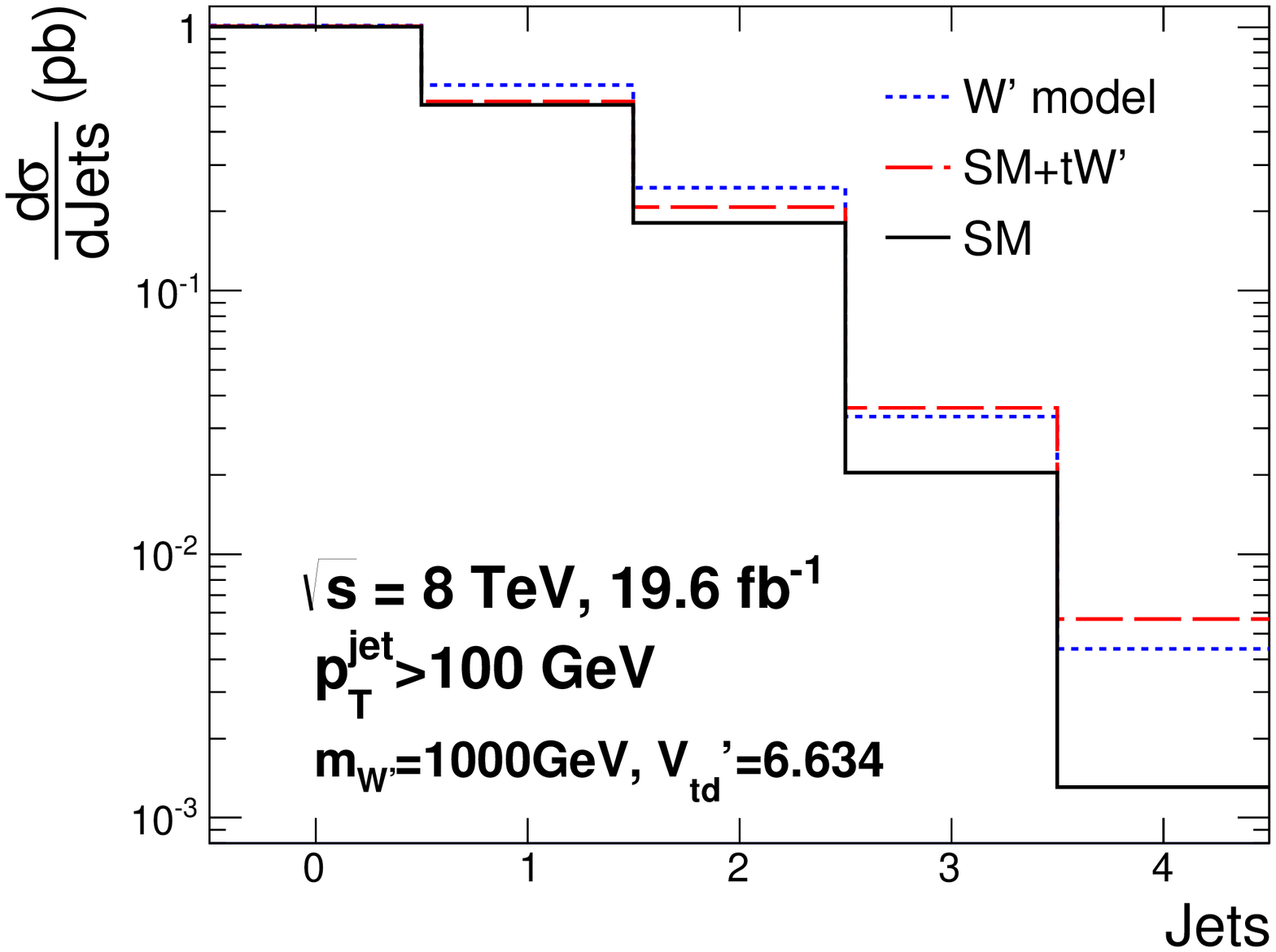}
\caption{The cross sections as a function of jet-multiplicity
for jets with $p_T>30$ GeV (left panels), $p_T>60$ GeV (middle panels), 
$p_T>100$ GeV (right panels).   The figures in the upper row are the results
for $m_{W^\prime}=500$ GeV and $V_{td}^\prime=4.195$.  The figures in the
lower row are the results for $m_{W^\prime}=1$ TeV and $V_{td}^\prime=6.634$.
The (red) dashed lines represent the results without the interference 
between the $tW^\prime+X$ and the SM $t\bar t +X$ production.  The (blue)
dotted lines are the results with interference included. 
\label{fig:wpint_ab} }
\end{figure*}

In Fig.\ \ref{fig:wpint_dis} we present the normalized multiplicity
distribution in order to compare with the CMS
data~\cite{CMS-PAS-TOP-12-041}.  On the other hand, the normalized
distribution tends to obscure some features of the $W^\prime$
contribution and the effects of interference.  In Fig.\
\ref{fig:wpint_ab}, we show instead the absolute cross sections as a
function of jet multiplicity for the SM+$tW^\prime$ process.  This
figure shows that the incoherent sum of the SM+$tW^\prime$ processes
is usually smaller than the complete calculation for
$n_{\mathrm{jets}}\leqslant3$, but it is larger than the result of the
complete calculation for $n_{\mathrm{jets}}\geqslant4$.  Thus, for a
light $W^\prime$ boson which has a relatively narrow width, including
the interference effect in studies of data on $tj$ resonance searches
will provide a stronger constraint on the $W^\prime$ model.  For a
heavy $W^\prime$ boson whose width is quite large, ignoring
interference in fits to the normalized jet-multiplicity data in $t\bar
t$ process, will lead to a constraint on the $W^\prime$ model that is
too strong.

\begin{figure*}[!hbtp]
\includegraphics[scale=0.29,clip]{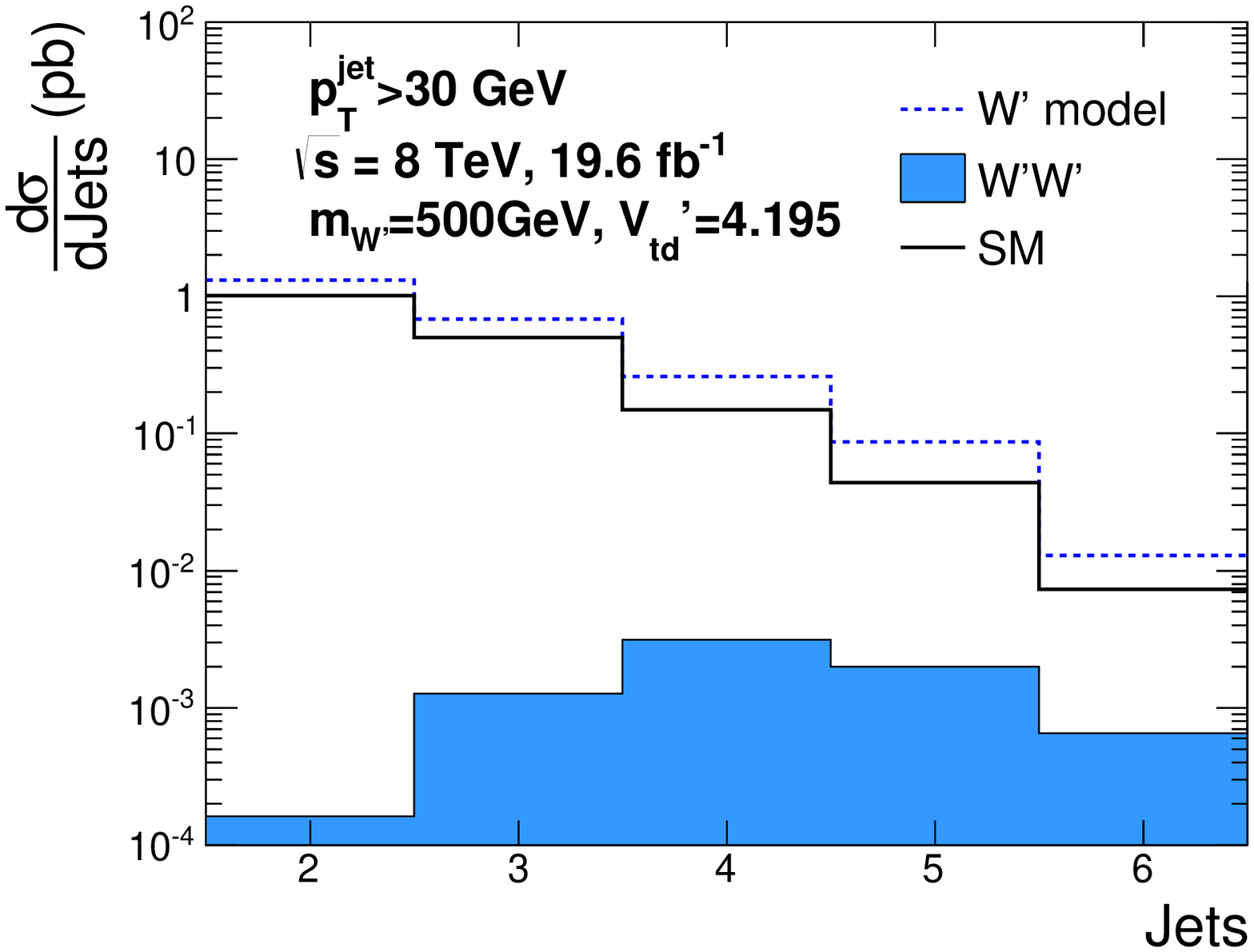}
\includegraphics[scale=0.29,clip]{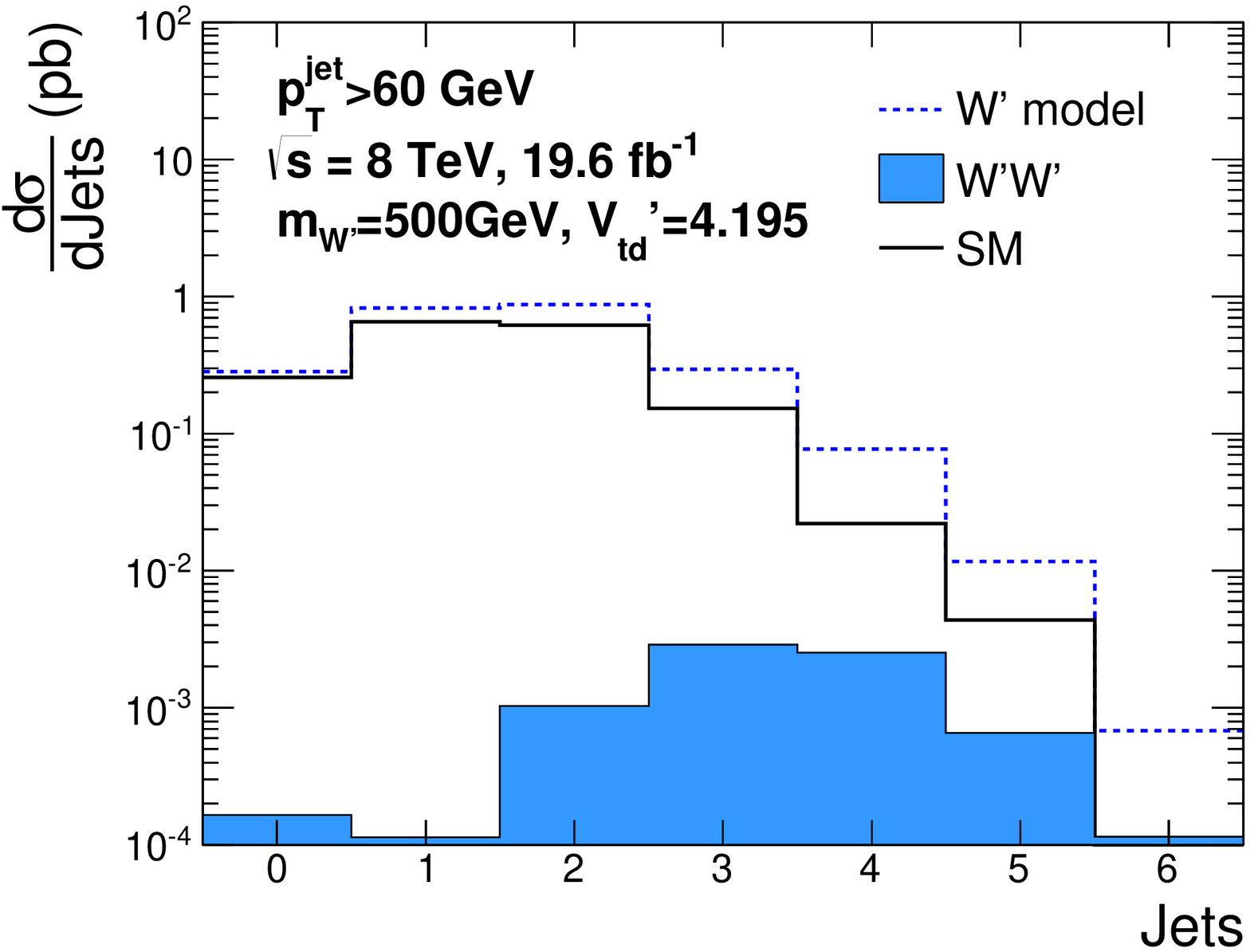}
\includegraphics[scale=0.29,clip]{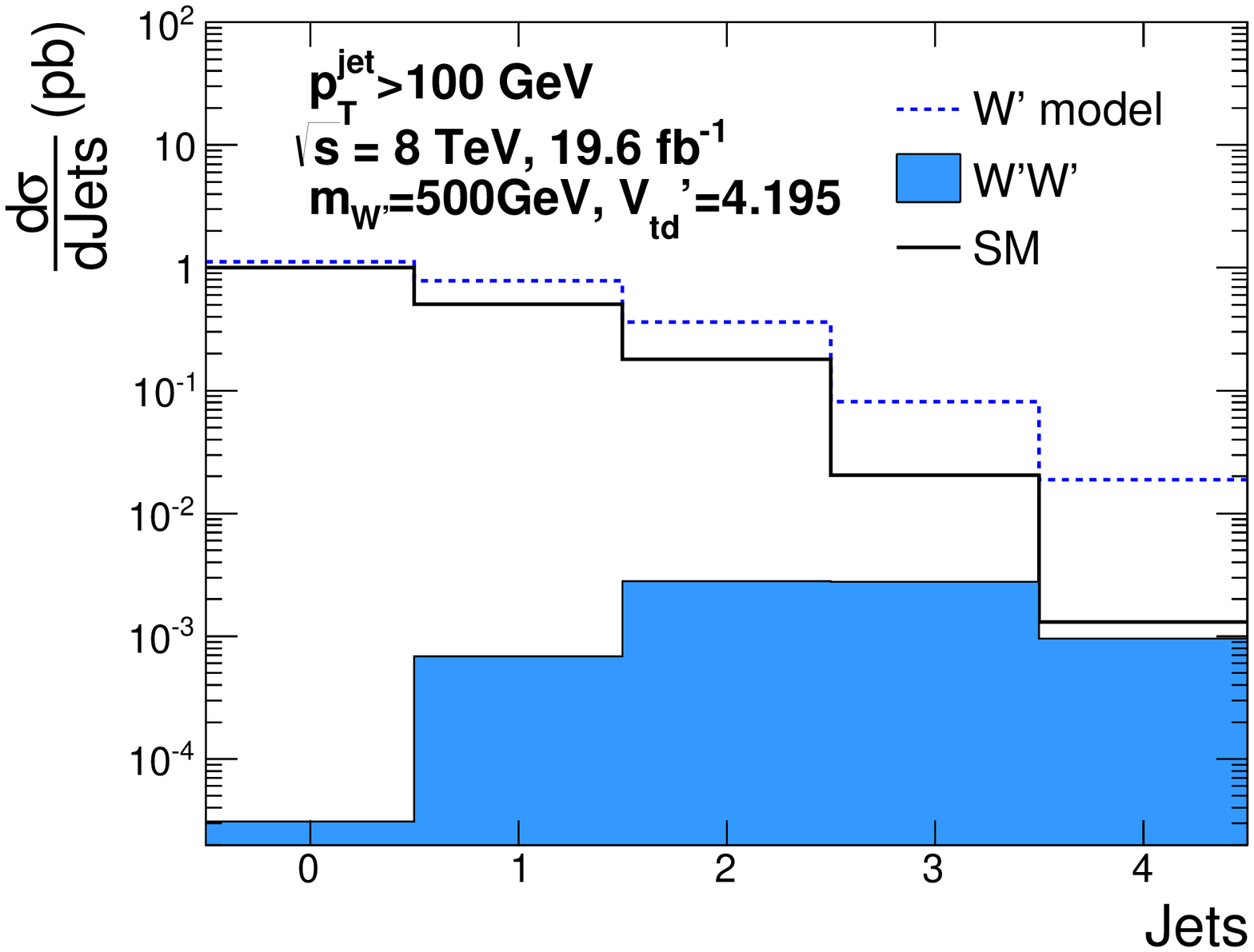}\\
\includegraphics[scale=0.29,clip]{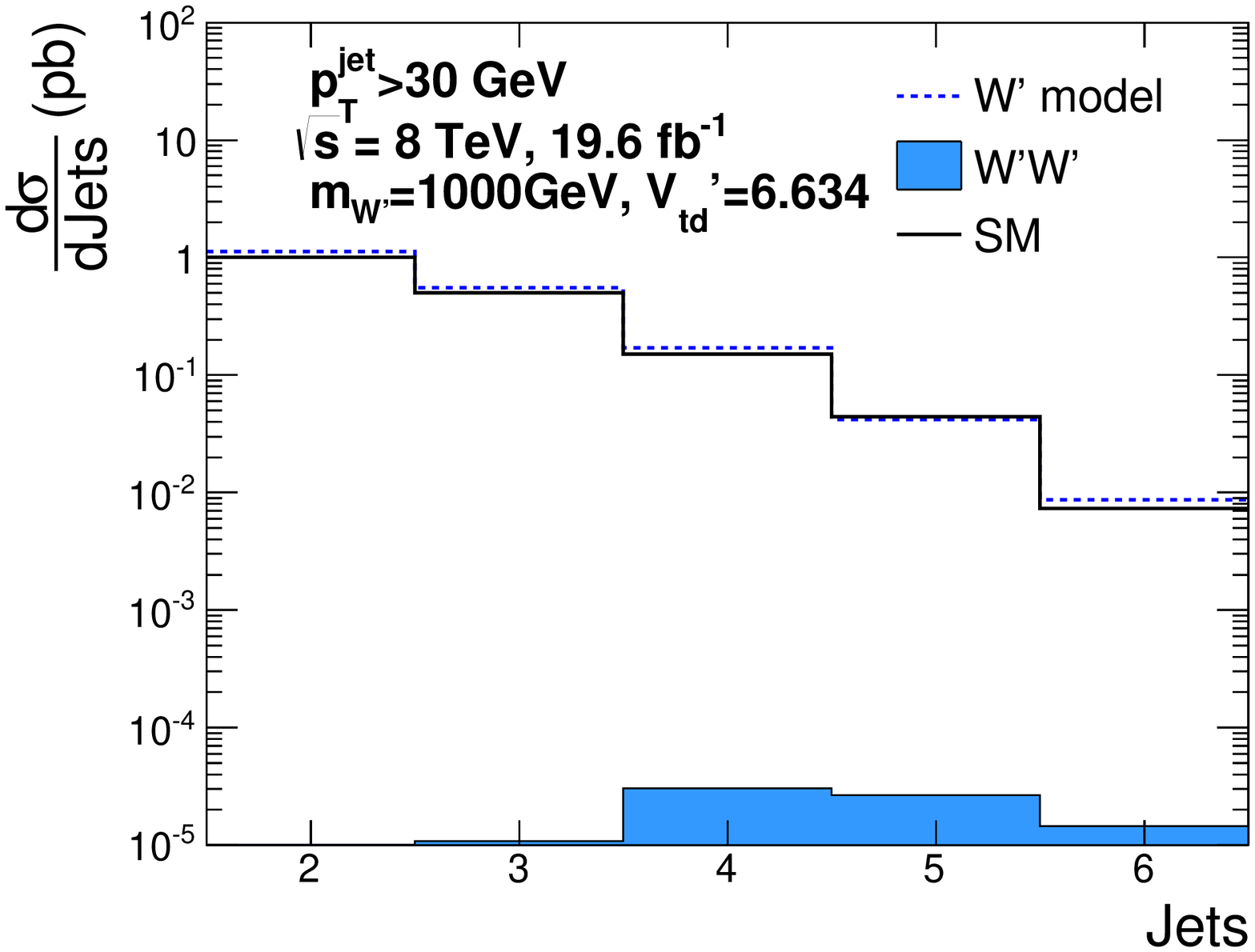}
\includegraphics[scale=0.29,clip]{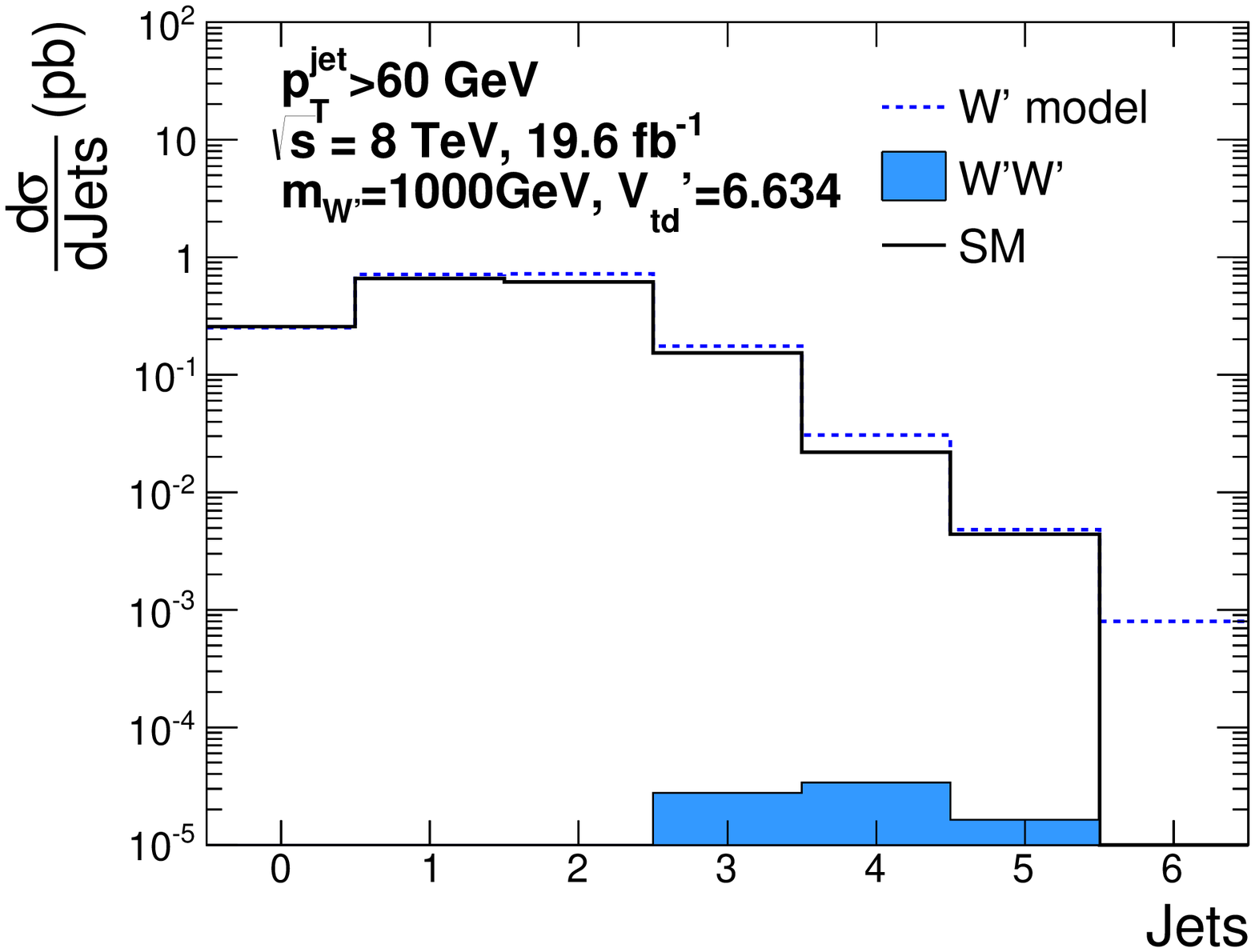}
\includegraphics[scale=0.29,clip]{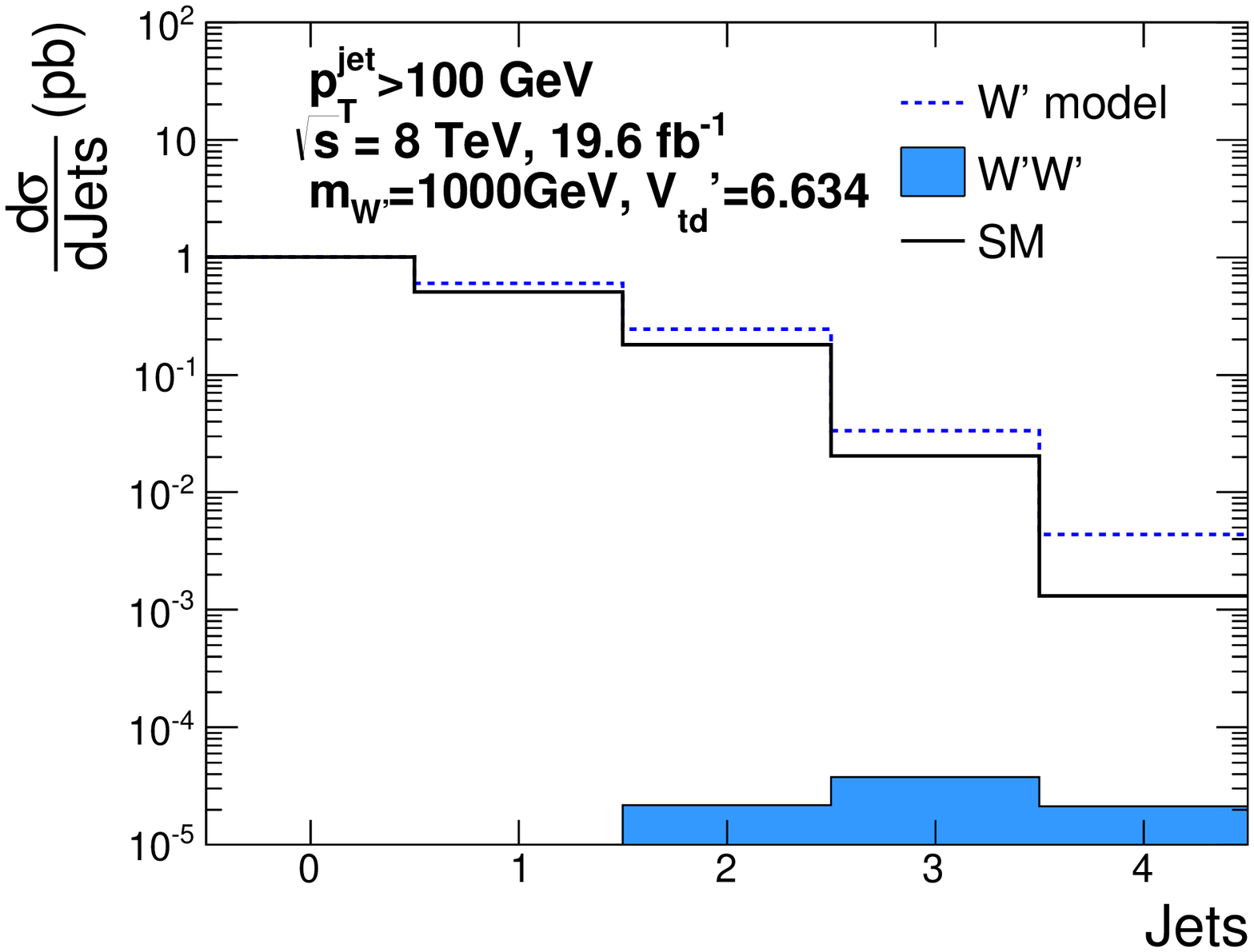}
\caption{Cross sections as a function of jet-multiplicity
for jets with $p_T>30$ GeV (left panels), $p_T>60$ GeV (middle panels), $p_T>100$
 GeV (right panels).   Results in the upper row are for $m_{W^\prime}=500$ GeV 
 and $V_{td}^\prime=4.195$, whereas  those in the lower row are for $m_{W^\prime}=1$ TeV 
 and $V_{td}^\prime=6.634$.  The (blue) dotted lines represent  
the complete calculation. The (sky blue) shadowed region is the contribution
from the $W^\prime W^\prime$ pair production process.  
\label{fig:wpwp_ab} }
\end{figure*}

In Fig.\ \ref{fig:wpwp_ab}, we show the absolute cross sections for
the $W^\prime W^\prime$ pair production process.  This process is
mediated by top-quark exchange and is fed by the $d \bar{d}$ parton
luminosity.  Its contribution to the jet multiplicity distribution is
typically several orders of magnitude smaller than the SM background,
as seen in Fig.\ \ref{fig:wpwp_ab}, becoming comparable when
$n_{\mathrm{jets}}\geqslant5$ for a light $W^\prime$ and large values of
the $p^{\mathrm{jet}}_T$ cut.  Overall, it is not an important component
of the complete contribution from the $W^\prime$ model in the regions
of parameter space explored in this paper.

\section{Summary and conclusions}
\label{sec:concl}

In this paper we investigate a model with right-handed coupling of a
$W^\prime$ boson to the first and third quark generations.  We fit for
values of the coupling constant $V^\prime_{td}$ consistent with
Tevatron data on the observed anomalously large top-quark
forward-backward asymmetry $\afb$ and $t\bar t$ cross section as a
function of $W^\prime$ mass (c.f., Fig.\ \ref{fig:1000gevtev}).  Our
theoretical expressions include higher-order $W^\prime$ loop
corrections whose contributions diminish the required best fit value
of the coupling strength compared to previous LO fits.

Given the model and our determination of its parameters, we then
investigate the consequences at the LHC.  For masses of the $W^\prime$
below 400~GeV, our previous comparison to early ATLAS data excluded
all relevant values of $V^\prime_{td}$ based on cross section rate
\cite{Duffty:2012zz}.  For larger masses, the predicted broader width
of the $W^\prime$ requires other strategies, and we focus on the
multiplicity distribution of jets accompanying a $t\bar{t}$ pair in
the full 8~TeV CMS data sample.  In the $W^\prime$ model, processes
such as associated $tW^\prime$ production and $W^\prime W^\prime$ pair
production, with $W^\prime \rightarrow \bar{t} d$, contribute to the
$t \bar{t}+\mathrm{n}j$ final state along with standard model QCD
production of $t \bar{t}+\mathrm{n}j$.

We simulate all $t\bar t+{\mathrm{n}}j$ processes including the
interference between the SM $t\bar t+{\mathrm{n}}j$ process and
inclusive $tW^\prime$ associated production; as well as contributions
from the $W^{\prime +}W^{\prime -}$ channel.  We examine the entire
mass range $200 < m_{W^\prime} < 1100$~GeV.  Our simulation includes
parton fragmentation and hadronization from
\textsc{PYTHIA6.4}~\cite{Sjostrand:2006za} and a detector simulation
using the \textsc{PGS} code~\cite{pgs}.  We compare our resulting jet
multiplicity distribution with data from the CMS
collaboration~\cite{CMS-PAS-TOP-12-041}.  We show that interference
plays a quantitatively significant role, altering the expected cross
sections and exclusion bounds.

The essential conclusions of our study are shown in Fig.\
\ref{fig:chi_ttj}.  Within the mass range $200 < m_{W^\prime} <
1100$~GeV, values of $V^\prime_{td}$ large enough to accommodate
$\afb$ observed at the Tevatron are incompatible with a good fit to
the jet multiplicity distribution at the LHC.

There are other new physics models proposed for the top-quark $\afb$
anomaly at Tevatron (for a more complete list of the references, cf.\
Ref.\ \cite{Duffty:2012zz}).  Many of them are disfavored or highly
constrained by LHC data and other direct or indirect experiments.  The
most studied of these models are $t$-channel $W^\prime$
\cite{Cheung:2009ch} and $Z^\prime$ \cite{Jung:2009jz}, and
$s$-channel axigluon models \cite{Antunano:2007da}.  The $W^\prime$
model is disfavored by this work.  The simplest $Z^\prime$ model is
highly constrained by the same-sign top-quark search at the LHC
\cite{Berger:2011ua,Aad:2012bb,Chatrchyan:2012sa}.  An updated
$Z^\prime$ model in which the $Z^\prime$ boson is not self-conjugate
\cite{Grinstein:2011yv}, so that there is no same-sign top-quark
signal at colliders, would also be strongly constrained by $t\bar
t+$jets data.  A heavy axigluon is constrained by dijet and $t\bar t$
resonance searches at the LHC \cite{Berger:2012tj}.  However, it is
still possible that a light axigluon ($\sim 300$ GeV) could explain
the $\afb$ anomaly \cite{Gresham:2012kv}.  Additional explanations
involving multiple Higgs doublets \cite{Han:2012dd,Wang:2012zv} that
are either composite \cite{Alvarez:2010js} or involve color-triplet
scalars remain open.

The difficulties encountered in constructing models of new physics
that can simultaneously accommodate the Tevatron asymmetry and LHC
observables motivate inquiry into the standard model QCD expectations
against which the data are compared.  We note that a simple change of
the renormalization scale brings the data and theory within $1
\sigma$.  This scale choice is similar to one that is used for the
forward-backward asymmetry in $e^-e^+\to \mu^-\mu^+$
\cite{Brodsky:2012ik}.  We look forward to the next stage of fully
differential NNLO calculations of $t\bar t$ production and decay that
should be incorporated into the understanding of experimental
acceptances, and allow for a full NLO prediction of $\afb$ after cuts.

\begin{acknowledgments}
The work of ELB and HZ at Argonne is supported in part by the U.S.\ 
DOE under Contract No.\ DE-AC02-06CH11357.  ZS and HZ are supported at
IIT by the DOE under Contract No.\ DE-SC0008347.  Part of this work
was done while ELB was visiting the Aspen Center for Physics and was
supported there in part by the National Science Foundation under
Contract No.\ PHYS-1066293.  ELB is pleased to recognize this support
and the hospitality of the Aspen Center for Physics.
\end{acknowledgments}

\bibliographystyle{apsrev}
\bibliography{wpttjlhc}

\begin{thebibliography}{58}
\expandafter\ifx\csname natexlab\endcsname\relax\def\natexlab#1{#1}\fi
\expandafter\ifx\csname bibnamefont\endcsname\relax
  \def\bibnamefont#1{#1}\fi
\expandafter\ifx\csname bibfnamefont\endcsname\relax
  \def\bibfnamefont#1{#1}\fi
\expandafter\ifx\csname citenamefont\endcsname\relax
  \def\citenamefont#1{#1}\fi
\expandafter\ifx\csname url\endcsname\relax
  \def\url#1{\texttt{#1}}\fi
\expandafter\ifx\csname urlprefix\endcsname\relax\def\urlprefix{URL }\fi
\providecommand{\bibinfo}[2]{#2}
\providecommand{\eprint}[2][]{\url{#2}}

\bibitem[{CMS(2013{\natexlab{a}})}]{CMS-PAS-B2G-12-010}
\bibinfo{type}{Tech. Rep.} \bibinfo{number}{CMS-PAS-B2G-12-010},
  \bibinfo{institution}{CERN}, \bibinfo{address}{Geneva}
  (\bibinfo{year}{2013}{\natexlab{a}}).

\bibitem[{ATL(2013)}]{ATLAS-CONF-2013-050}
\bibinfo{type}{Tech. Rep.} \bibinfo{number}{ATLAS-CONF-2013-050},
  \bibinfo{institution}{CERN}, \bibinfo{address}{Geneva}
  (\bibinfo{year}{2013}).

\bibitem[{\citenamefont{Duffty and Sullivan}(2012)}]{Duffty:2012rf}
\bibinfo{author}{\bibfnamefont{D.}~\bibnamefont{Duffty}} \bibnamefont{and}
  \bibinfo{author}{\bibfnamefont{Z.}~\bibnamefont{Sullivan}},
  \bibinfo{journal}{Phys.\ Rev.} \textbf{\bibinfo{volume}{D86}},
  \bibinfo{pages}{075018} (\bibinfo{year}{2012}), \eprint{1208.4858}.

\bibitem[{\citenamefont{Duffty and Sullivan}(2013)}]{Duffty:2013aba}
\bibinfo{author}{\bibfnamefont{D.}~\bibnamefont{Duffty}} \bibnamefont{and}
  \bibinfo{author}{\bibfnamefont{Z.}~\bibnamefont{Sullivan}}
  (\bibinfo{year}{2013}), \eprint{1307.1820}.

\bibitem[{\citenamefont{Cheung et~al.}(2009)\citenamefont{Cheung, Keung, and
  Yuan}}]{Cheung:2009ch}
\bibinfo{author}{\bibfnamefont{K.}~\bibnamefont{Cheung}},
  \bibinfo{author}{\bibfnamefont{W.-Y.} \bibnamefont{Keung}}, \bibnamefont{and}
  \bibinfo{author}{\bibfnamefont{T.-C.} \bibnamefont{Yuan}},
  \bibinfo{journal}{Phys.\ Lett.} \textbf{\bibinfo{volume}{B682}},
  \bibinfo{pages}{287} (\bibinfo{year}{2009}), \eprint{0908.2589}.

\bibitem[{\citenamefont{Barger et~al.}(2010)\citenamefont{Barger, Keung, and
  Yu}}]{Barger:2010mw}
\bibinfo{author}{\bibfnamefont{V.}~\bibnamefont{Barger}},
  \bibinfo{author}{\bibfnamefont{W.-Y.} \bibnamefont{Keung}}, \bibnamefont{and}
  \bibinfo{author}{\bibfnamefont{C.-T.} \bibnamefont{Yu}},
  \bibinfo{journal}{Phys.\ Rev.} \textbf{\bibinfo{volume}{D81}},
  \bibinfo{pages}{113009} (\bibinfo{year}{2010}), \eprint{1002.1048}.

\bibitem[{\citenamefont{Cao et~al.}(2010)\citenamefont{Cao, McKeen, Rosner,
  Shaughnessy, and Wagner}}]{Cao:2010zb}
\bibinfo{author}{\bibfnamefont{Q.-H.} \bibnamefont{Cao}},
  \bibinfo{author}{\bibfnamefont{D.}~\bibnamefont{McKeen}},
  \bibinfo{author}{\bibfnamefont{J.~L.} \bibnamefont{Rosner}},
  \bibinfo{author}{\bibfnamefont{G.}~\bibnamefont{Shaughnessy}},
  \bibnamefont{and} \bibinfo{author}{\bibfnamefont{C.~E.}
  \bibnamefont{Wagner}}, \bibinfo{journal}{Phys.\ Rev.}
  \textbf{\bibinfo{volume}{D81}}, \bibinfo{pages}{114004}
  (\bibinfo{year}{2010}), \eprint{1003.3461}.

\bibitem[{\citenamefont{Cheung and Yuan}(2011)}]{Cheung:2011qa}
\bibinfo{author}{\bibfnamefont{K.}~\bibnamefont{Cheung}} \bibnamefont{and}
  \bibinfo{author}{\bibfnamefont{T.-C.} \bibnamefont{Yuan}},
  \bibinfo{journal}{Phys.\ Rev.} \textbf{\bibinfo{volume}{D83}},
  \bibinfo{pages}{074006} (\bibinfo{year}{2011}), \eprint{1101.1445}.

\bibitem[{\citenamefont{Shelton and Zurek}(2011)}]{Shelton:2011hq}
\bibinfo{author}{\bibfnamefont{J.}~\bibnamefont{Shelton}} \bibnamefont{and}
  \bibinfo{author}{\bibfnamefont{K.~M.} \bibnamefont{Zurek}},
  \bibinfo{journal}{Phys.\ Rev.} \textbf{\bibinfo{volume}{D83}},
  \bibinfo{pages}{091701} (\bibinfo{year}{2011}), \eprint{1101.5392}.

\bibitem[{\citenamefont{Gresham
  et~al.}(2011{\natexlab{a}})\citenamefont{Gresham, Kim, and
  Zurek}}]{Gresham:2011dg}
\bibinfo{author}{\bibfnamefont{M.~I.} \bibnamefont{Gresham}},
  \bibinfo{author}{\bibfnamefont{I.-W.} \bibnamefont{Kim}}, \bibnamefont{and}
  \bibinfo{author}{\bibfnamefont{K.~M.} \bibnamefont{Zurek}},
  \bibinfo{journal}{Phys.\ Rev.} \textbf{\bibinfo{volume}{D84}},
  \bibinfo{pages}{034025} (\bibinfo{year}{2011}{\natexlab{a}}),
  \eprint{1102.0018}.

\bibitem[{\citenamefont{Barger et~al.}(2011)\citenamefont{Barger, Keung, and
  Yu}}]{Barger:2011ih}
\bibinfo{author}{\bibfnamefont{V.}~\bibnamefont{Barger}},
  \bibinfo{author}{\bibfnamefont{W.-Y.} \bibnamefont{Keung}}, \bibnamefont{and}
  \bibinfo{author}{\bibfnamefont{C.-T.} \bibnamefont{Yu}},
  \bibinfo{journal}{Phys.\ Lett.} \textbf{\bibinfo{volume}{B698}},
  \bibinfo{pages}{243} (\bibinfo{year}{2011}), \eprint{1102.0279}.

\bibitem[{\citenamefont{Bhattacherjee et~al.}(2011)\citenamefont{Bhattacherjee,
  Biswal, and Ghosh}}]{Bhattacherjee:2011nr}
\bibinfo{author}{\bibfnamefont{B.}~\bibnamefont{Bhattacherjee}},
  \bibinfo{author}{\bibfnamefont{S.~S.} \bibnamefont{Biswal}},
  \bibnamefont{and} \bibinfo{author}{\bibfnamefont{D.}~\bibnamefont{Ghosh}},
  \bibinfo{journal}{Phys.\ Rev.} \textbf{\bibinfo{volume}{D83}},
  \bibinfo{pages}{091501} (\bibinfo{year}{2011}), \eprint{1102.0545}.

\bibitem[{\citenamefont{Craig et~al.}(2011)\citenamefont{Craig, Kilic, and
  Strassler}}]{Craig:2011an}
\bibinfo{author}{\bibfnamefont{N.}~\bibnamefont{Craig}},
  \bibinfo{author}{\bibfnamefont{C.}~\bibnamefont{Kilic}}, \bibnamefont{and}
  \bibinfo{author}{\bibfnamefont{M.~J.} \bibnamefont{Strassler}},
  \bibinfo{journal}{Phys.\ Rev.} \textbf{\bibinfo{volume}{D84}},
  \bibinfo{pages}{035012} (\bibinfo{year}{2011}), \eprint{1103.2127}.

\bibitem[{\citenamefont{Gresham
  et~al.}(2011{\natexlab{b}})\citenamefont{Gresham, Kim, and
  Zurek}}]{Gresham:2011pa}
\bibinfo{author}{\bibfnamefont{M.~I.} \bibnamefont{Gresham}},
  \bibinfo{author}{\bibfnamefont{I.-W.} \bibnamefont{Kim}}, \bibnamefont{and}
  \bibinfo{author}{\bibfnamefont{K.~M.} \bibnamefont{Zurek}},
  \bibinfo{journal}{Phys.\ Rev.} \textbf{\bibinfo{volume}{D83}},
  \bibinfo{pages}{114027} (\bibinfo{year}{2011}{\natexlab{b}}),
  \eprint{1103.3501}.

\bibitem[{\citenamefont{Chen et~al.}(2011)\citenamefont{Chen, Law, and
  Li}}]{Chen:2011mga}
\bibinfo{author}{\bibfnamefont{C.-H.} \bibnamefont{Chen}},
  \bibinfo{author}{\bibfnamefont{S.~S.} \bibnamefont{Law}}, \bibnamefont{and}
  \bibinfo{author}{\bibfnamefont{R.-H.} \bibnamefont{Li}},
  \bibinfo{journal}{J.\ Phys.} \textbf{\bibinfo{volume}{G38}},
  \bibinfo{pages}{115008} (\bibinfo{year}{2011}), \eprint{1104.1497}.

\bibitem[{\citenamefont{Krohn et~al.}(2011)\citenamefont{Krohn, Liu, Shelton,
  and Wang}}]{Krohn:2011tw}
\bibinfo{author}{\bibfnamefont{D.}~\bibnamefont{Krohn}},
  \bibinfo{author}{\bibfnamefont{T.}~\bibnamefont{Liu}},
  \bibinfo{author}{\bibfnamefont{J.}~\bibnamefont{Shelton}}, \bibnamefont{and}
  \bibinfo{author}{\bibfnamefont{L.-T.} \bibnamefont{Wang}},
  \bibinfo{journal}{Phys.\ Rev.} \textbf{\bibinfo{volume}{D84}},
  \bibinfo{pages}{074034} (\bibinfo{year}{2011}), \eprint{1105.3743}.

\bibitem[{\citenamefont{Gresham
  et~al.}(2012{\natexlab{a}})\citenamefont{Gresham, Kim, and
  Zurek}}]{Gresham:2011fx}
\bibinfo{author}{\bibfnamefont{M.~I.} \bibnamefont{Gresham}},
  \bibinfo{author}{\bibfnamefont{I.-W.} \bibnamefont{Kim}}, \bibnamefont{and}
  \bibinfo{author}{\bibfnamefont{K.~M.} \bibnamefont{Zurek}},
  \bibinfo{journal}{Phys.\ Rev.} \textbf{\bibinfo{volume}{D85}},
  \bibinfo{pages}{014022} (\bibinfo{year}{2012}{\natexlab{a}}),
  \eprint{1107.4364}.

\bibitem[{\citenamefont{Cao et~al.}(2012)\citenamefont{Cao, Hikasa, Wang, Wu,
  and Yang}}]{Cao:2011hr}
\bibinfo{author}{\bibfnamefont{J.}~\bibnamefont{Cao}},
  \bibinfo{author}{\bibfnamefont{K.}~\bibnamefont{Hikasa}},
  \bibinfo{author}{\bibfnamefont{L.}~\bibnamefont{Wang}},
  \bibinfo{author}{\bibfnamefont{L.}~\bibnamefont{Wu}}, \bibnamefont{and}
  \bibinfo{author}{\bibfnamefont{J.~M.} \bibnamefont{Yang}},
  \bibinfo{journal}{Phys.\ Rev.} \textbf{\bibinfo{volume}{D85}},
  \bibinfo{pages}{014025} (\bibinfo{year}{2012}), \eprint{1109.6543}.

\bibitem[{\citenamefont{Yan et~al.}(2012)\citenamefont{Yan, Wang, Shao, and
  Li}}]{Yan:2011tf}
\bibinfo{author}{\bibfnamefont{K.}~\bibnamefont{Yan}},
  \bibinfo{author}{\bibfnamefont{J.}~\bibnamefont{Wang}},
  \bibinfo{author}{\bibfnamefont{D.~Y.} \bibnamefont{Shao}}, \bibnamefont{and}
  \bibinfo{author}{\bibfnamefont{C.~S.} \bibnamefont{Li}},
  \bibinfo{journal}{Phys.\ Rev.} \textbf{\bibinfo{volume}{D85}},
  \bibinfo{pages}{034020} (\bibinfo{year}{2012}), \eprint{1110.6684}.

\bibitem[{\citenamefont{Berger et~al.}(2011{\natexlab{a}})\citenamefont{Berger,
  Cao, Chen, Yu, and Zhang}}]{Berger:2011pu}
\bibinfo{author}{\bibfnamefont{E.~L.} \bibnamefont{Berger}},
  \bibinfo{author}{\bibfnamefont{Q.-H.} \bibnamefont{Cao}},
  \bibinfo{author}{\bibfnamefont{C.-R.} \bibnamefont{Chen}},
  \bibinfo{author}{\bibfnamefont{J.-H.} \bibnamefont{Yu}}, \bibnamefont{and}
  \bibinfo{author}{\bibfnamefont{H.}~\bibnamefont{Zhang}}
  (\bibinfo{year}{2011}{\natexlab{a}}), \eprint{1111.3641}.

\bibitem[{\citenamefont{Knapen et~al.}(2012)\citenamefont{Knapen, Zhao, and
  Strassler}}]{Knapen:2011hu}
\bibinfo{author}{\bibfnamefont{S.}~\bibnamefont{Knapen}},
  \bibinfo{author}{\bibfnamefont{Y.}~\bibnamefont{Zhao}}, \bibnamefont{and}
  \bibinfo{author}{\bibfnamefont{M.~J.} \bibnamefont{Strassler}},
  \bibinfo{journal}{Phys.\ Rev.} \textbf{\bibinfo{volume}{D86}},
  \bibinfo{pages}{014013} (\bibinfo{year}{2012}), \eprint{1111.5857}.

\bibitem[{\citenamefont{Berger et~al.}(2012)\citenamefont{Berger, Cao, Chen,
  Yu, and Zhang}}]{Berger:2012nw}
\bibinfo{author}{\bibfnamefont{E.~L.} \bibnamefont{Berger}},
  \bibinfo{author}{\bibfnamefont{Q.-H.} \bibnamefont{Cao}},
  \bibinfo{author}{\bibfnamefont{C.-R.} \bibnamefont{Chen}},
  \bibinfo{author}{\bibfnamefont{J.-H.} \bibnamefont{Yu}}, \bibnamefont{and}
  \bibinfo{author}{\bibfnamefont{H.}~\bibnamefont{Zhang}},
  \bibinfo{journal}{Phys.\ Rev.\ Lett.} \textbf{\bibinfo{volume}{108}},
  \bibinfo{pages}{072002} (\bibinfo{year}{2012}), \eprint{1201.1790}.

\bibitem[{\citenamefont{Duffty et~al.}(2012)\citenamefont{Duffty, Sullivan, and
  Zhang}}]{Duffty:2012zz}
\bibinfo{author}{\bibfnamefont{D.}~\bibnamefont{Duffty}},
  \bibinfo{author}{\bibfnamefont{Z.}~\bibnamefont{Sullivan}}, \bibnamefont{and}
  \bibinfo{author}{\bibfnamefont{H.}~\bibnamefont{Zhang}},
  \bibinfo{journal}{Phys.\ Rev.} \textbf{\bibinfo{volume}{D85}},
  \bibinfo{pages}{094027} (\bibinfo{year}{2012}), \eprint{1203.4489}.

\bibitem[{\citenamefont{Adelman et~al.}(2013)\citenamefont{Adelman, Ferrando,
  and White}}]{Adelman:2012py}
\bibinfo{author}{\bibfnamefont{J.}~\bibnamefont{Adelman}},
  \bibinfo{author}{\bibfnamefont{J.}~\bibnamefont{Ferrando}}, \bibnamefont{and}
  \bibinfo{author}{\bibfnamefont{C.}~\bibnamefont{White}},
  \bibinfo{journal}{J.\ High Energy Phys.} \textbf{\bibinfo{volume}{1302}},
  \bibinfo{pages}{091} (\bibinfo{year}{2013}), \eprint{1206.5731}.

\bibitem[{\citenamefont{Endo and Iwamoto}(2013)}]{Endo:2012mi}
\bibinfo{author}{\bibfnamefont{M.}~\bibnamefont{Endo}} \bibnamefont{and}
  \bibinfo{author}{\bibfnamefont{S.}~\bibnamefont{Iwamoto}},
  \bibinfo{journal}{Phys.\ Lett.} \textbf{\bibinfo{volume}{B718}},
  \bibinfo{pages}{1070} (\bibinfo{year}{2013}), \eprint{1207.5900}.

\bibitem[{\citenamefont{Berger et~al.}(2013)\citenamefont{Berger, Cao, Chen,
  and Zhang}}]{Berger:2012tj}
\bibinfo{author}{\bibfnamefont{E.~L.} \bibnamefont{Berger}},
  \bibinfo{author}{\bibfnamefont{Q.-H.} \bibnamefont{Cao}},
  \bibinfo{author}{\bibfnamefont{C.-R.} \bibnamefont{Chen}}, \bibnamefont{and}
  \bibinfo{author}{\bibfnamefont{H.}~\bibnamefont{Zhang}},
  \bibinfo{journal}{Phys.\ Rev.} \textbf{\bibinfo{volume}{D88}},
  \bibinfo{pages}{014033} (\bibinfo{year}{2013}), \eprint{1209.4899}.

\bibitem[{\citenamefont{Berger et~al.}(2011{\natexlab{b}})\citenamefont{Berger,
  Cao, Yu, and Yuan}}]{Berger:2011xk}
\bibinfo{author}{\bibfnamefont{E.~L.} \bibnamefont{Berger}},
  \bibinfo{author}{\bibfnamefont{Q.-H.} \bibnamefont{Cao}},
  \bibinfo{author}{\bibfnamefont{J.-H.} \bibnamefont{Yu}}, \bibnamefont{and}
  \bibinfo{author}{\bibfnamefont{C.-P.} \bibnamefont{Yuan}},
  \bibinfo{journal}{Phys.\ Rev.} \textbf{\bibinfo{volume}{D84}},
  \bibinfo{pages}{095026} (\bibinfo{year}{2011}{\natexlab{b}}),
  \eprint{1108.3613}.

\bibitem[{\citenamefont{Aaltonen et~al.}(2013)}]{Aaltonen:2012it}
\bibinfo{author}{\bibfnamefont{T.}~\bibnamefont{Aaltonen}} \bibnamefont{et~al.}
  (\bibinfo{collaboration}{CDF Collaboration}), \bibinfo{journal}{Phys.\ Rev.}
  \textbf{\bibinfo{volume}{D87}}, \bibinfo{pages}{092002}
  (\bibinfo{year}{2013}), \eprint{1211.1003}.

\bibitem[{\citenamefont{Abazov et~al.}(2011)}]{Abazov:2011rq}
\bibinfo{author}{\bibfnamefont{V.~M.} \bibnamefont{Abazov}}
  \bibnamefont{et~al.} (\bibinfo{collaboration}{D0 Collaboration}),
  \bibinfo{journal}{Phys.\ Rev.} \textbf{\bibinfo{volume}{D84}},
  \bibinfo{pages}{112005} (\bibinfo{year}{2011}), \eprint{1107.4995}.

\bibitem[{ATL(2011)}]{ATLAS-CONF-2011-100}
\bibinfo{type}{Tech. Rep.} \bibinfo{number}{ATLAS-CONF-2011-100},
  \bibinfo{institution}{CERN}, \bibinfo{address}{Geneva}
  (\bibinfo{year}{2011}).

\bibitem[{\citenamefont{Chatrchyan
  et~al.}(2012{\natexlab{a}})}]{Chatrchyan:2012su}
\bibinfo{author}{\bibfnamefont{S.}~\bibnamefont{Chatrchyan}}
  \bibnamefont{et~al.} (\bibinfo{collaboration}{CMS Collaboration}),
  \bibinfo{journal}{Phys.\ Lett.} \textbf{\bibinfo{volume}{B717}},
  \bibinfo{pages}{351} (\bibinfo{year}{2012}{\natexlab{a}}),
  \eprint{1206.3921}.

\bibitem[{\citenamefont{Aad et~al.}(2012{\natexlab{a}})}]{Aad:2012em}
\bibinfo{author}{\bibfnamefont{G.}~\bibnamefont{Aad}} \bibnamefont{et~al.}
  (\bibinfo{collaboration}{ATLAS Collaboration}), \bibinfo{journal}{Phys.\
  Rev.} \textbf{\bibinfo{volume}{D86}}, \bibinfo{pages}{091103}
  (\bibinfo{year}{2012}{\natexlab{a}}), \eprint{1209.6593}.

\bibitem[{\citenamefont{Li et~al.}(2005)\citenamefont{Li, Mishima, and
  Sanda}}]{Li:2005kt}
\bibinfo{author}{\bibfnamefont{H.-n.} \bibnamefont{Li}},
  \bibinfo{author}{\bibfnamefont{S.}~\bibnamefont{Mishima}}, \bibnamefont{and}
  \bibinfo{author}{\bibfnamefont{A.}~\bibnamefont{Sanda}},
  \bibinfo{journal}{Phys.\ Rev.} \textbf{\bibinfo{volume}{D72}},
  \bibinfo{pages}{114005} (\bibinfo{year}{2005}), \eprint{hep-ph/0508041}.

\bibitem[{\citenamefont{Gresham
  et~al.}(2012{\natexlab{b}})\citenamefont{Gresham, Kim, Tulin, and
  Zurek}}]{Gresham:2012wc}
\bibinfo{author}{\bibfnamefont{M.~I.} \bibnamefont{Gresham}},
  \bibinfo{author}{\bibfnamefont{I.-W.} \bibnamefont{Kim}},
  \bibinfo{author}{\bibfnamefont{S.}~\bibnamefont{Tulin}}, \bibnamefont{and}
  \bibinfo{author}{\bibfnamefont{K.~M.} \bibnamefont{Zurek}},
  \bibinfo{journal}{Phys.\ Rev.} \textbf{\bibinfo{volume}{D86}},
  \bibinfo{pages}{034029} (\bibinfo{year}{2012}{\natexlab{b}}),
  \eprint{1203.1320}.

\bibitem[{\citenamefont{Beenakker et~al.}(1994)\citenamefont{Beenakker, Denner,
  Hollik, Mertig, Sack et~al.}}]{Beenakker:1993yr}
\bibinfo{author}{\bibfnamefont{W.}~\bibnamefont{Beenakker}},
  \bibinfo{author}{\bibfnamefont{A.}~\bibnamefont{Denner}},
  \bibinfo{author}{\bibfnamefont{W.}~\bibnamefont{Hollik}},
  \bibinfo{author}{\bibfnamefont{R.}~\bibnamefont{Mertig}},
  \bibinfo{author}{\bibfnamefont{T.}~\bibnamefont{Sack}}, \bibnamefont{et~al.},
  \bibinfo{journal}{Nucl.\ Phys.} \textbf{\bibinfo{volume}{B411}},
  \bibinfo{pages}{343} (\bibinfo{year}{1994}).

\bibitem[{\citenamefont{Gabrielli and Raidal}(2011)}]{Gabrielli:2011jf}
\bibinfo{author}{\bibfnamefont{E.}~\bibnamefont{Gabrielli}} \bibnamefont{and}
  \bibinfo{author}{\bibfnamefont{M.}~\bibnamefont{Raidal}},
  \bibinfo{journal}{Phys.\ Rev.} \textbf{\bibinfo{volume}{D84}},
  \bibinfo{pages}{054017} (\bibinfo{year}{2011}), \eprint{1106.4553}.

\bibitem[{\citenamefont{Martin et~al.}(2009)\citenamefont{Martin, Stirling,
  Thorne, and Watt}}]{Martin:2009iq}
\bibinfo{author}{\bibfnamefont{A.}~\bibnamefont{Martin}},
  \bibinfo{author}{\bibfnamefont{W.}~\bibnamefont{Stirling}},
  \bibinfo{author}{\bibfnamefont{R.}~\bibnamefont{Thorne}}, \bibnamefont{and}
  \bibinfo{author}{\bibfnamefont{G.}~\bibnamefont{Watt}},
  \bibinfo{journal}{Eur.\ Phys.\ J.} \textbf{\bibinfo{volume}{C63}},
  \bibinfo{pages}{189} (\bibinfo{year}{2009}), \eprint{0901.0002}.

\bibitem[{D0-(2012)}]{D0-NOTE-6363}
\bibinfo{type}{Tech. Rep.} \bibinfo{number}{D0 Note 6363},
  \bibinfo{institution}{FERMILAB}, \bibinfo{address}{Batavia}
  (\bibinfo{year}{2012}).

\bibitem[{\citenamefont{Kuhn and Rodrigo}(2012)}]{Kuhn:2011ri}
\bibinfo{author}{\bibfnamefont{J.~H.} \bibnamefont{Kuhn}} \bibnamefont{and}
  \bibinfo{author}{\bibfnamefont{G.}~\bibnamefont{Rodrigo}},
  \bibinfo{journal}{J.\ High Energy Phys.} \textbf{\bibinfo{volume}{1201}},
  \bibinfo{pages}{063} (\bibinfo{year}{2012}), \eprint{1109.6830}.

\bibitem[{\citenamefont{Aaltonen et~al.}(2009)}]{Aaltonen:2009iz}
\bibinfo{author}{\bibfnamefont{T.}~\bibnamefont{Aaltonen}} \bibnamefont{et~al.}
  (\bibinfo{collaboration}{CDF Collaboration}), \bibinfo{journal}{Phys.\ Rev.\
  Lett.} \textbf{\bibinfo{volume}{102}}, \bibinfo{pages}{222003}
  (\bibinfo{year}{2009}), \eprint{0903.2850}.

\bibitem[{\citenamefont{Campbell et~al.}(2013)\citenamefont{Campbell, Ellis,
  and Williams}}]{mcfm}
\bibinfo{author}{\bibfnamefont{J.}~\bibnamefont{Campbell}},
  \bibinfo{author}{\bibfnamefont{K.}~\bibnamefont{Ellis}}, \bibnamefont{and}
  \bibinfo{author}{\bibfnamefont{C.}~\bibnamefont{Williams}}
  (\bibinfo{year}{2013}), \urlprefix\url{http://mcfm.fnal.gov/}.

\bibitem[{\citenamefont{Alwall et~al.}(2011)\citenamefont{Alwall, Herquet,
  Maltoni, Mattelaer, and Stelzer}}]{Alwall:2011uj}
\bibinfo{author}{\bibfnamefont{J.}~\bibnamefont{Alwall}},
  \bibinfo{author}{\bibfnamefont{M.}~\bibnamefont{Herquet}},
  \bibinfo{author}{\bibfnamefont{F.}~\bibnamefont{Maltoni}},
  \bibinfo{author}{\bibfnamefont{O.}~\bibnamefont{Mattelaer}},
  \bibnamefont{and} \bibinfo{author}{\bibfnamefont{T.}~\bibnamefont{Stelzer}},
  \bibinfo{journal}{J.\ High Energy Phys.} \textbf{\bibinfo{volume}{1106}},
  \bibinfo{pages}{128} (\bibinfo{year}{2011}), \eprint{1106.0522}.

\bibitem[{\citenamefont{Conway et~al.}(2013)}]{pgs}
\bibinfo{author}{\bibfnamefont{J.}~\bibnamefont{Conway}} \bibnamefont{et~al.}
  (\bibinfo{year}{2013}),
  \urlprefix\url{http://www.physics.ucdavis.edu/~conway/research/software/pgs/%
pgs4-olympics.htm}.

\bibitem[{CMS(2013{\natexlab{b}})}]{CMS-PAS-TOP-12-041}
\bibinfo{type}{Tech. Rep.} \bibinfo{number}{CMS-PAS-TOP-12-041},
  \bibinfo{institution}{CERN}, \bibinfo{address}{Geneva}
  (\bibinfo{year}{2013}{\natexlab{b}}).

\bibitem[{\citenamefont{Sjostrand et~al.}(2006)\citenamefont{Sjostrand, Mrenna,
  and Skands}}]{Sjostrand:2006za}
\bibinfo{author}{\bibfnamefont{T.}~\bibnamefont{Sjostrand}},
  \bibinfo{author}{\bibfnamefont{S.}~\bibnamefont{Mrenna}}, \bibnamefont{and}
  \bibinfo{author}{\bibfnamefont{P.~Z.} \bibnamefont{Skands}},
  \bibinfo{journal}{J.\ High Energy Phys.} \textbf{\bibinfo{volume}{0605}},
  \bibinfo{pages}{026} (\bibinfo{year}{2006}), \eprint{hep-ph/0603175}.

\bibitem[{\citenamefont{Mangano et~al.}(2007)\citenamefont{Mangano, Moretti,
  Piccinini, and Treccani}}]{Mangano:2006rw}
\bibinfo{author}{\bibfnamefont{M.~L.} \bibnamefont{Mangano}},
  \bibinfo{author}{\bibfnamefont{M.}~\bibnamefont{Moretti}},
  \bibinfo{author}{\bibfnamefont{F.}~\bibnamefont{Piccinini}},
  \bibnamefont{and} \bibinfo{author}{\bibfnamefont{M.}~\bibnamefont{Treccani}},
  \bibinfo{journal}{J.\ High Energy Phys.} \textbf{\bibinfo{volume}{0701}},
  \bibinfo{pages}{013} (\bibinfo{year}{2007}), \eprint{hep-ph/0611129}.

\bibitem[{\citenamefont{Aaltonen et~al.}(2012)}]{Aaltonen:2012qn}
\bibinfo{author}{\bibfnamefont{T.}~\bibnamefont{Aaltonen}} \bibnamefont{et~al.}
  (\bibinfo{collaboration}{CDF Collaboration}), \bibinfo{journal}{Phys.\ Rev.\
  Lett.} \textbf{\bibinfo{volume}{108}}, \bibinfo{pages}{211805}
  (\bibinfo{year}{2012}), \eprint{1203.3894}.

\bibitem[{\citenamefont{Jung et~al.}(2010)\citenamefont{Jung, Murayama, Pierce,
  and Wells}}]{Jung:2009jz}
\bibinfo{author}{\bibfnamefont{S.}~\bibnamefont{Jung}},
  \bibinfo{author}{\bibfnamefont{H.}~\bibnamefont{Murayama}},
  \bibinfo{author}{\bibfnamefont{A.}~\bibnamefont{Pierce}}, \bibnamefont{and}
  \bibinfo{author}{\bibfnamefont{J.~D.} \bibnamefont{Wells}},
  \bibinfo{journal}{Phys.\ Rev.} \textbf{\bibinfo{volume}{D81}},
  \bibinfo{pages}{015004} (\bibinfo{year}{2010}), \eprint{0907.4112}.

\bibitem[{\citenamefont{Antunano et~al.}(2008)\citenamefont{Antunano, Kuhn, and
  Rodrigo}}]{Antunano:2007da}
\bibinfo{author}{\bibfnamefont{O.}~\bibnamefont{Antunano}},
  \bibinfo{author}{\bibfnamefont{J.~H.} \bibnamefont{Kuhn}}, \bibnamefont{and}
  \bibinfo{author}{\bibfnamefont{G.}~\bibnamefont{Rodrigo}},
  \bibinfo{journal}{Phys.\ Rev.} \textbf{\bibinfo{volume}{D77}},
  \bibinfo{pages}{014003} (\bibinfo{year}{2008}), \eprint{0709.1652}.

\bibitem[{\citenamefont{Berger et~al.}(2011{\natexlab{c}})\citenamefont{Berger,
  Cao, Chen, Li, and Zhang}}]{Berger:2011ua}
\bibinfo{author}{\bibfnamefont{E.~L.} \bibnamefont{Berger}},
  \bibinfo{author}{\bibfnamefont{Q.-H.} \bibnamefont{Cao}},
  \bibinfo{author}{\bibfnamefont{C.-R.} \bibnamefont{Chen}},
  \bibinfo{author}{\bibfnamefont{C.~S.} \bibnamefont{Li}}, \bibnamefont{and}
  \bibinfo{author}{\bibfnamefont{H.}~\bibnamefont{Zhang}},
  \bibinfo{journal}{Phys.\ Rev.\ Lett.} \textbf{\bibinfo{volume}{106}},
  \bibinfo{pages}{201801} (\bibinfo{year}{2011}{\natexlab{c}}),
  \eprint{1101.5625}.

\bibitem[{\citenamefont{Aad et~al.}(2012{\natexlab{b}})}]{Aad:2012bb}
\bibinfo{author}{\bibfnamefont{G.}~\bibnamefont{Aad}} \bibnamefont{et~al.}
  (\bibinfo{collaboration}{ATLAS Collaboration}), \bibinfo{journal}{J.\ High
  Energy Phys.} \textbf{\bibinfo{volume}{1204}}, \bibinfo{pages}{069}
  (\bibinfo{year}{2012}{\natexlab{b}}), \eprint{1202.5520}.

\bibitem[{\citenamefont{Chatrchyan
  et~al.}(2012{\natexlab{b}})}]{Chatrchyan:2012sa}
\bibinfo{author}{\bibfnamefont{S.}~\bibnamefont{Chatrchyan}}
  \bibnamefont{et~al.} (\bibinfo{collaboration}{CMS Collaboration}),
  \bibinfo{journal}{J.\ High Energy Phys.} \textbf{\bibinfo{volume}{1208}},
  \bibinfo{pages}{110} (\bibinfo{year}{2012}{\natexlab{b}}),
  \eprint{1205.3933}.

\bibitem[{\citenamefont{Grinstein et~al.}(2011)\citenamefont{Grinstein, Kagan,
  Trott, and Zupan}}]{Grinstein:2011yv}
\bibinfo{author}{\bibfnamefont{B.}~\bibnamefont{Grinstein}},
  \bibinfo{author}{\bibfnamefont{A.~L.} \bibnamefont{Kagan}},
  \bibinfo{author}{\bibfnamefont{M.}~\bibnamefont{Trott}}, \bibnamefont{and}
  \bibinfo{author}{\bibfnamefont{J.}~\bibnamefont{Zupan}},
  \bibinfo{journal}{Phys.\ Rev.\ Lett.} \textbf{\bibinfo{volume}{107}},
  \bibinfo{pages}{012002} (\bibinfo{year}{2011}), \eprint{1102.3374}.

\bibitem[{\citenamefont{Gresham et~al.}(2013)\citenamefont{Gresham, Shelton,
  and Zurek}}]{Gresham:2012kv}
\bibinfo{author}{\bibfnamefont{M.}~\bibnamefont{Gresham}},
  \bibinfo{author}{\bibfnamefont{J.}~\bibnamefont{Shelton}}, \bibnamefont{and}
  \bibinfo{author}{\bibfnamefont{K.~M.} \bibnamefont{Zurek}},
  \bibinfo{journal}{J.\ High Energy Phys.} \textbf{\bibinfo{volume}{1303}},
  \bibinfo{pages}{008} (\bibinfo{year}{2013}), \eprint{1212.1718}.

\bibitem[{\citenamefont{Han et~al.}(2012)\citenamefont{Han, Liu, Wu, Yang, and
  Zhang}}]{Han:2012dd}
\bibinfo{author}{\bibfnamefont{C.}~\bibnamefont{Han}},
  \bibinfo{author}{\bibfnamefont{N.}~\bibnamefont{Liu}},
  \bibinfo{author}{\bibfnamefont{L.}~\bibnamefont{Wu}},
  \bibinfo{author}{\bibfnamefont{J.~M.} \bibnamefont{Yang}}, \bibnamefont{and}
  \bibinfo{author}{\bibfnamefont{Y.}~\bibnamefont{Zhang}}
  (\bibinfo{year}{2012}), \eprint{1212.6728}.

\bibitem[{\citenamefont{Wang and Han}(2012)}]{Wang:2012zv}
\bibinfo{author}{\bibfnamefont{L.}~\bibnamefont{Wang}} \bibnamefont{and}
  \bibinfo{author}{\bibfnamefont{X.-F.} \bibnamefont{Han}},
  \bibinfo{journal}{J.\ High Energy Phys.} \textbf{\bibinfo{volume}{1205}},
  \bibinfo{pages}{088} (\bibinfo{year}{2012}), \eprint{1203.4477}.

\bibitem[{\citenamefont{Alvarez et~al.}(2011)\citenamefont{Alvarez, Da~Rold,
  and Szynkman}}]{Alvarez:2010js}
\bibinfo{author}{\bibfnamefont{E.}~\bibnamefont{Alvarez}},
  \bibinfo{author}{\bibfnamefont{L.}~\bibnamefont{Da~Rold}}, \bibnamefont{and}
  \bibinfo{author}{\bibfnamefont{A.}~\bibnamefont{Szynkman}},
  \bibinfo{journal}{J.\ High Energy Phys.} \textbf{\bibinfo{volume}{1105}},
  \bibinfo{pages}{070} (\bibinfo{year}{2011}), \eprint{1011.6557}.

\bibitem[{\citenamefont{Brodsky and Wu}(2012)}]{Brodsky:2012ik}
\bibinfo{author}{\bibfnamefont{S.~J.} \bibnamefont{Brodsky}} \bibnamefont{and}
  \bibinfo{author}{\bibfnamefont{X.-G.} \bibnamefont{Wu}},
  \bibinfo{journal}{Phys.\ Rev.} \textbf{\bibinfo{volume}{D85}},
  \bibinfo{pages}{114040} (\bibinfo{year}{2012}), \eprint{1205.1232}.

\end{thebibliography}
  
\end{document}